\documentclass{aa}
\usepackage{natbib}
\bibpunct{(}{)}{;}{a}{}{,} 

\usepackage[english]{babel}           
\usepackage{graphicx,epsfig}
\usepackage{txfonts}
\usepackage{mathrsfs}
\usepackage{lscape}
\usepackage{eucal}
\usepackage{pifont}
\usepackage{rotating}
\usepackage{multirow}
\usepackage{tikz}
\usepackage{hyperref}
\usepackage{longtable}
\usepackage[fleqn]{amsmath}
\usepackage{breqn}

\def \cc    {\ifmmode{\,{\rm cm}^{-3}}\else{$\,{\rm cm}^{-3}$}\fi}
\def \cq    {\ifmmode{\,{\rm cm}^{-2}}\else{$\,{\rm cm}^{-2}$}\fi}
\def \mic   {\ifmmode{\,\mu{\rm m}}\else{$\mu$m}\fi}
\def \eccs  {\ifmmode{\,{\rm erg}\,{\rm cm}^{-3} {\rm s}^{-1}}\else{$\,{\rm erg}\,{\rm cm}^{-3} {\rm s}^{-1}$}\fi}
\def \ecc   {\ifmmode{\,{\rm erg}\,{\rm cm}^{-3}}\else{$\,{\rm erg}\,{\rm cm}^{-3}$}\fi}
\def \ecqs  {\ifmmode{\,{\rm erg}\,{\rm cm}^{-2}\,{\rm s}^{-1}\,{\rm 
             sr}^{-1}}\else{$\,{\rm erg}\,{\rm cm}^{-2}\,{\rm s}^{-1}\,{\rm sr}^{-1}$}\fi}
\def \ecss  {\ifmmode{\,{\rm erg}\,{\rm cm}^{-2}\,{\rm s}^{-1}}\else{$\,{\rm erg}\,{\rm cm}^{-2}\,{\rm s}^{-1}$}\fi}
\def \deg   {\ifmmode{^{\circ}}\else{$^{\circ}$}\fi} 
\def \pc    {\ifmmode{\,{\rm pc}}\else{$\,{\rm pc}$}\fi} 
\def \kms   {\ifmmode{\,{\rm km}\,{\rm s}^{-1}}\else{km s$^{-1}$}\fi} 
\def \kmspc {\ifmmode{\,{\rm km}\,{\rm s}^{-1}\,{\rm pc}^{-1}}\else{km s$^{-1}$ pc$^{-1}$}\fi} 
\def \MJysr {\ifmmode{\,{\rm MJy\,sr}^{-1}}\else{$\,{\rm MJy\,sr}^{-1}$}\fi} 
\def \Kkms  {\ifmmode{\,{\rm K\,km\,s}^{-1}}\else{$\,{\rm K\,km\,s}^{-1}$}\fi}
\def \epso{\ifmmode{\overline{\varepsilon}_{\rm obs}}\else{$\overline{\varepsilon}_{\rm obs}$}\fi}
\def \utM{\ifmmode{u_{\theta,{\rm M}}}\else{$u_{\theta,{\rm M}}$}\fi}
\def \urM{\ifmmode{u_{r,{\rm M}}}\else{$u_{r,{\rm M}}$}\fi}
\def \twCO{\ifmmode{\rm ^{12}CO}\else{$\rm^{12}CO$}\fi} 
\def \thCO{\ifmmode{\rm ^{13}CO}\else{$\rm^{13}CO$}\fi} 
\def \CeiO{\ifmmode{\rm C^{18}O}\else{$\rm C^{18}O$}\fi} 
\def \twCN{\ifmmode{\rm ^{12}CN}\else{$\rm^{12}CN$}\fi} 
\def \thCN{\ifmmode{\rm ^{13}CN}\else{$\rm^{13}CN$}\fi} 
\def \HdCO{\ifmmode{\rm H_{2}CO}\else{$\rm H_{2}CO$}\fi} 
\def \twHdCO{\ifmmode{\rm ^{12}H_{2}CO}\else{$\rm^{12}H_{2}CO$}\fi} 
\def \thHdCO{\ifmmode{\rm ^{13}H_{2}CO}\else{$\rm^{13}H_{2}CO$}\fi} 
\def \twC{\ifmmode{\rm ^{12}C}\else{$\rm^{12}C$}\fi} 
\def \thC{\ifmmode{\rm ^{13}C}\else{$\rm^{13}C$}\fi} 
\def \Hp{\ifmmode{\rm H^+}\else{$\rm H^+$}\fi} 
\def \Cp{\ifmmode{\rm C^+}\else{$\rm C^+$}\fi} 
\def \Sp{\ifmmode{\rm S^+}\else{$\rm S^+$}\fi} 
\def \Op{\ifmmode{\rm O^+}\else{$\rm O^+$}\fi} 
\def \CFp{\ifmmode{\rm CF^+}\else{$\rm CF^+$}\fi}
\def \CHp{\ifmmode{\rm CH^+}\else{$\rm CH^+$}\fi}
\def \CHdp{\ifmmode{\rm CH_2^+}\else{$\rm CH_2^+$}\fi}
\def \CHtp{\ifmmode{\rm CH_3^+}\else{$\rm CH_3^+$}\fi} 
\def \SHp{\ifmmode{\rm SH^+}\else{$\rm SH^+$}\fi}
\def \SHdp{\ifmmode{\rm SH_2^+}\else{$\rm SH_2^+$}\fi}
\def \SHtp{\ifmmode{\rm SH_3^+}\else{$\rm SH_3^+$}\fi}
\def \twCHp{\ifmmode{\rm ^{12}CH^+}\else{$\rm^{12}CH^+$}\fi}
\def \thCHp{\ifmmode{\rm ^{13}CH^+}\else{$\rm^{13}CH^+$}\fi}
\def \CtH{\ifmmode{\rm C_2H}\else{$\rm C_2H$}\fi} 
\def \CthHt{\ifmmode{\rm C_3H_2}\else{$\rm C_3H_2$}\fi} 
\def \Htp{\ifmmode{\rm H_3^+}\else{$\rm H_3^+$}\fi} 
\def \COp{\ifmmode{\rm CO^+}\else{$\rm CO^+$}\fi} 
\def \HCOp{\ifmmode{\rm HCO^+}\else{$\rm HCO^+$}\fi} 
\def \HtOp{\ifmmode{\rm H_3O^+}\else{$\rm H_3O^+$}\fi} 
\def \HCfiN{\ifmmode{\rm HC_5N}\else{$\rm HC_5N$}\fi} 
\def \wat{\ifmmode{\rm H_2O}\else{$\rm H_2O$}\fi} 
\def \HdO{\ifmmode{\rm H_2O}\else{$\rm H_2O$}\fi} 
\def \OHp{\ifmmode{\rm OH^+}\else{$\rm OH^+$}\fi} 
\def \HdOp{\ifmmode{\rm H_2O^+}\else{$\rm H_2O^+$}\fi} 
\def \HtOp{\ifmmode{\rm H_3O^+}\else{$\rm H_3O^+$}\fi} 
\def \NHd{\ifmmode{\rm NH_2}\else{$\rm NH_2$}\fi} 
\def \NHtrois{\ifmmode{\rm NH_3}\else{$\rm NH_3$}\fi} 
\def \oxy{\ifmmode{\rm O_2}\else{$\rm O_2$}\fi} 
\def \HH{\ifmmode{\rm H_2}\else{$\rm H_2$}\fi}
\def \Jone{\ifmmode{\rm {(J=1--0)}}\else{{(J=1--0)}}\fi} 
\def \Jtwo{\ifmmode{\rm {(J=2--1)}}\else{{(J=2--1)}}\fi} 
\def \Jthr{\ifmmode{\rm {(J=3--2)}}\else{{(J=3--2)}}\fi} 
\def \Jfou{\ifmmode{\rm {(J=4--3)}}\else{{(J=4--3)}}\fi} 
\def \Jfiv{\ifmmode{\rm {J=4--3}}\else{{J=4--3}}\fi} 
\def \Ta{\ifmmode{\rm T_A}\else{$\rm T_A$}\fi} 
\def \Tas{\ifmmode{\rm T_A^*}\else{$\rm T_A^*$}\fi} 
\def \Tmb{\ifmmode{\rm T_{mb}}\else{$\rm T_{mb}$}\fi} 
\def \Tr{\ifmmode{\rm T_r}\else{$\rm T_r$}\fi} 
\def \Trs{\ifmmode{\rm T_r^*}\else{$\rm T_r^*$}\fi}
\def \NHt{\ifmmode{N_{\rm H}}\else{$N_{\rm H}$}\fi}
\def \NH{\ifmmode{N({\rm H})}\else{$N({\rm H})$}\fi}
\def \NH2{\ifmmode{N({\rm H}_2)}\else{$N({\rm H}_2)$}\fi}
\def \NCH{\ifmmode{N({\rm CH})}\else{$N({\rm CH})$}\fi}
\def \NHF{\ifmmode{N({\rm HF})}\else{$N({\rm HF})$}\fi}
\def \dens{\ifmmode{n_{\rm H}}\else{$n_{\rm H}$}\fi}
\def \nCO{\ifmmode{n({\rm CO})}\else{$n({\rm CO})$}\fi}
\def \nHF{\ifmmode{n({\rm HF})}\else{$n({\rm HF})$}\fi}
\def \nH2{\ifmmode{n({\rm H}_2)}\else{$n({\rm H}_2)$}\fi}

\begin{document}

\title{Models of irradiated molecular shocks}

\author{
  B. Godard            \inst{1}, 
  G. Pineau des Forêts \inst{1,2}, 
  P. Lesaffre          \inst{1}, 
  A. Lehmann           \inst{1}, 
  A. Gusdorf           \inst{1}, \and
  E. Falgarone         \inst{1}
}

\institute{
Observatoire de Paris, École normale supérieure, Université PSL, Sorbonne Université, CNRS, LERMA, F-75005, Paris, France 
\and
Institut d’Astrophysique Spatiale, Université Paris-Saclay, Orsay, 91405 Cedex, France
  }

 \date{Received 14 September 2018 / Accepted 20 November 2018}

\abstract{The recent discovery of excited molecules in starburst galaxies 
observed with ALMA and the Herschel space telescope has highlighted the necessity to 
understand the relative contributions of radiative and mechanical energies in the formation 
of molecular lines and explore the conundrum of turbulent gas bred in the wake of galactic 
outflows.}
{The goal of the paper is to present a detailed study of the propagation of low velocity 
(5 to 25 \kms) stationary molecular shocks in environments illuminated by an external
ultraviolet (UV) radiation field. In particular, we intend to show how the structure, 
dynamics, energetics, and chemical properties of shocks are modified by 
UV photons and to estimate how efficiently shocks can produce line emission.}
{We implemented several key physico-chemical processes in the Paris-Durham shock code  to improve the treatment of the radiative transfer and its impact on dust and 
gas particles. We propose a new integration algorithm to find the steady-state solutions 
of magnetohydrodynamics equations in a range of parameters in which the fluid evolves from 
a supersonic to a subsonic regime. We explored the resulting code over a wide range 
of physical conditions, which encompass diffuse interstellar clouds and hot and dense 
photon-dominated regions (PDR).}
{We find that C-type shock conditions 
cease to exist as soon as $G_0 > 0.2\,\, (\dens/ \cc)^{1/2}$. Such conditions trigger the 
emergence of another category of stationary solutions, called C*-type and CJ-type 
shocks, in which the shocked gas is momentarily subsonic along its trajectory. These 
solutions are shown to be unique for a given set of physical conditions and correspond 
to dissipative structures in which the gas is heated up to temperatures comprised between 
those found in C-type and adiabatic J-type shocks. High temperatures combined with the 
ambient UV field favour the production or excitation of a few molecular species to the detriment
of others, hence leading to specific spectroscopic tracers such as rovibrational 
lines of \HH\ and rotational lines of CH$^+$. Unexpectedly, the rotational lines of 
CH$^+$ may carry as much as several percent of the shock kinetic energy.}
{Ultraviolet photons are found to strongly modify the way the mechanical energy of interstellar 
shocks is processed and radiated away.
In spite of what intuition dictates, a strong external UV radiation field boosts the 
efficiency of low velocity interstellar shocks in the production of several molecular
lines which become evident tracers of turbulent dissipation.}


\authorrunning{B. Godard et al.}
\titlerunning{Models of irradiated molecular shocks} 
\maketitle

\section{Introduction}

The interstellar medium would be a nice, quiet, and somehow boring place to study if it 
were not constantly perturbed by strong dynamical events such as protostellar outflows, 
cloud collisions, supernovae, or galactic outflows. By interacting with the ambient 
medium, those events inject a large amount of mechanical energy in their surrounding 
environments at scales far larger than the diffusion length scales. A turbulent
cascade therefore develops in which part of the initial kinetic energy is processed and 
transferred to all scales in a distribution of lower velocity dynamical structures that
may carry a substantial fraction of the total kinetic energy.

The reprocessing of the initial available kinetic energy into a turbulent cascade is 
particularly well illustrated in the Stephan's Quintet galaxy collisions. Colliding 
galaxies at relative velocities of $\sim$$1000$ \kms\ not only trigger a large-scale 
initial shock clearly identified in X-rays \citep{Trinchieri2003} but also a myriad 
of structures at far lower velocities (shocks, shears, and vortices). These structures 
carry the kinetic signature of the large-scale collision and radiate in the 
rovibrational lines of \HH\ a total power that exceeds that seen in X-rays 
\citep{Appleton2006,Guillard2009}. This example depicts a very generic modus 
operandi of the interstellar medium. Low velocity dissipative structures such as 
low velocity shocks allow the production of specific molecules that are usually not 
abundant in ambient gas. The lines of these molecules carry valuable 
information regarding the event driving the injection of mechanical energy and 
the way this energy is distributed in the gas and radiated away 
\citep{Lesaffre2013,Lehmann2016}.

The physics and signatures of non-irradiated molecular shocks have been the subject 
of numerous theoretical works, describing the dynamics and thermochemistry of 
gas and dust particles in shock waves (e.g. \citealt{Hollenbach1979,Draine1980,Kaufman1996,
Flower2003,Walmsley2005,Chapman2006,Guillet2007,Flower2010a}). All these works have led to 
valuable predictions and opened a wide observational field to study the properties and 
track the evolution of a great variety of galactic and extragalactic environments, including young 
stellar objects and molecular outflows (e.g. \citealt{Gusdorf2008a,Gusdorf2008,Flower2013,
Nisini2015,Tafalla2015}), dense environments (e.g. \citealt{Pon2016}),
supernovae remnants (e.g. \citealt{Burton1990,reach2005}), and giant molecular 
clouds and filaments (e.g. \citealt{Pon2012,Louvet2016,Lee2016}). 
In many cases, the direct comparison of theoretical predictions and observations has 
proven to be a powerful tool to understand the nature of astrophysical sources and has given 
access, for instance, to their lifetimes, mass ejection rates, typical densities, and
magnetization.

Despite these successes, recent observations have revealed several chemical discrepancies that 
challenge the scope of the current models of molecular shocks. The relative emissions 
of oxygen-bearing species detected in the Orion H$_2$ peak 1 \citep{Melnick2015}, in 
the vicinity of low-mass protostars \citep{Kristensen2013,Kristensen2017,Karska2014}, 
in jets embedded in massive star-forming regions \citep{Leurini2015,Gusdorf2017}, or 
in supernovae remnants \citep{Snell2005,Hewitt2009} all show significant disagreements 
with the intensity ratios predicted in non-irradiated low velocity shocks. It is usually 
proposed that these discrepancies could be the trace of shocks illuminated by ultraviolet 
photons, emitted either by an external source of radiation or by the shock itself.

Several models have been developed in the past to follow the propagation and chemistry of 
self-irradiated molecular shocks (e.g. \citealt{Shull1979,Neufeld1989,Hollenbach1989}). 
In contrast, few theoretical works have been devoted so far to shocks irradiated by an 
external UV field, and more generally to the formation of shock waves in PDRs. In 
particular, while extensive studies of irradiated shocks have been performed both in 
diffuse interstellar clouds \citep{Monteiro1988,Lesaffre2013} and dense environments
\citep{Melnick2015}, only low irradiation conditions were explored (up to ten times
the standard interstellar radiation field).

The question of the impact of a strong UV radiation field on interstellar shocks is now 
magnified by the recent discovery of \CHp\ in submillimetre starburst galaxies at the peak 
of the star formation history \citep{Falgarone2017}. It is inferred that the broad line 
profiles seen in emission ($\sim$$1000$ \kms) likely trace the turbulent gas stirred up by 
galactic outflows. However, the fact that broad line wings appear only in \CHp\ 
and not in other molecular tracers (e.g. CO, H$_2$O; \citealt{Swinbank2010,Omont2013}) 
suggests that the chemistry at play in 
these regions of turbulent dissipation is peculiar and may be influenced by the strong 
radiation emitted by massive star-forming regions. The possible entwinement 
of radiative and mechanical energies raises the broader question of how these energies are 
actually processed in the interstellar medium. Building models capable of studying such 
environments has therefore become paramount to establish the energy budget of external 
galaxies and understand the relative importance of mechanical and radiative sources in 
the formation and excitation of molecules.

In this paper we present a detailed study of low velocity ($\leqslant 25$ \kms) molecular 
shocks irradiated 
by an external source of ultraviolet photons. The numerical method and physical 
processes taken into account are described in Sect. \ref{Sect-Numeric}. The specific 
dynamical, thermal, and chemical properties of irradiated shocks are presented in Sects. 
\ref{Sect-dynamics} and \ref{Sect-chemistry}. Open questions and perspectives
are addressed in Sects. \ref{Sect-discussion} and \ref{Sect-conclusion}.

\section{Framework and physical ingredients} \label{Sect-Numeric}

\begin{figure}
\begin{center}
\includegraphics[width=9.0cm,angle=0]{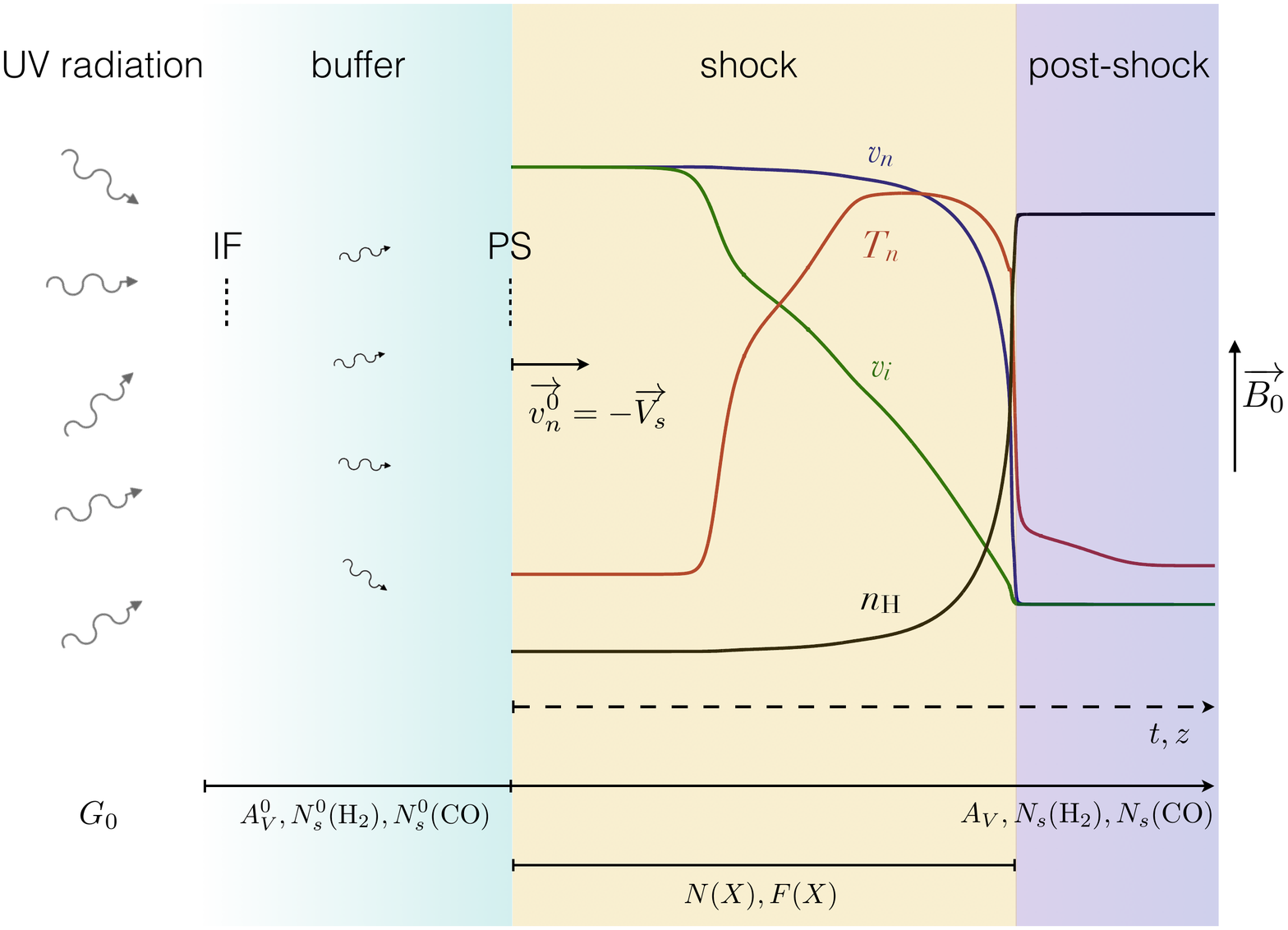}
\caption{Schematic view of the shock model geometrical assumptions in the shock 
frame. The positions of the ionization front and of the 
pre-shock are indicated as IF and PS.}
\label{Fig-scheme-shock}
\end{center}
\end{figure}

The model presented in this work is based on the Paris-Durham shock code 
\citep{Flower2003}, a public numerical tool\footnote{available on the ISM plateform 
\url{https://ism.obspm.fr}.} designed to compute the dynamical, thermal, and chemical 
evolution of interstellar matter in a steady-state plane-parallel shock wave. We 
further improve the version recently developed by \citet{Lesaffre2013}, which is built to
follow the effects of moderate ultraviolet irradiation, by including additional 
fundamental processes of PDR physics to study the propagation and chemistry 
of molecular shocks over a broad range of irradiation conditions.

\subsection{Geometry and main parameters}

A schematic view of the plane-parallel geometry adopted in the model is shown on Fig. 
\ref{Fig-scheme-shock}. Following the prescription of \citet{Lesaffre2013}, we assume 
that a plane-parallel shock wave propagates with a speed ${\bf V}_{\rm s}$ with respect
to the pre-shock gas, in the direction 
perpendicular to the illuminated surface of the gas and to the ambient magnetic field
${\bf B_0}$. The shock wave is irradiated 
upstream by an isotropic flux of UV photons equal to the standard interstellar radiation 
field \citep{Mathis1983} scaled with a parameter $G_0$. A plane-parallel layer of 
gas and dust (hereafter designated as the "buffer") is also assumed to stand upstream 
of the shock reducing the UV photon flux that reaches the pre-shock medium over a distance 
set by its visible extinction $A_V^0$ or equivalently its total hydrogen column density 
$N_{\rm H}^0$.

The last point of the buffer corresponds to the pre-shock medium and marks the origin of 
time $t$ and distance $z$ for the computation of the shock. The visible extinction $A_V$ 
and the self-shielding column densities of \HH\ and CO in the shock, $N_s(\HH)$ and 
$N_s({\rm CO})$, are integrated over the entire structure, including the buffer,
to account correctly for the total absorption of UV photons. However, 
the output column density $N({\rm X})$ and line flux $F({\rm X})$ of 
any species X shown throughout the paper are integrated over the shock only, ignoring 
both the buffer and post-shock medium. 
Details on the computation of the timescale 
$t_{\rm s}$ and size $z_{\rm s}$ of the shock are given in Sect. \ref{Sect-size}.

In practice, we first compute the radiative transfer and chemical and thermal 
structures of the buffer with the Paris-Durham shock code by following, in a Lagrangian
frame, a fluid particle moving away from the ionization front at a small and constant 
velocity until it reaches the position PS (see Fig. \ref{Fig-scheme-shock})
where it enters the shock.
Using the physical properties of the pre-shock gas as initial conditions, we then run 
the code in its classical configuration and compute the time-dependent dynamical, 
chemical, and thermal evolution of matter in the shock.

The main parameters of the model and the range of values explored in this work are 
given in Table \ref{Tab-main}. Compared to Table 1 of \citet{Lesaffre2013}, we
removed $N_s^0(\HH)$ and $N_s^0({\rm CO})$ from the list of parameters because
those quantities, which correspond to the column densities of \HH\ and CO in the
buffer layer, are now self-consistently calculated by the code when computing the
properties of the pre-shock gas.


\begin{table*}
\begin{center}
\caption{Main parameters of the shock code, standard model, and range of values explored in this work.}
\label{Tab-main}
\begin{tabular}{l r r r l}
\hline
name & standard & range & unit & definition \\
\hline
$n_{\rm H}$    &                 $10^4$ & $10^2$ $-$ $10^5$    & cm$^{-3}$   & pre-shock proton density$^a$ \\
$G_0$          &                      1 & 0 $-$ $10^4$         &             & radiation field scaling factor$^b$ \\
$A_V^0$        & $10^{-2}$ \& $10^{-1}$ & $10^{-2}$ $-$ $10^1$ &             & pre-shock visual extinction \\
$u_{\rm turb}$ &                      1 & $-$                  & km s$^{-1}$ & turbulent velocity \\
$\zeta_{\HH}$  &    $3 \times 10^{-17}$ & $-$                  & s$^{-1}$    & \HH\ cosmic-ray ionization rate \\
$V_{\rm s}$    &                     10 & 5 $-$ 25             & km s$^{-1}$ & shock velocity \\
$b$            &                      1 & 0.1 $-$ 10           &             & magnetic parameter$^c$ \\
\hline
\end{tabular}
\begin{list}{}{}
(a) defined as $\dens=n({\rm H}) + 2n(\HH)$.\\
(b) the scaling factor $G_0$ is applied to the standard ultraviolet radiation field of \citet{Mathis1983}.\\
(c) sets the initial transverse magnetic field $\frac{B_0}{1 \mu{\rm G}} = b \times (\frac{\dens}{1 \cc})^{0.5}$.
\end{list}
\end{center}
\end{table*}

\subsection{Grains and PAHs}

As described in Appendix \ref{Append-grains}, we assume that the grains at the 
ionization front (IF in Fig. \ref{Fig-scheme-shock}) follow
a power-law size distribution. Such distribution is subsequently modified in the
buffer and in the shock as erosion, adsorption, and desorption processes take place. 
The initial range of sizes considered, the assumed mass densities of the grain cores 
and mantles, and the chemical composition of the cores are all given in 
Table \ref{Tab-grain}. Conversely, we model PAHs as single particles of size 6.4 
\AA\ and chemical composition C$_{54}$H$_{18}$. To allow comparisons with 
the model developed by \citet{Lesaffre2013}, a fixed abundance $n({\rm PAH})/\dens
=10^{-6}$ is adopted throughout the entire paper\footnote{\citet{Draine2007} 
find that a PAH-to-dust mass ratio of 4.6\% is required to explain the observed 
galactic mid-infrared emission. For a dust-to-gas mass ratio of 0.01, this 
corresponds to a PAH abundance of $9.7 \times 10^{-7}$, which is close to the value
adopted in this work.}.

\subsection{Ultraviolet continuum radiative transfer}

The attenuation of the UV photons flux throughout the geometry shown in Fig. 
\ref{Fig-scheme-shock} is solved using a simplified model of radiative transfer. 
We only consider absorption and neglect all emission and diffusion processes
by gas and dust. Under these approximations, the specific monochromatic intensity 
$I$ at a wavelength $\lambda$ propagating in the cloud satisfies a reduced radiative
transfer equation
\begin{equation}
{\rm cos}(\theta) \frac{\partial I(z,\theta,\lambda)}{\partial z} = - \kappa(z,\lambda) I(z,\theta,\lambda)
,\end{equation}
where $z$ is the distance to the border of the cloud, $\theta$ is the angle of the
propagation direction (with respect to the direction perpendicular to the cloud surface), 
and $\kappa = \kappa_D + \kappa_G$ is the sum of the dust and gas absorption coefficients.
Integrating this equation over the distance $z$ leads to a simple relation
\begin{equation} \label{Eq-radtran}
I(z,\theta,\lambda) = I_0(\theta,\lambda) \,\, {\rm exp} \left[ -\int_0^z 
\frac{\kappa(s,\lambda)}{{\rm cos}(\theta)} ds \right]
\end{equation}
between $I$ and $I_0$, the specific monochromatic intensity at the border of the
cloud. We define
\begin{equation}
\tau_\lambda = \int_0^z \frac{\kappa(s,\lambda)}{{\rm cos}(\theta)} ds
\end{equation}
and
\begin{equation}
A_\lambda = 2.5 \, {\rm log}_{10}(e) \, \tau_\lambda
,\end{equation}
respectively, as the optical depth and the extinction along the direction $\theta$.
The monochromatic mean intensity, averaged over all directions, can be written 
\begin{equation}
J(z,\lambda) = \frac{1}{2} \int_0^{\pi/2} I(z,\theta,\lambda) \, {\rm sin}(\theta) \, d\theta
\end{equation}
and the UV photon flux integrated between $\lambda_{\rm min}=911$ \AA\ and 
$\lambda_{\rm max}=2400$ \AA\
\begin{equation}
F(z) = 2 \pi \int_0^{\pi/2} \int_{\lambda_{\rm min}}^{\lambda_{\rm max}} \frac{\lambda}{hc} 
I(z,\theta,\lambda) \, {\rm sin}(\theta) \, d\theta \, d\lambda
.\end{equation}

Absorption of UV photons by grain 
particles is calculated considering grains with density $n_g$ and a single size $a_g$,
where $n_g$ is derived from the size distribution while
\begin{equation}
a_g = \sqrt{\langle r_m^2 \rangle}
\end{equation}
is chosen as the size representative of the cross section of grains covered by ice mantles 
(see Appendix \ref{Append-grains}). Within this framework, the dust attenuation coefficient 
at position $z$ is therefore 
\begin{equation}
\kappa_D(z,\lambda) = \pi \, a_g^2(z) \, Q_{\rm abs}(a_g,\lambda) \, n_g,
\end{equation}
where $Q_{\rm abs}(a_g,\lambda)$ is the absorption coefficient of the grains. For
simplicity, we adopt the absorption coefficients of spherical graphite grains
of radius $a_g$ derived by \citet{Draine1984} and \citet{Laor1993}.

The absorption of UV photons by the gas is finally taken into account considering
only continuum processes, i.e. the photodissociation and photoionization of atoms 
and molecules. The gas attenuation coefficients are thus
\begin{equation}
\kappa_G(z,\lambda) = \sum_i \sigma({\rm X}_i,\lambda) \, n({\rm X}_i)
,\end{equation}
where $\sigma({\rm X}_i,\lambda)$ is the photodissociation or photoionization cross
section of species ${\rm X}_i$. The code is written to potentially include all cross 
sections available in the Leiden database\footnote{\url{http://www.strw.leidenuniv.nl/~ewine/photo}}
and recently updated by \citet{Heays2017}. In this paper, however, and to reduce the 
complexity of the problem, $\kappa_G(z,\lambda)$ is computed including only the
photoionization of C, S, C$_2$, CH, OH, and H$_2$O, and the photodissociation of 
C$_2$, C$_3$, CH, OH, and H$_2$O. We chose this approach to treat properly the ionization
of the main atomic compounds and the destruction of several carbon and oxygen bearing 
species, which are discussed in Sect. \ref{Sect-chemistry}. All the other photodestruction 
processes are implemented in the chemistry alone and not in the radiative transfer.

With these settings, the radiative transfer equation (Eq. \ref{Eq-radtran}) is
solved along 20 directions spread between $\theta=0$ and $\theta=\pi/2$.
The discretization leads to an error in the calculation of the UV photon flux of about 
8\% for an isotropic radiation field (at the ionization front) and of less than 10\% 
deeper in the cloud. We 
present in Appendix \ref{Append-PDR} a detailed comparison between our approach 
and the predictions of the Meudon PDR code \citep{Le-Petit2006} for the same 
environment. With the grain parameters described in Appendix \ref{Append-grains}, 
the prescriptions adopted in this work are able to capture the physical structure and 
chemical transitions obtained in state-of-the art PDR codes\footnote{Quantitatively, 
there is a deviation in the computation of the UV photon flux of less than 50 \% for 
$A_V \leqslant 3$.} 
\citep{Le-Petit2006,Le-Bourlot2012}.

\subsection{Photodestruction processes} \label{Sect-photodes}

Photodestruction of atoms and molecules can involve continuum processes, i.e. occur 
through the absorption of photons over a continuous range of energy, or line processes, 
i.e. occur through the absorption of photons in atomic or molecular lines at given 
wavelengths (e.g. \citealt{van-Dishoeck1988,van-Dishoeck2006}). In this work we take
advantage of the computation of the radiation field spectrum to perform detailed 
treatments of both processes.

Following the approach of the Meudon PDR code \citep{Le-Petit2006}, the continuum 
absorption mechanisms are treated in two different ways. If the photodestruction cross 
section of a species ${\rm X}_i$ is known and included in the code, the corresponding 
photoreaction rate is computed by direct integration over the radiation field intensity
\begin{equation}
k_\gamma({\rm X}_i,z) = 2 \pi \int_0^{\pi/2} \int_{\lambda_{\rm min}}^{\lambda_{\rm th}} 
\sigma ({\rm X}_i,\lambda) \, \frac{\lambda}{hc} I(z,\theta,\lambda) \, {\rm sin}(\theta) 
\, d\theta \, d\lambda
,\end{equation}
where $\lambda_{\rm th}$ is the ionization or dissociation threshold (expressed in 
wavelength). If not, the photoreaction rate is calculated using the analytical 
fits provided by \citet{Heays2017}. In PDR physics, it is customary to express such
rates as functions of the visual extinction $A_V$ with the form
\begin{equation} \label{Eq-photodiss}
k_\gamma({\rm X}_i,z) = G_0 \, \alpha \, {\rm exp} (-\beta A_V)
,\end{equation}
where $\alpha$ and $\beta$ are constant coefficients. Such prescription, however, can
only be applied if the visual extinction, used in this case as a proxy for the absorption of UV
photons, has the same significance as the visual extinction deduced from the
radiative transfer, or, in other words, if the underlying dust extinction curve used
in Eq. \ref{Eq-photodiss} and that used in the radiative transfer are identical.
To allow a coherent treatment of the photochemistry and the extinction based solely
on the dust composition and size distribution chosen in input, we substitute 
relation \ref{Eq-photodiss} by a more versatile expression
\begin{equation} \label{Eq-photodiss2}
k_\gamma({\rm X}_i,z) = G_0 \, \alpha \, \left( \frac{F(z)}{F_{\rm ISRF}} \right)^{\frac{\beta}{2.8}},
\end{equation}
where $F_{\rm ISRF}$ is the UV photon flux associated with the isotropic standard interstellar radiation field  (ISRF) for
which the $\alpha$ and $\beta$ coefficients are computed, and 
the factor $2.8$ is deduced from a power-law fit of the photon flux $F(z)$ as a 
function of $A_V$ for $0 \leqslant A_V \leqslant 3$ assuming a galactic extinction 
curve.

A different treatment is required for the photodissociations of \HH\ and CO 
that occur in lines and therefore involve self- and cross-shielding processes 
\citep{Abgrall1992,Lee1996,Le-Petit2002}. In the previous version of the code, 
these aspects were taken into account using the analytical shielding expressions 
tabulated by \citet{Lee1996} and \citet{Draine1996} as functions of the column 
densities of \HH\ and CO at the current point \citep{Lesaffre2013}. To improve
the treatment of the photodissociation of \HH, we part from this approach 
and compute the absorption of UV photons by electronic lines of molecular hydrogen 
and the subsequent dissociation probabilities. This treatment is performed using 
the FGK approximation \citep{Federman1979} and including all the UV lines of \HH\ 
in the Lyman and Werner bands. At each point $z$ the code computes the central optical 
depth of each of those lines by integrating the abundance of \HH\ towards the
radiation source, taking into account the thermal broadening
and microturbulent motions (set with a parameter $u_{\rm turb}$, see Table 
\ref{Tab-main}) in the Doppler spreading of the line profiles. These
optical depths are then used to attenuate the radiation field at the line
central wavelengths, hence reducing the \HH\ dissociation rate.

Admittedly, large velocity gradients such as those found in shock waves 
could have an impact on the computation of optical depths and therefore on 
the photodissociation rate of \HH\ \citep{Monteiro1988}. This effect is,
however, not yet included in the model.

\subsection{Chemical network} \label{Sect-ChemNet}

The treatment of chemistry is carried out starting with the network of \citet{Flower2015}. This network incorporates 139 species, including gas-phase and solid-phase compounds, and
about a thousand reactions. Regarding gas-phase reactions, both the sulphur and 
nitrogen chemistries are modified to take into account the recent updates of the 
reaction rates introduced in Table D1 of \citet{Neufeld2015} and Tables B1-B3
of \citet{Le-Gal2014}. 
Regarding interactions with grains, we expand the network to model thermal desorption 
and photodesorption processes in addition to the sputtering, cosmic-ray and secondary 
photons desorption mechanisms already included in the previous versions.

Following \citet{Hollenbach2009}, the thermal desorption rate per atom or molecule of 
a species ${\rm X}_i^*$ stuck on grain mantles is modelled as 
\begin{equation}
k_{\rm thd}({\rm X}_i^*,z) = 1.6 \times 10^{11} \sqrt{ \frac{E_i}{k} \frac{m_{\rm H}}{m_i} } 
\, {\rm exp} \left(-\frac{E_i}{k T_g(z)} \right),
\end{equation}
where $m_i$ is the mass of ${\rm X}_i^*$, $E_i$ its adsorption energy taken from
\citet{Hasegawa1993} and \citet{Aikawa1996}, and $T_g(z)$ is the grain temperature.
The photodesorption rate of ${\rm X}_i^*$ is written 
\begin{equation}
k_{\rm phd}({\rm X}_i^*,z) = Y_i \, F(z) \, \pi \, \langle r_m^2 \rangle \, n_g \, f_i,
\end{equation}
where $Y_i$ is the photodesorption yield set to $10^{-3}$ for all species 
\citep{Hollenbach2009}, and $f_i$ is the fraction of the surface adsorption 
sites occupied by ${\rm X}_i^*$.

The calculation of the desorption rates by interaction with secondary photons is
finally modified to be coherent with the expression used for grain ionization. 
Adopting the formalism of \citet{Flower2003}, we model the secondary 
photon desorption rate of species ${\rm X}_i^*$ as
\begin{equation}
k_{\rm scd}({\rm X}_i^*,z) = 0.15 \, \left[ \zeta_{\HH} n({\rm \HH}) + \zeta_{\rm H} n({\rm H}) \right] 
                          \frac{n({\rm \HH})}{n({\rm H}) + n({\rm \HH})} \, Y_i \, f_i.
\end{equation}
The values $\zeta_{\HH}$ and $\zeta_{\rm H}$ are the primary cosmic-ray ionization rates 
of \HH\ and H. $0.15 \, \left[ \zeta_{\HH} n({\rm \HH}) + \zeta_{\rm H} n({\rm H}) 
\right] \frac{n({\rm \HH})}{n({\rm H}) + n({\rm \HH})}$ is thus the number of secondary 
UV photons produced per second and per unit volume through the excitation of the Rydberg
states of \HH\ by cosmic-ray induced electrons.

Because our analysis is limited to low velocity shocks ($V_S \leqslant 25$ \kms) and low
pre-shock densities ($\dens \leqslant 10^5$ \cc), we neglect, throughout this paper, 
the effects of grain-grain interactions, such as shattering, vaporization, and 
coagulation (e.g. \citealt{Jones1996}). This approximation is supported by recent 
models of grain dynamics which show that the shattering and vaporization of large 
grains have a significant impact on the grain size distribution only for high 
velocity shocks ($V_S > 20$ \kms) propagating in dense media ($\dens > 10^4$ \cc)
(see Fig. 7 of \citealt{Guillet2009} and Fig. 4 of \citealt{Guillet2011}).
Similarly, the grain coagulation is found to marginally affect the grain size 
distribution in low density molecular shocks (see Fig. 14 of \citealt{Guillet2011}).

\subsection{H$_2$ radiative pumping} \label{sect_radpump}

Besides the thermal and chemical evolution of the gas, the code also computes the 
time-dependent populations of the rovibrational levels of \HH\ \citep{Flower2003}.
In addition to inelastic collisions with H, He, \HH, and H$^+$ and the probability 
of exciting \HH\ at formation on grain surfaces, which are two processes already taken into 
account, we include the excitation by radiative pumping of the electronic 
lines of molecular hydrogen followed by fluorescence.

The radiative pumping is computed following the approach used to estimate the 
self-shielding and photodissociation rate (see Sect. \ref{Sect-photodes}). At each time 
step, the absorption of UV photons by the discrete lines of the Lyman (${\rm B} ^{1} 
\Sigma_u^+ - {\rm X} ^{1}\Sigma_g^+$) and Werner (${\rm C} ^{1}\Pi_u - {\rm X} ^{1}
\Sigma_g^+$) band systems and the subsequent cascades in the rovibrational levels of 
the fundamental electronic state are calculated. The line opacities, which reduce the
intensity of the radiation field capable of pumping the electronic states, are integrated 
over the column density $N_s(\HH)$ (see Fig. \ref{Fig-scheme-shock}) using the FGK
approximation (\citealt{Federman1979}).

\subsection{Dust charge and temperature}

Grains are considered to be either neutral, negatively, or positively charged. As 
done for the radiative transfer, the abundances of these different states are 
computed assuming grains with a single size $a_g$. Photoelectric ejection, 
electron attachment, and charge exchange between ions and grains are modelled 
adopting the formalism of \citet{Bakes1994}. The photoelectric ejection rate of a 
grain of charge $Z$ is thus written
\begin{equation}
k_{\rm phe}(Z,z) = \int_\Omega \int_{\lambda} 
Y_{\rm ion}(a_g,Z) \, \pi \, a_g^2(z) \, Q_{\rm abs}(a_g,\lambda) \, \frac{\lambda}{hc} 
I(z,\Omega,\lambda) d\Omega \, d\lambda,
\end{equation}
where $Y_{\rm ion}(a_g,Z)$ is the ionization yield, and the charge attachment rate
\begin{equation}
k_{\rm att}(Z,z) = n({\rm X_i}) \left(\frac{8kT_i}{\pi m_i} \right)^{1/2}
\pi a_g^2 \tilde{J}\left(\tau = \frac{a_g k T_i}{Z({\rm X_i})^2}, \nu = \frac{Ze}{Z({\rm X_i})} 
\right)
,\end{equation}
where $n({\rm X_i})$ and $Z({\rm X_i})$ are the abundance and charge of the colliding 
particle ${\rm X_i}$, and $T_i$ is the ion temperature. The function $\tilde{J}$ is 
described by \citet{Draine1987} and accounts for Coulomb interactions of the colliding 
system.

The heating rate of the gas induced by the photoelectric effect is modelled using the
efficiencies deduced by \citet{Bakes1994} for very small graphitic grains and PAHs
(Eqs. 42 and 43 of their paper).

At last, the temperature $T_d$ of grains of size $a_g$ is computed from the equilibrium 
between the radiative energy absorbed by the grain, the thermal exchange induced by 
collisions between grains and gas (at temperature $T_n$), and infrared radiative emission,
\begin{equation} \label{Eq-Td}
E_{\rm abs} + E_{\rm gg} = E_{\rm em}
,\end{equation}
where
\begin{equation}
E_{\rm abs} = \int_\Omega \int_{\lambda} \pi \, a_g^2(z) Q_{\rm abs}(a_g,\lambda) I(z,\Omega,\lambda) d\Omega \, d\lambda,
\end{equation}
\begin{equation}
E_{\rm gg} =  3.5 \times 10^{-34} \, \sqrt{T_n} \, (T_n - T_d) \, n_{\rm H}^2 / n_g,
\end{equation}
and
\begin{equation}
E_{\rm em} = \int_\Omega \int_{\lambda} \pi \, a_g^2(z) Q_{\rm abs}(a_g,\lambda) B_\lambda(T_d) d\Omega \, d\lambda,
\end{equation}
assuming that grains radiate as black-bodies with an intensity equal to the Planck 
function $B_\lambda(T_d)$. In the code, $T_d$ is computed at each time step by solving 
Eq. \ref{Eq-Td} using a Newton-Raphson scheme.

\subsection{Cooling}
Following \citet{Flower2003} and \citet{Lesaffre2013}, the cooling of the gas
is computed taking into account the excitation
of the fine structure lines and metastable lines of C, N, O, S, Si, C$^+$, N$^+$, 
O$^+$, S$^+$, and Si$^+$, and the rovibrational lines of \HH, OH, H$_2$O, NH$_3$, 
$^{12}$CO, and $^{13}$CO. The cooling resulting from the atomic lines and 
rovibrational lines of \HH\ is calculated in the optically thin limit. In the 
framework of a 1D plane-parallel geometry, this assumption is found to hold 
for the main coolant of the gas (namely \HH, C$^+$, and O) at a level better
than a few percent. The cooling through the rovibrational lines of OH and NH$_3$ is
calculated analytically using the prescription of \citet{Le-Bourlot2002}, while the 
cooling induced by the molecular lines of H$_2$O, $^{12}$CO, and $^{13}$CO is taken 
from the values tabulated by \citet{Neufeld1993}, which include opacity effects.

\section{The models: C, C*, CJ, and J stationary shocks} \label{Sect-dynamics}

The influence of the external UV field on the propagation and properties of interstellar
shocks is studied through several grids of models\footnote{The grids have been run on the
computing cluster Totoro funded by the ERC Advanced Grant MIST.} $-$ 3120 runs in total 
$-$ covering a broad range of physical
conditions. These include an initial pre-shock density varying between $10^2$ and $10^5$ \cc, an 
ambient UV radiation field with a scaling factor $G_0$ between $0$ and $10^4$, and a total 
hydrogen column density in the buffer between $2 \times 10^{19}$ and $2 \times 10^{22}$ 
cm$^{-2}$ (see Table \ref{Tab-main}). For reasons we detail below (see Sect. 
\ref{Sect-limits}), we limit our 
analysis to low velocity and moderately magnetized shocks, i.e. shocks with a speed 
comprised between 5 and 25 \kms\ and a pre-shock transverse magnetic field parameterized 
by $b$ (see Table \ref{Tab-main}) varying between 0.1 and 10.

\subsection{Pre-shock conditions}

\begin{figure*}[!ht]
\begin{center}
\includegraphics[width=16.0cm,trim = 1cm 2cm 1cm 1cm, clip,angle=0]{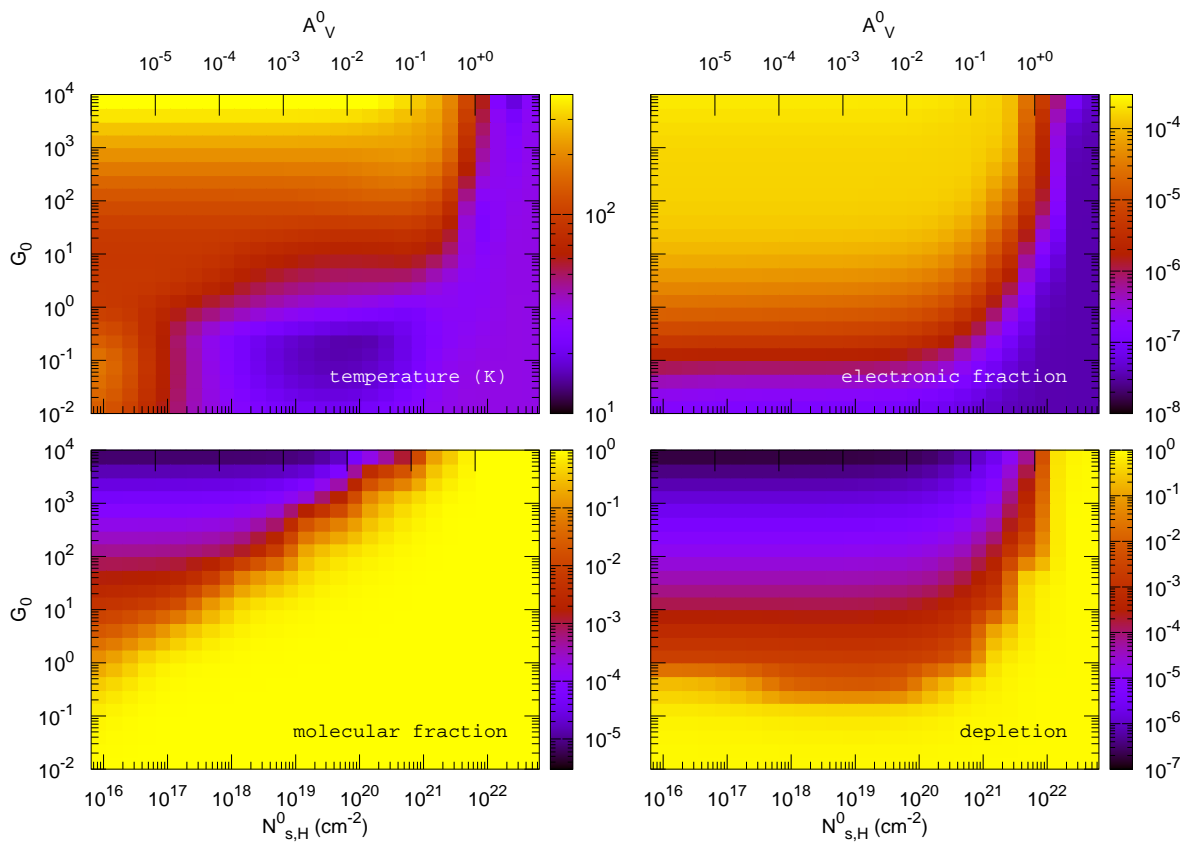}
\caption{Temperature (top left), electronic fraction (i.e. $n(e)/\dens$, top right), 
molecular fraction (i.e. $2n(\HH)/\dens$, bottom left), and depletion (bottom right) 
in the pre-shock medium as functions of $G_0$ and $N_{s,{\rm H}}^0$ (or $A^0_V$) for 
a density $\dens=10^4$ \cc. The depletion is computed as the fraction of the abundance 
of heavy elements locked up onto grain mantles.}
\label{Fig-preshock}
\end{center}
\end{figure*}

The range of pre-shock conditions explored in this work are summarized in Fig. 
\ref{Fig-preshock}, which shows the temperature, electronic fraction, molecular
fraction, and depletion of elements on grain surfaces in the pre-shock gas as 
functions of the input radiation field and size of the buffer. Since the 
amount of elements locked up onto grain surfaces in well-shielded environments
is highly uncertain \citep{Gibb2000}, two formalisms are considered: if there is 
no radiation field ($G_0=0$), we switch off all adsorption and desorption processes 
for the computation of the pre-shock conditions and adopt the chemical composition of grain 
mantles given in Table 2 of \citet{Flower2003}; conversely, if a radiation field with any 
strength is illuminating the gas and dust (as in Fig. \ref{Fig-preshock}), both adsorption 
and desorption processes are included. We note that this last assumption may lead 
to strong depletion at low radiation fields (bottom right panel of Fig. 
\ref{Fig-preshock}), which may be unrealistic compared to observations.

Fig. \ref{Fig-preshock} shows the well-known transition regions 
of PDRs discussed in many studies (i.e. \citealt{Hollenbach1999,Rollig2007}).
Within the FGK approximation (see Sect. \ref{sect_radpump}), the treatment of 
the radiative pumping of \HH\ rovibrational levels has a small influence on the 
general properties of the gas and on the pre-shock conditions shown in Fig.
\ref{Fig-preshock}. The most noticeable impact arises at large 
radiation fields ($G_0 \geqslant 10^3$) where the
pre-shock temperature and electronic fraction increase by more than 10 \% when
the pumping is included. While inconsequential over a wide range of parameters, 
this process is however paramount for shocks propagating in hot and dense PDRs
(i.e. $G_0 = 10^4$ and $\dens = 10^5$ \cc) in which the UV pumping strongly limits 
the cooling efficiency of \HH\ and induces a raise 
of the pre-shock temperature (see Fig. \ref{Fig-temp-add} 
of Appendix \ref{Append-add-figures}).


As shown in Fig. \ref{Fig-preshock}, the range of parameters chosen in this work 
allows us to explore shocks propagating in a wide variety of environments, spanning 
atomic to molecular gas, low and high ionization conditions, and cold and hot
pre-shock medium (from $\sim 20$ K to $\sim 1000$ K).
While extensive grids of shock models have been run over the 
entire parameter space, most of the analysis below is carried out varying one parameter
at a time around two standard models defined as follows: $\dens=10^4$ \cc, $G_0=1$, 
$V_S=10$ \kms, $b=1$, and $A_V^0=10^{-2}$ or $10^{-1}$. In order to highlight the 
impacts of the UV radiation field on the dynamics of the gas, a buffer visual 
extinction $A_V^0=10^{-2}$ is assumed throughout all Sect. \ref{Sect-dynamics}.
Conversely, the predictions on atomic and molecular emission presented in Sect.
\ref{Sect-chemistry} are obtained assuming $A_V^0=10^{-1}$. Such choice of standard 
models not only opens the exploration
of all the conditions shown in Fig. \ref{Fig-preshock} with a limited number of
runs but is also found to be sufficient to present most of our results on the physics 
of irradiated shocks in a synthetic manner. More exhaustive results are given in 
Appendix \ref{Append-add-figures} where we lay out additional figures and predictions 
in a blooming potpourri.

\subsection{Stationary magnetized molecular shocks} \label{Sect-phys-shocks}

\begin{figure*}[!ht]
\begin{center}
\includegraphics[width=16.0cm,trim = 2cm 2cm 2cm 2cm,angle=0]{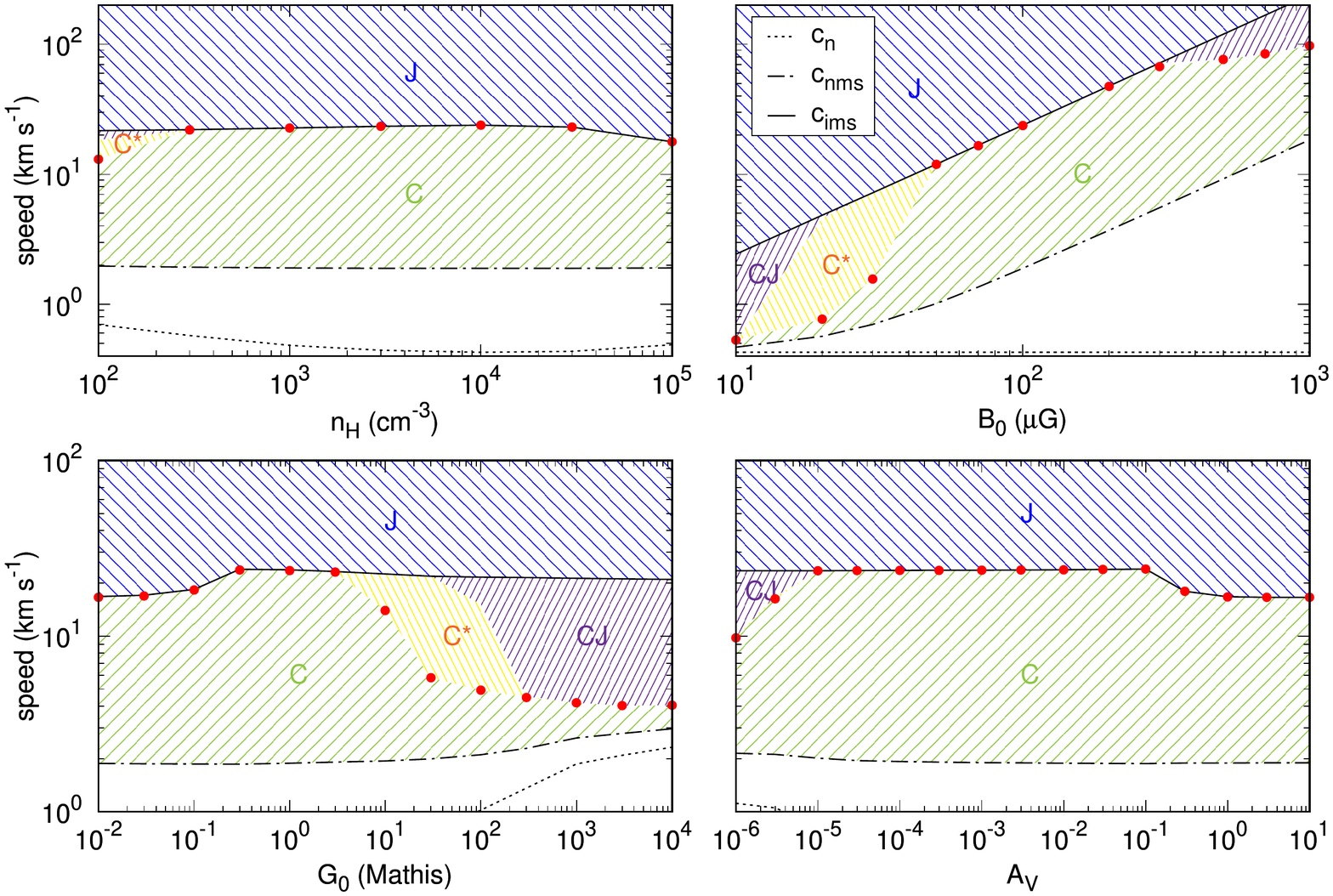}
\caption{Neutral sound speed (dotted) and neutral (dot-dashed) and ion (solid) 
magnetosonic speeds as functions of the density (upper left), transverse magnetic 
field intensity (upper right), UV radiation field intensity (lower left), and visual extinction (lower right). All the non-varying parameters are set to their 
standard values (see Table \ref{Tab-main}), and the standard value of the buffer visual 
extinction is set to $A_V^0=10^{-2}$. The critical velocities above which C-type shocks
cannot exist are shown with red points, while coloured areas highlight the range of
velocities at which C- (green), C*- (yellow), CJ- (violet), and J-type (blue) shocks
develop.}
\label{Fig-zermatt}
\end{center}
\end{figure*}

We recall that the standard model adopted in this section is defined by a
buffer visual extinction $A_V^0=10^{-2}$.
The physics of magnetized molecular shocks and the steady-state solutions 
allowed to propagate in the interstellar medium have been described in many studies 
(e.g. \citealt{Mullan1971,Draine1980,Kennel1990,Dopita2003}). 
As prescribed by the Paris-Durham code, we limit our study to stationary plane-parallel 
shock waves where the magnetic field is assumed to be 
transverse to the direction of propagation. It follows that the influence of UV 
photons on time-dependent shocks \citep{Chieze1998,Lesaffre2004} and on slow or 
intermediate steady-state shocks \citep{Draine1993,Lehmann2016a} are not addressed 
in this work.

A shock is a pressure-driven instability of fluid dynamics in which the kinetic energy 
of the flow is transferred from large scales to small scales where it dissipates primarily 
into heat. At steady-state, a shock structure connects two thermodynamical equilibrium points: 
the upstream unstable state and downstream stable state. The nature of the instability, 
the scales involved, and the forces at play in the evolution of the fluid between these 
two points depend on the velocity of the flow compared to its characteristic wave speeds. 
Although the expressions of those speeds are different depending on the physical properties 
of the gas, it is customary to discuss the physics of shocks in terms of the wave speeds 
derived for an isothermal magnetohydrodynamic fluid in the limit of weak and strong 
coupling between ionized and neutral species, i.e. the sound speeds and magnetosonic 
speeds in the ions ($c_{\rm i}$ and $c_{\rm ims}$) and in the neutrals ($c_{\rm n}$ and 
$c_{\rm nms}$)\footnote{We note that the characteristic speeds always verify $c_{\rm ims} > 
c_{\rm nms} > c_{\rm n}$.}
\begin{equation}
c_{\rm i}   = \left( \gamma_i P_i / \rho_i \right)^{1/2},
\end{equation}
\begin{equation}
c_{\rm n}   = \left( \gamma_n P_n / \rho_n \right)^{1/2},
\end{equation}
\begin{equation} \label{eq-cims}
c_{\rm ims} = \left( c_{\rm i}^2 + B^2 / 4 \pi \rho_i \right)^{1/2}, \quad {\rm and}
\end{equation}
\begin{equation}
c_{\rm nms} = \left( c_{\rm n}^2 + B^2 / 4 \pi \rho_n \right)^{1/2},
\end{equation}
where $\gamma_i$, $\gamma_n$, $P_i$, $P_n$, $\rho_i$, and $\rho_n$ are the heat capacity
ratio, pressure, and mass density 
of the ions and the neutrals and $B$ is the strength of the magnetic field. As we explain below (see Sect. \ref{Sect-gr-pahs}), we assume that all grains are coupled 
with the magnetic field: it follows that $\rho_i$ not only includes the mass density of 
ionized species but also that of dust particles.
Interstellar shocks occur when the upstream flow is dynamically unstable, which happens
when $V_S \geqslant c_{\rm n}$ (resp. $c_{\rm nms})$ in the limit of weak (resp. strong) 
coupling between ions and neutrals.

Assuming a fixed ionization fraction, the nature of a shock can be understood by solely 
discussing the impact of the magnetic field and gas cooling.

In very weak magnetic field environments, 
ions and neutrals are 
fully coupled with each other. In the frame of the shock, both fluids are decelerated 
through the combined actions 
of viscous stresses and the thermal pressure gradient induced by viscous heating. The 
gas quickly jumps from a supersonic to a subsonic regime, with respect to the neutral 
sound speed, over a distance of about 
a few mean free paths. The resulting structure is referred to as a J-type shock. Without 
magnetic field, the gas pressure of the post-shock flow is only thermal, which implies
that the downstream fluid is subsonic. If a magnetic field is present, magnetic pressure
increases in the post-shock. In this case, the final velocity of the flow may be supersonic 
if the magnetic pressure is large enough to stabilize the downstream flow.

When the magnetic field is sufficiently strong such that $c_{\rm ims} > V_S$, 
a magnetic precursor 
can develop: the Lorentz force decelerates the ions which decouple from the neutrals in the 
upstream flow. In turn, this decoupling induces a drag force that slows down the neutral
fluid. If the coupling between ions and neutrals is weak, the resulting effect is too
small to smooth the neutral velocity gradient:\ the viscous stresses and thermal 
pressure gradient remain the dominant forces involved in the motion of the neutrals. Two 
characteristic lengths therefore  appear: the magnetic precursor scale \citep{Draine1980} 
over which the ions decelerate continuously, and the viscous length where the neutrals
velocity jumps. Such a structure is called a CJ-type shock\footnote{We recall that all
the shocks described in this paper are stationary structures. The CJ-type shocks explored in this
work should thus not be identified to non-stationary CJ-type shocks often discussed in
the literature (e.g. \citealt{Chieze1998}).}. While the neutral fluid 
necessarily becomes subsonic at the jump, the downstream flow may be supersonic (for
reasons given above). In this case, the neutral fluid crosses the sonic speed twice,
at the jump position and somewhere along the downstream trajectory.

Increasing the magnetic field enlarges the size of the magnetic precursor. The drag
force applies for a longer time, smoothing the neutral velocity gradient and thermal
pressure gradient, thus reducing the strength of viscous stresses which may become 
negligible over the entire trajectory. When this happens, the thermal pressure gradient
and the ion-neutral friction slow down the neutrals continuously to the downstream 
point. Those kinds of shocks are separated into two classes. If the neutral fluid becomes 
subsonic along its trajectory, the structure is referred to as a C*-type shock, if not as 
a C-type shock. It follows that the downstream flow of a C-type shock is necessarily
supersonic. It is worth noting that the differentiation between C*- and C-type shocks 
is one of semantics and numerics only. These two types of shocks formally and physically correspond to
the same kind of structure with similar thermochemical evolutions (see Sect. 
\ref{Sect-profils}).

The impact of the cooling described by \citet{Chernoff1987} can be understood starting 
with a CJ-type shock. Increasing the cooling in this situation reduces the thermal 
pressure gradient and neutral velocity Laplacian, hence the strength of the viscous 
stresses. As in the previous case, if the viscous forces become small enough compared
to the drag force and thermal pressure gradient, either a C*-type or a C-type 
structure develops.

The nature of an interstellar shock is therefore the result of a subtle interplay between
the strength of the magnetic field and the ionization degree $-$ which control the size of 
the magnetic precursor and its capacity to slow down and heat the neutrals $-$ and the cooling 
rate. In practice single-fluid J-type shocks or bi-fluid C-type shocks are computed 
using a forward integration technique and starting from the pre-shock conditions (e.g.
\citealt{Flower2015}). Since this method is numerically unstable as soon as the neutral gas become
subsonic, a more sophisticated algorithm is used to compute CJ-type and C*-type shocks.
This algorithm, fully described in Appendix \ref{Append-method}, is based on the works 
of \citet{Chernoff1987} and \citet{Roberge1990} and combines forward integration techniques
with shooting methods.

\subsection{Existence of J-, CJ-, C*-, and C-type shocks}

The characteristic speeds $c_{\rm n}$, $c_{\rm ims}$, and $c_{\rm nms}$ computed in 
the pre-shock fluid and the domains of existence of the different kinds of stationary 
shocks are shown in Fig. \ref{Fig-zermatt} as functions of the pre-shock density, 
magnetic field, and irradiation conditions. Each parameter is explored around 
the standard model for a buffer visual extinction $A_V^0=0.01$ (see Table. \ref{Tab-main}).

\subsubsection{Critical speeds}

The minimal speed required for a shock to propagate is a problem of non-ideal 
magnetohydrodynamic. In the limit of low coupling between ions and neutrals, a shock develops if the 
perturbation applied to the medium travels faster than the neutral sound speed. Conversely, 
in the limit of strong coupling, a shock exists only if the perturbation travels faster 
than the neutral magnetosonic speed. While the intermediate case has been treated in
several numerical studies (e.g. \citealt{Balsara1996}), no analytical formula has ever 
been derived. To simplify, we identify in Fig. \ref{Fig-zermatt} the neutral magnetosonic 
speed $c_{\rm nms}$ as the minimal speed required to induce a shock. We do so because 
we found that C-type shocks below this limit ($\sim$$2$ \kms\ for the 
standard model) induce a relative velocity difference between the upstream and 
downstream flows smaller than 10\%.
Above this limit, the parameter space shown in Fig. \ref{Fig-zermatt} 
is divided in four different regions which set the domains of existence of J-, CJ-, C*-, 
and C-type shocks.

The limit between J-type and other kinds of molecular shocks is given by the ions 
magnetosonic speed, hence by the strength of the magnetic field and the mass 
density of the ionized flow in the pre-shock medium (Eq. \ref{eq-cims}). Assuming that 
all grains -- including the neutrals -- contribute to the inertia of the charged fluid 
(see \citealt{Guillet2007,Lesaffre2013}, and Sect. \ref{Sect-gr-pahs}), the mass density
of the ionized flow is dominated by the grains. With an initial transverse magnetic 
field $B_0$ proportional to $\sqrt{\dens}$ (see Table \ref{Tab-main}), $c_{\rm ims}$ 
therefore linearly depends on $B_0$, does not depend on the gas density, and softly 
depends on the dust-to-gas mass ratio. Without depletion, i.e. at large radiation 
fields or weak extinctions (bottom panels of Fig. \ref{Fig-preshock}), the mass of grains 
lies in the cores which is set by their composition and grain size 
distribution (see Table \ref{Tab-grain} and Appendix \ref{Append-grains}). Conversely, 
if the depletion becomes important, the mantles contribute to the grain mass; in the
limit of high depletion, i.e. at low radiation fields or large extinctions, the dust-to-gas 
mass ratio is increased by a factor of two and the ion magnetosonic speed is divided by 
a factor 1.4.

The range of existence of C-, C*-, and CJ-type shocks below the ions magnetosonic speed
can be explained in the light of the driving mechanisms of molecular shocks described in 
Sect. \ref{Sect-phys-shocks}. The impact of the radiation field on a C-type shock 
is understood by two main processes: the enhancement of the photoionization mechanisms 
and the increase of the photodissociation of \HH. Indeed, as the electronic fraction 
increases, so does the coupling between the ions and the neutrals: the size of the 
shock therefore decreases and the compressive heating rate increases. Concurrently, 
the diminution of both the self-shielding and the abundance of \HH\ strongly reduces 
the cooling rate of the gas. If the cooling is insufficient to prevent the neutrals 
from crossing the sound speed, the structure becomes a C* shock. If the cooling is even 
lower, i.e. too low to repress the impact of viscous stresses, the structure becomes a 
CJ shock.

The role of the magnetic field is also straightforward. Low magnetic field intensities
reduce the size of the magnetic precursor, hence the size of the shock, favouring 
the existence of C* and CJ shocks (see Sect. \ref{Sect-phys-shocks}). Conversely, large 
magnetic field intensities increase the ions magnetosonic speed, hence the maximal velocity for 
which a magnetic precursor can develop. When $c_{\rm ims}$ is larger than the critical 
speed required to dissociate \HH\ by collisions \citep{Le-Bourlot2002}, a domain of 
existence of CJ shocks appears (top right panel of Fig. \ref{Fig-zermatt}).

\subsubsection{Influence of the UV field}

Fig. \ref{Fig-zermatt} shows that the structure of interstellar stationary 
shocks is mostly governed by the intensity of the magnetic field and strength of 
the UV radiation field. The magnetic field tunes the magnitude of the coupling between 
ions and neutrals and therefore controls the range of velocities over which a shock can 
exist and a magnetic precursor can develop. Once those limits are set, the radiation 
field becomes the most decisive quantity:  its impact on the ionization fraction and 
the cooling efficiency (magnified by the pumping of \HH) determines the range of 
existence of C, C*, and CJ shocks.

Most shocks propagating in dense or low-illuminated environments ($\dens \geqslant 
10^4$ \cc\ or $G_0 \leqslant 1$) appear to be either J or C. This ceases to be true, 
however, in more diffuse or illuminated media (see Figs. \ref{Fig-zermatt} and 
\ref{Fig-zermatt-add}) where the radiation field drastically reduces the domain 
of existence of C-type shocks in favour of C* and CJ-type 
structures. As a rule of thumb we find that those kinds of shocks become dominant as 
soon as 
\begin{equation}
G_0 > 20 \times \left(\frac{\dens}{10^4\,\,\cc} \right)^{1/2}.
\end{equation}
It follows that only low velocity 
C-type shocks ($V_S \leqslant 5$ \kms) can propagate in prototypical PDR ($\dens 
\sim 10^4$, $G_0 \sim 100$) or in very diffuse clouds ($\dens \sim 30$ \cc, $G_0 \sim 1$).
This limit on the maximal velocity of C-type shocks is in agreement with the work
of \citet{Melnick2015} who found similar breakdown velocities in irradiated
environments.

\subsubsection{Influence of grains and PAHs} \label{Sect-gr-pahs}

\begin{figure*}[!ht]
\begin{center}
\includegraphics[width=18.0cm,trim = 1cm 1cm 0cm 0cm, clip,angle=0]{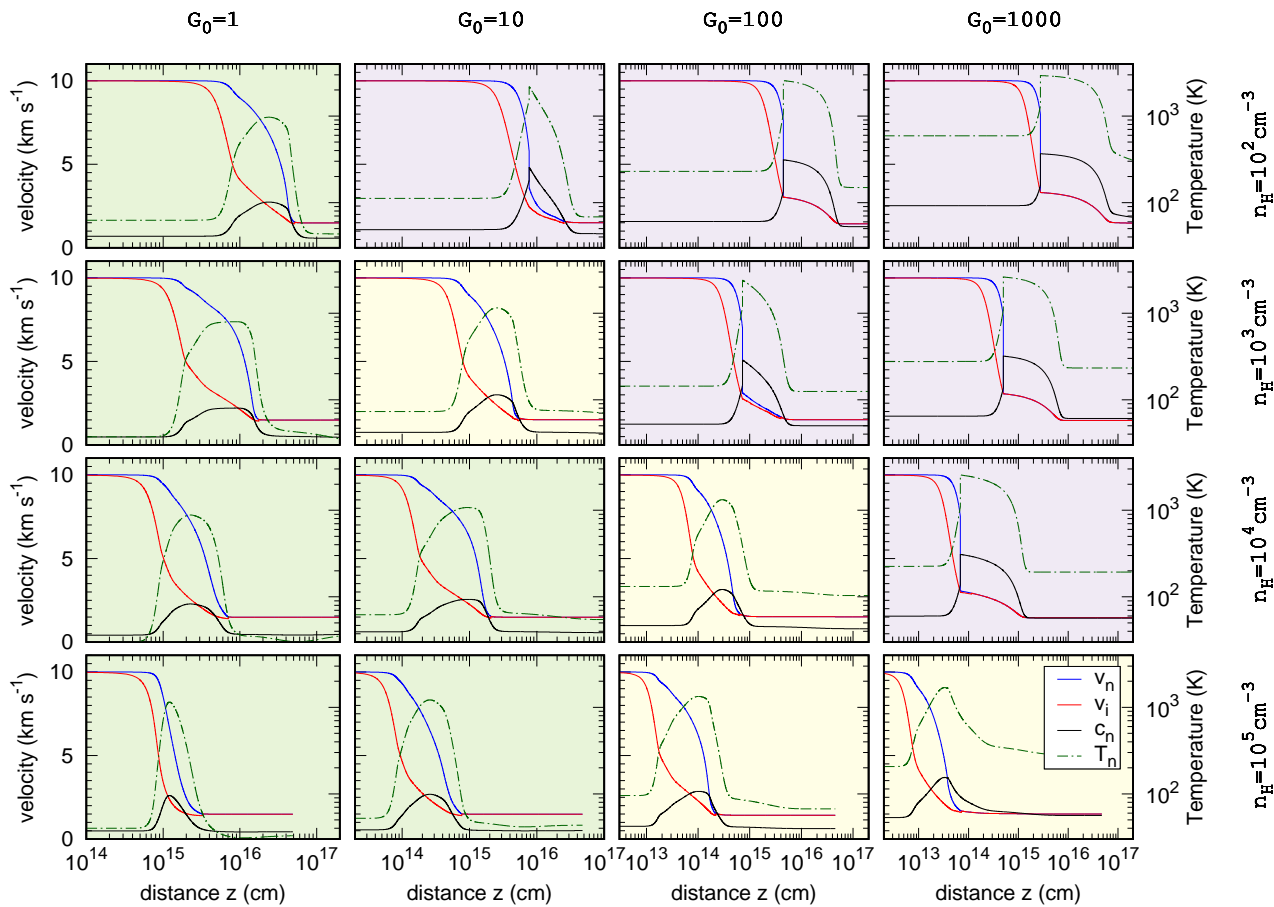}
\caption{Profiles of the neutral sound speed (solid black), velocity of the neutrals 
(solid blue) and that of the ions (solid red) in the frame of the shock, and temperature 
of the neutrals (dot-dashed 
green) computed in stationary shocks propagating in media with different densities and 
radiation field intensities. All the other parameters are set to their standard values 
(see Table \ref{Tab-main}), and the standard value of the buffer visual extinction is set 
to $A_V^0=10^{-2}$. C-type, C*-type, and CJ-type shocks are highlighted in
green, yellow, and violet, respectively.}
\label{Fig-velo-profils}
\end{center}
\end{figure*}

The cold neutral medium is a partially ionized environment with a very low ionization
fraction. Its elemental composition in the solar neighbourhood implies that the fraction 
of mass contained in gas-phase ionized species is below $3 \times 10^{-3}$ for diffuse
and transluscent clouds and quickly drops in dark clouds. With a dust-to-gas mass ratio 
of about 1\% and efficient ionization and recombination processes, charged dust grains 
are thus a dominant contributor to the inertia of the ionized flow, and therefore 
strongly influence its dynamics.

The impact of dust on magnetized shocks has been the subject of many studies that have 
successively described the effects of the dust decoupling (e.g. \citealt{Ciolek2002,
Ciolek2004}), 
their size distribution (e.g. \citealt{Chapman2006,Guillet2009}), and the fluctuation 
of their charge \citep{Guillet2007} on the structure of the shock. Two results are of 
particular interest in this paper. Firstly, as the ionization and recombination of grains 
is fast, the charge of grains rapidly fluctuates. In particular, the survival timescale 
of a neutral grain is found to be very small compared to its dynamical timescale \citep{Guillet2008},
i.e. the time it needs to reach the neutral velocity. This  
result holds for all grain sizes and all irradiation conditions. It follows 
that, for a given size of grains, both neutral and charged grains can be considered as 
a single fluid. Secondly, whether this fluid is coupled with the neutral flow or 
with the ionized flow solely depends on the Hall factor $\Gamma,$ which measures the ratio 
of the grain drag to the grain gyration timescales and depends on the grain elastic cross 
section $\sigma$, grain charge $q$, strength of the magnetic field $B$, 
mass density of the neutrals $\rho_n$, and velocity drift between grains and 
neutrals $\delta v$ as (\citealt{Guillet2007}, Eqs. 9 to 11)
\begin{equation}
\Gamma = \frac{\tau_{\rm drag}}{\tau_{\rm gyr}} \propto \frac{|q| B}{\rho_n \langle \sigma \delta v \rangle}.
\end{equation}

The results shown here (including Fig. \ref{Fig-zermatt}) are obtained adopting a PAH 
relative abundance of $\sim 10^{-6}$ and assuming that all grains are coupled to the 
magnetic field, i.e. $\Gamma > 1$. Large grains in irradiated environments are expected to be multiply 
charged (e.g. \citealt{Draine2002}). Because the Hall factor is inversely proportional 
to the square of the grain size, we estimate that this assumption is always valid for small grains ($a 
\leqslant 0.1$ $\mu$m) and holds for large grains as long as $G_0/\dens \geqslant 10^{-4}$ cm$^3$ 
and $b\geqslant 1$. In turn, it most probably fails in dense and dark environments 
($G_0/\dens \leqslant 10^{-5}$ cm$^3$) or in media with a very low magnetic field ($b \sim 0.1$), 
where the Hall factor computed for grains larger than 0.2 $\mu$m falls below unity, which means 
that a fraction of the dust mass is coupled to the neutral fluid \citep{Guillet2007}. 
The main impact of PAHs is to reduce the abundance of electrons in the gas phase. In 
irradiated environments (resp. dark clouds), where grains are dominantly positive
(resp. negative), PAHs therefore lead to an increase (resp. decrease) of the Hall 
factor. The rather large value of PAHs abundance adopted in this work therefore strengthens 
our hypothesis on the coupling between grains and the magnetic field in irradiated media 
but weakens it in dark environments.

As a whole, the  outcome of this assumption ($\Gamma > 1$) is a minimization of the ions 
magnetosonic velocity. This, in turn, reduces the range of existence of C, C*, and 
CJ shocks. In dark clouds with classical magnetization ($b=1$) for instance, 
where it could be debated, we find that these kinds of shocks propagate at 
velocities necessarily smaller than  $\sim 25$ \kms\ (see Fig. \ref{Fig-zermatt-add} of 
Appendix \ref{Append-add-figures}); this value is significantly lower than the 40 \kms\ derived 
in most of the previous studies (e.g. \citealt{Kaufman1996,Flower2003,Flower2010a,Melnick2015}) 
and also arguable. All these results show that a detailed treatment of grains and PAHs can be 
critical for the nature of interstellar shocks. For the sake of simplicity, we adopt a 
formalism which works over most of the parameter domain and extend it to environments 
where it may not be justified. An accurate treatment would allow us to study magnetized
shocks in dark environments at velocities larger than the maximal value we 
infer. In practice, such a study requires us to adopt a multi-fluid approach
to treat each size of grains as a separate flow with its own momentum and thermodynamical
evolution \citep{Ciolek2002,Chapman2006,Guillet2007}. This is beyond the scope of this paper.

\subsection{Shock profiles and energetics} \label{Sect-profils}

\begin{figure}
\begin{center}
\includegraphics[width=8.5cm,trim = 1.5cm 1cm 1cm 1cm, clip,angle=0]{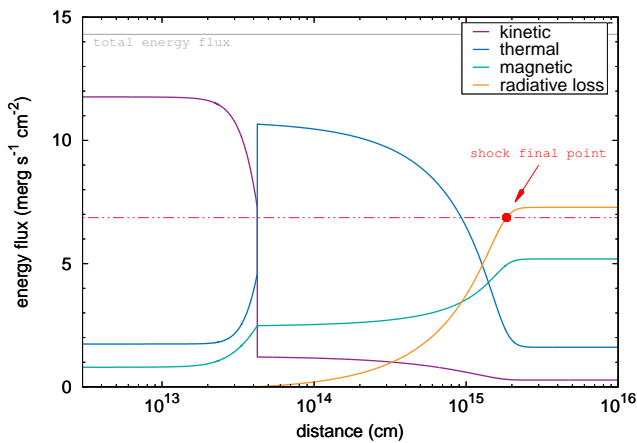}
\caption{Kinetic, magnetic, and thermal energy fluxes compared to the integrated 
radiative loss in a CJ-type shock propagating at 10 \kms\ in a medium of density 
$10^4$ \cc\ and illuminated by a strong radiation field $G_0 = 10^4$. All the other 
parameters are set to their standard values (see Table \ref{Tab-main}), and the standard 
value of the buffer visual extinction is set to $A_V^0=10^{-2}$. The light grey line
corresponds to the total conserved energy flux. The red point highlights the criterion
used to separate the shock from the post-shock region (Eq. \ref{Eq-size}).}
\label{Fig-shock-nrj}
\end{center}
\end{figure}

The main physical properties of molecular shocks are shown in Fig. \ref{Fig-velo-profils}, 
which shows, in the frame of the shock, the ion and neutral velocity profiles and 
the neutral temperature 
profiles computed in different environments for shocks propagating at 10 \kms. Starting 
from the upstream supersonic state, a gas cell is progressively 
slowed down, compressed, heated up through compression, ion-neutral friction and 
viscosity, and eventually cooled down through line emission. The thermodynamical path 
followed by the cell and the final state reached downstream are simply the outcome 
of the mandatory conversion of the initial kinetic energy into magnetic and  
thermal energies and radiative losses (see Fig. \ref{Fig-shock-nrj}).

In weakly magnetized J-type shocks with large compression ratios, the maximum temperature
reached by the gas reduces to a simple analytical formula \citep{Lesaffre2013}
\begin{equation} \label{Eq-Tmax}
T_{\rm max} = 53\,\,(V_S / 1 \kms)^2 \,\, {\rm K},
\end{equation}
which gives 5300 K for a shock at 10 \kms.
This estimation corresponds to the extreme case where most of the initial kinetic energy
is adiabatically converted into thermal energy at the shock front. In magnetized shocks
or C-type shocks, part of the kinetic energy is used to compress the magnetic field, 
hence increase the magnetic pressure. The rest of the initial energy is converted into 
heat over a characteristic length controlled by both the size of the magnetic precursor 
and the cooling rate (see Sect. \ref{Sect-phys-shocks}). The maximum temperature of 
those shocks is thus necessarily smaller than that obtained in the adiabatic case and 
decreases with the strength of the magnetic field and strength of the cooling. 
Such a behaviour has been abundantly confirmed with models of low radiation field 
environments or in dark clouds, where the temperature profiles of C-type shocks are
found to be considerably smoother and broader than the equivalent profiles obtained 
in weakly magnetized J-type shocks with the same velocities (e.g. \citealt{Flower2010,
Flower2013,Lesaffre2013,Melnick2015}). For instance, these authors report typical 
maximal temperatures of 1000 K for 10 \kms\ C-type shocks, a value five times lower 
than that found in the weakly magnetized adiabatic case.

The impact of an external UV radiation field can be understood in this framework. 
Increasing the radiation field increases the ionization fraction of the pre-shock 
medium, hence the strength of the coupling between the neutrals and the ions. As the 
magnetic precursor shrinks, the kinetic energy is converted into heat over a smaller 
distance, which leads to a larger maximal temperature of C-type shocks (see Fig. 
\ref{Fig-velo-profils}, Fig. \ref{Fig-temp-add} of Appendix \ref{Append-add-figures} 
and Fig. 2 of \citealt{Melnick2015}) that eventually become C*-type shocks. As shown 
in Fig. \ref{Fig-temp-add} of Appendix \ref{Append-add-figures}, this effect is  
moderate and appears to be stronger for low velocity shocks and low density media. In 
particular, the thermodynamical profiles of C-type and C*-type shocks are found to be 
very similar and to differ only by their size and the fact that C*-type shocks are 
partly subsonic. This whole picture is, however, strongly modified at large radiation 
fields, when the flux of UV photons is strong enough to photodissociate \HH. The reduction 
of the cooling rate of the gas leads to larger temperatures and eventually induces 
an adiabatic jump along the trajectory (CJ-type shock). As the radiation field increases, 
the jump arises sooner, instantly\footnote{an adiabatic jump lasts less than six months 
for $V_S = 10$ \kms\ and $\dens \geqslant 10^2$ \cc.} 
converting a large fraction of the kinetic energy into 
heat and magnetic compression. Depending on the velocity and the environment considered, 
the maximal temperature of CJ-type shocks can be two to ten times larger than that 
obtained in C-type shocks and persist over a large tail set by the long cooling 
timescale of the gas (see Fig \ref{Fig-temp-add}).

\subsection{Shock size and timescale} \label{Sect-size}

\begin{figure*}
\begin{center}
\includegraphics[width=16.0cm,trim = 1cm 3cm 0cm 1cm, clip,angle=0]{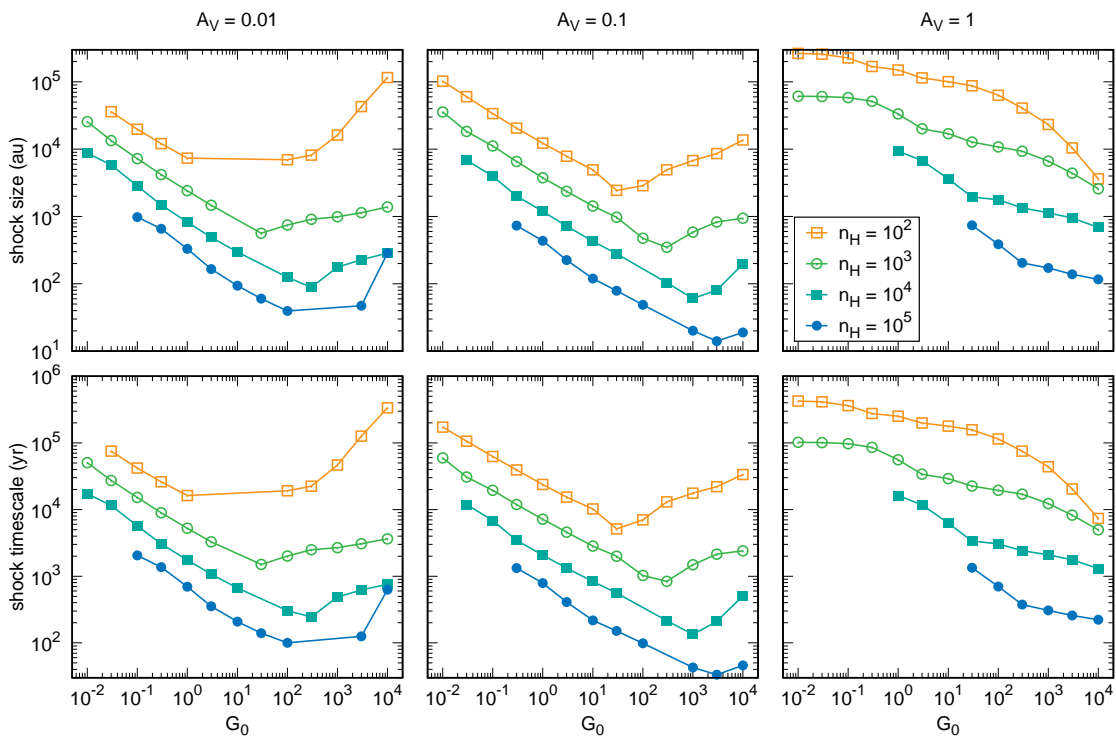}
\caption{Shock sizes (top panels) and timescales (bottom panels) as functions of
the UV radiation field and different visual extinction $A_V$ computed for shocks 
propagating at 10 \kms\ in media of densities of $10^2$ (empty squares), $10^3$
(empty circles), $10^4$ (filled squares), and $10^5$ (filled circles) \cc. The 
non-varying parameters are set to their standard values (see Table \ref{Tab-main}).
J-type shocks propagating in dense and weakly irradiated clouds are deliberately removed
from the figure (see main text), hence the missing points on the left side of each
panel.}
\label{Fig-shock-size}
\end{center}
\end{figure*}

The determination of the shock size and timescale, i.e. the limit marking the 
beginning of the post-shock medium (see Fig. \ref{Fig-scheme-shock}), is an important 
issue that is not only useful to interpret the physics of the shock but can also 
have a strong influence on the output quantities predicted by the models. Indeed, as 
shown for instance by \citet{Gusdorf2008a}, \citet{Leurini2014}, and \citet{Lehmann2017}, 
the inclusion of the post-shock gas (see Fig. \ref{Fig-scheme-shock}) can modify by orders of magnitude the column densities 
of numerous species, and hence have a strong impact for the comparison with observations. 
Several definitions have been proposed so far, including criteria on the abundance 
profiles, temperature profile, or ion-neutral coupling length \citep{Draine1980,
Wardle1999,Lesaffre2013,Melnick2015,Lehmann2016a}. Unfortunately, while these propositions 
may be useful to study a specific problem, they do not provide a reliable definition for 
comparison of shock length in different models and they cannot be applied indifferently 
to all kinds of shocks. In addition, any criterion based on the sole comparison of pre-shock 
and post-shock properties is bound to fail when strong external radiation fields are 
considered and if the post-shock medium conditions are nowhere near those of the pre-shock.

To overcome these issues, we adopt a definition of the shock size based on energetic 
considerations. As illustrated in Fig. \ref{Fig-shock-nrj}, a stationary shock propagating 
in the interstellar medium is a structure in which the mechanical energy flux is progressively 
converted into magnetic and thermal energy fluxes and is partially radiated 
away through line emission. The transition between the shock and the post-shock medium can 
therefore be designated as the point at which most of the radiation induced by the shock
has been emitted. In the following, we thus define the shock size $z_s$ as 
the distance verifying
\begin{equation} \label{Eq-size}
\frac{\Upsilon(z) - \Upsilon(0)}{\Upsilon(\infty)-\Upsilon(0)} \geqslant 95 \%
,\end{equation}
where
\begin{equation}
\Upsilon(z) = \mathscr{F}_{\rm K}(z) + \mathscr{F}_{\rm B}(z) + \mathscr{F}_{\rm T}(z)
\end{equation}
is the sum of the kinetic, 
\begin{equation}
\mathscr{F}_{\rm K}(z) = \frac{1}{2} \left( \rho_n v_n^3 + \rho_i v_i^3 \right),
\end{equation}
magnetic, 
\begin{equation}
\mathscr{F}_{\rm B}(z) = \frac{B_0^2}{4\pi} \, \frac{V_S^2}{v_i},
\end{equation}
and thermal,
\begin{equation}
\mathscr{F}_{\rm T}(z) = \frac{5}{2} k_B \left( \rho_n v_n T_n / \mu_n +  \rho_i v_i T_i / \mu_i \right),
\end{equation}
energy fluxes, and $\mu_n$ and $\mu_i$ are the mean molecular masses of the neutrals 
and of the ions. The shock 
timescale is then set as $t_s = t(z_s)$, i.e. as the time required for a fluid 
particle to reach $z_s$. Because they are built on a conservative quantity, these 
definitions have the advantage of capturing key aspects of shock physics. Moreover
they offer a more universal criterion that can be applied to any kind of structures, 
including C-, C*-, CJ-, or J-type shocks, propagating in dark or highly illuminated
environments. An example of this criterion applied to an irradiated molecular shock
is shown in Fig. \ref{Fig-shock-nrj}.

The values of $z_s$ and $t_s$ obtained for a 10 \kms\ shock propagating in various 
environments are shown in Fig. \ref{Fig-shock-size}. To simplify,
we deliberately remove from Fig. \ref{Fig-shock-size} the models at low radiation
fields and large densities that correspond to J-type shocks propagating in a 
medium in which all the heavy elements are stuck onto grains. Indeed, while 
academically interesting, these models may be unrealistic for the study of 
dense and dark clouds, for which standard models usually assume a substantial 
fraction of heavy elements in the gas phase (e.g. \citealt{Flower2003,Gusdorf2008a,
Gusdorf2008,Guillet2009,Flower2010a,Anderl2013}). Because the dark clouds are not 
the subject of this paper, these models are not studied.

Each panel of Fig. \ref{Fig-shock-size} can be separated in two regions 
depending on $G_0$ where the size and timescale of the shock either decrease or 
increase as functions of the strength 
of the UV field. (1) The first region, at low radiation field intensity, corresponds 
to the domain of existence of C-type and C*-type shocks. In this domain, the shock
size is mainly given by the length of the magnetic precursor, which is inversely 
proportional to the electron density (see Eq. 36.44 of \citealt{Draine2011}).
If the UV photons are the main ionization source of the gas, i.e. at low extinction, 
it is easy to show that the electron density writes 
$n_e \propto (G_0 \dens)^{1/2}$: the shock size is therefore proportional to 
$G_0^{-1/2} \dens^{-1/2}$ as observed on the left and middle panels of Fig. 
\ref{Fig-shock-size}. Conversely, if the UV photons are a secondary ionization 
process, i.e. at high extinction, the dependance on $G_0$
is progressively lost and the size of the shock becomes simply proportional to 
$\dens^{-1/2}$ as confirmed on the right panels of Fig. \ref{Fig-shock-size}.
(2) The second region, where the size of the shock increases with the strength 
of the UV field, corresponds to the domain of existence of CJ-type shocks. In this
domain, the ionization fraction has reached a maximum and the shock size is 
given by the large temperature tail induced by a reduced cooling efficiency 
due to the photodissociation of \HH. If \HH\ is the major coolant of the gas,
the shock size is inversely proportional to $n(\HH)$, thus directly proportional
to $G_0$ ; if \HH\ is a secondary coolant, the dependence of $z_s$ on $G_0$ is
weak or lost. All these cases can be identified in the left and middle panels of 
Fig. \ref{Fig-shock-size}.

Overall, Fig. \ref{Fig-shock-size} shows that the shock sizes and timescales cover
a broad range of values, spread over three orders of magnitude, depending on the
medium in which they propagate. Interestingly, if J-type shocks are removed from
the analysis, the size of interstellar shocks appears to be weakly dependent on 
the shock velocity. Exploring the entire grid of models shows that similar sizes 
are found for $V_S$ varying between 5 and 20 \kms, except at very low radiation 
fields ($G_0/\dens \sim 10^{-5}$ cm$^3$) where sizes differ by a factor of four and at large 
radiation fields ($G_0/\dens \geqslant 10^{-1}$ cm$^3$) where they differ by a factor of three.

The crossing time represented in the bottom panels of Fig. \ref{Fig-shock-size} can be 
interpreted as the time required for the shock to reach its steady-state configuration 
\citep{Chieze1998,Lesaffre2004}. Scaling almost exactly as the shock size, this critical
dynamical timescale is found to strongly depend on the medium considered, varying from
a few hundred years in dense and moderately irradiated environments to a few tens of
thousand years in diffuse gas. In most cases, these values are considerably smaller 
than the turnover timescales of the corresponding environment, suggesting that 
steady-state shocks are relevant structures for the study of interstellar medium.
This is true, in particular, for diffuse and dense environments where the linewidth-size
relation leads to dynamical timescales 100 times larger than the shock crossing time
(e.g. \citealt{Hennebelle2012}). This conclusion weakens, however, in outflows and jets
(e.g. \citealt{Gusdorf2008}) and in dense PDRs
in which the photoevaporation timescale becomes comparable or smaller than the shock 
crossing time \citep{Storzer1998}. Comparing the results of Fig. \ref{Fig-shock-size} 
to the propagation timescale of ionization fronts \citep{Gorti2002} shows that 
shocks can no longer be considered at steady state as soon as $\dens \geqslant 10^5$
\cc, $G_0 \geqslant 10^3$ and $A_V \leqslant 10^{-2}$, i.e. at the border of our
grid of models.

\section{Molecules in irradiated shocks} \label{Sect-chemistry}

\begin{figure*}
\begin{center}
\includegraphics[width=6.3cm,trim = 1cm 3.5cm 12.0cm 2.5cm, clip,angle=0]{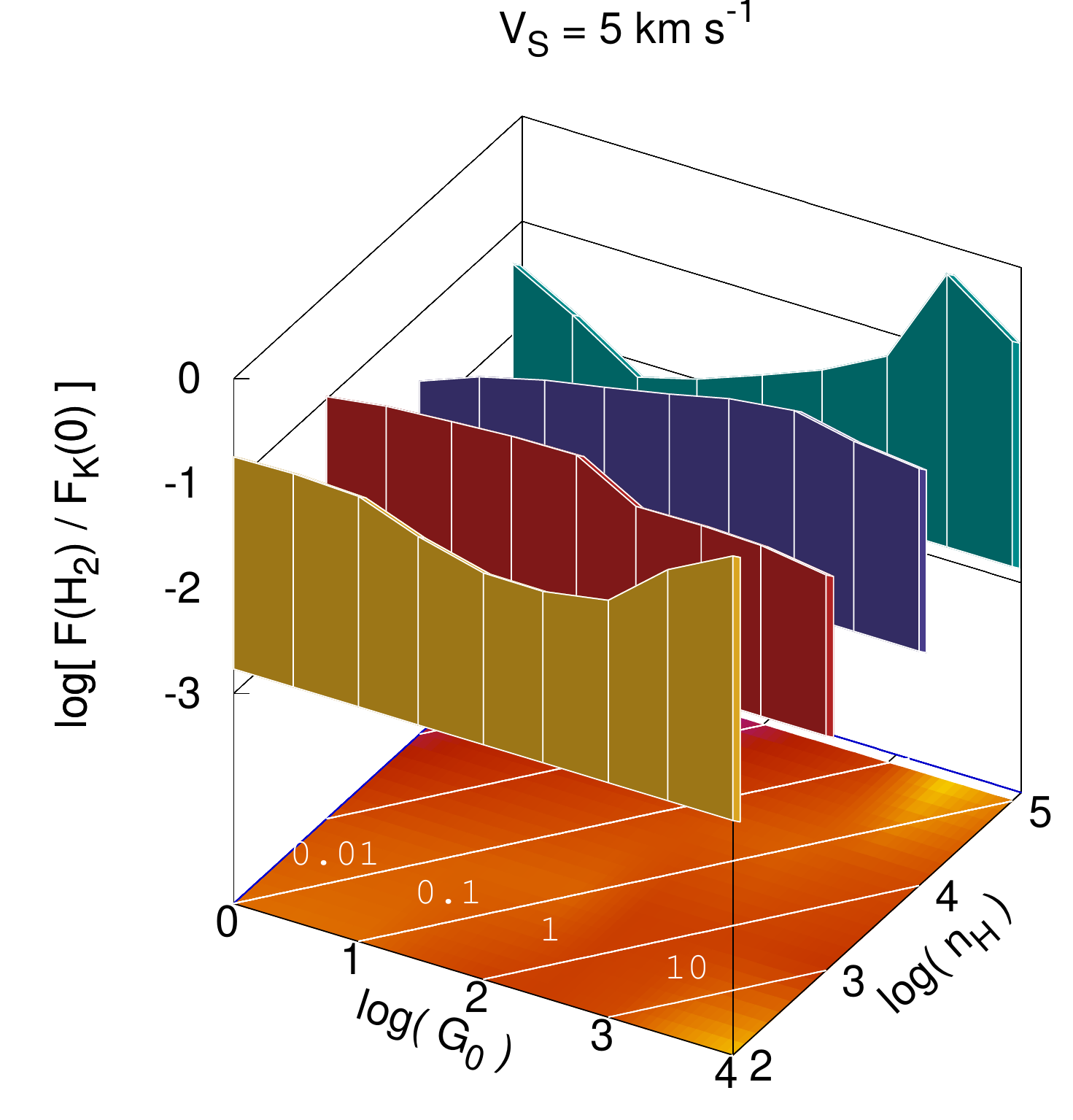}
\includegraphics[width=5.5cm,trim = 3cm 3.5cm 12.0cm 2.5cm, clip,angle=0]{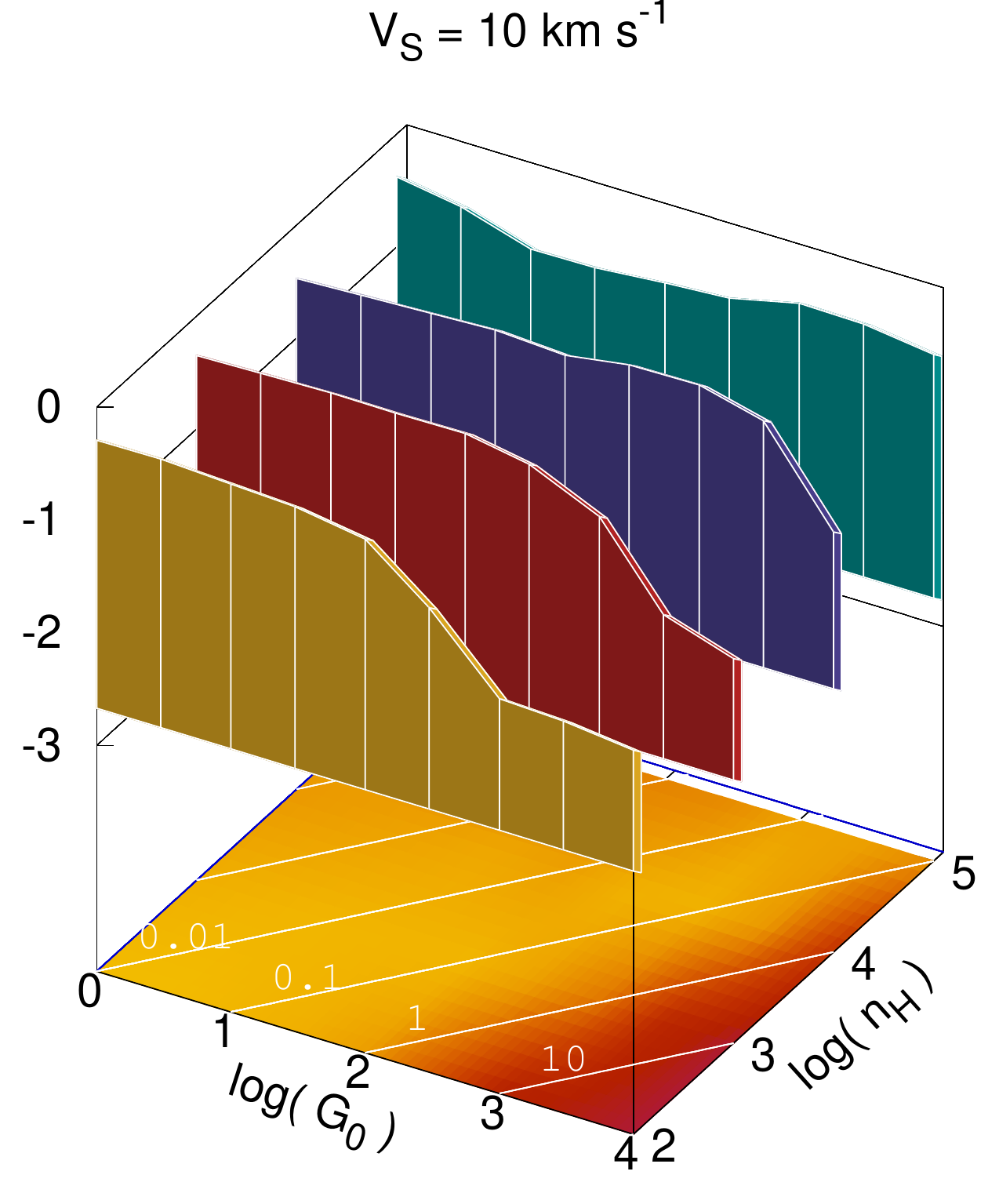}
\includegraphics[width=6.3cm,trim = 3cm 3.5cm 10.0cm 2.5cm, clip,angle=0]{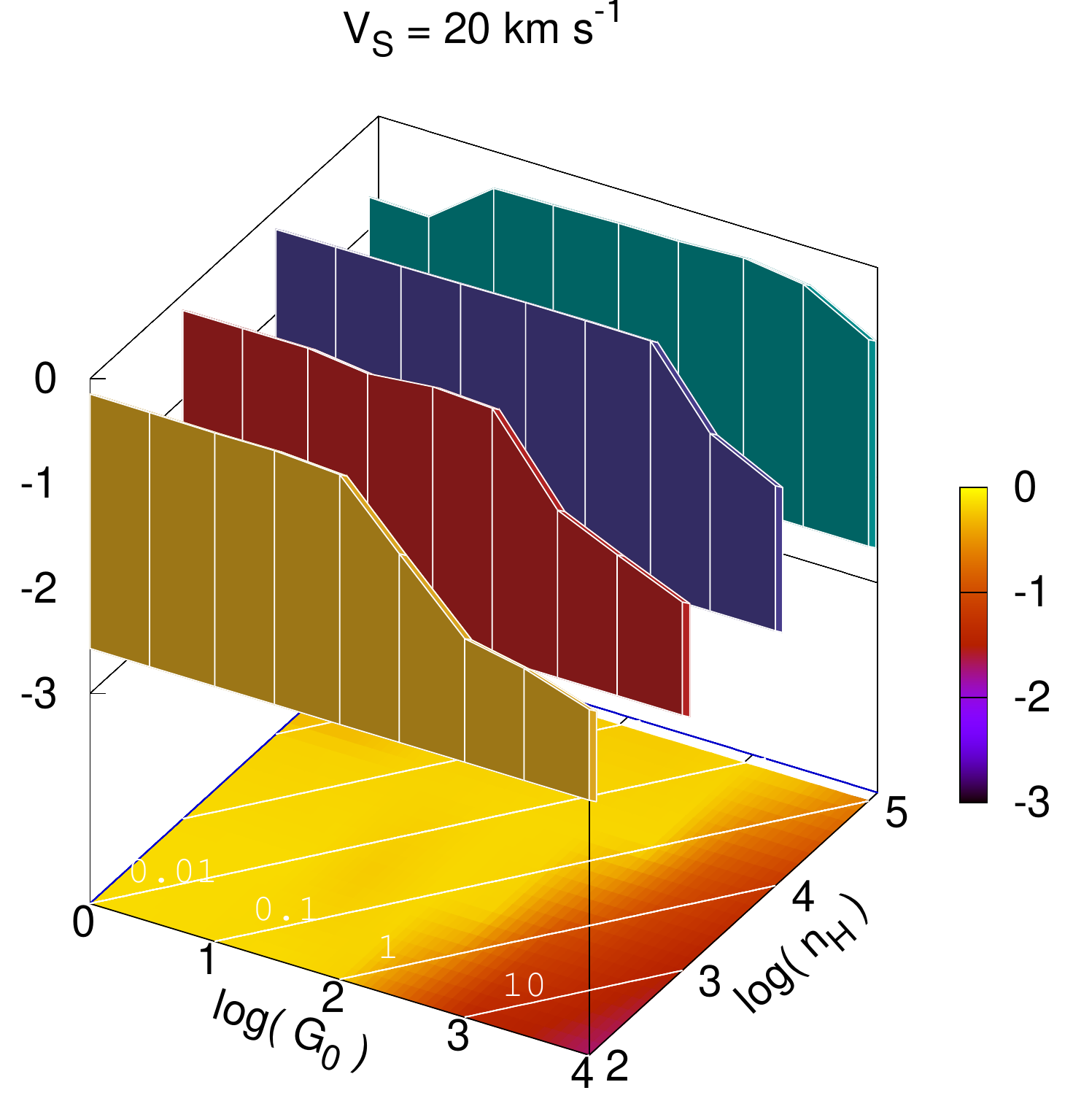}
\includegraphics[width=6.3cm,trim = 1cm 3.5cm 12.0cm 4.0cm, clip,angle=0]{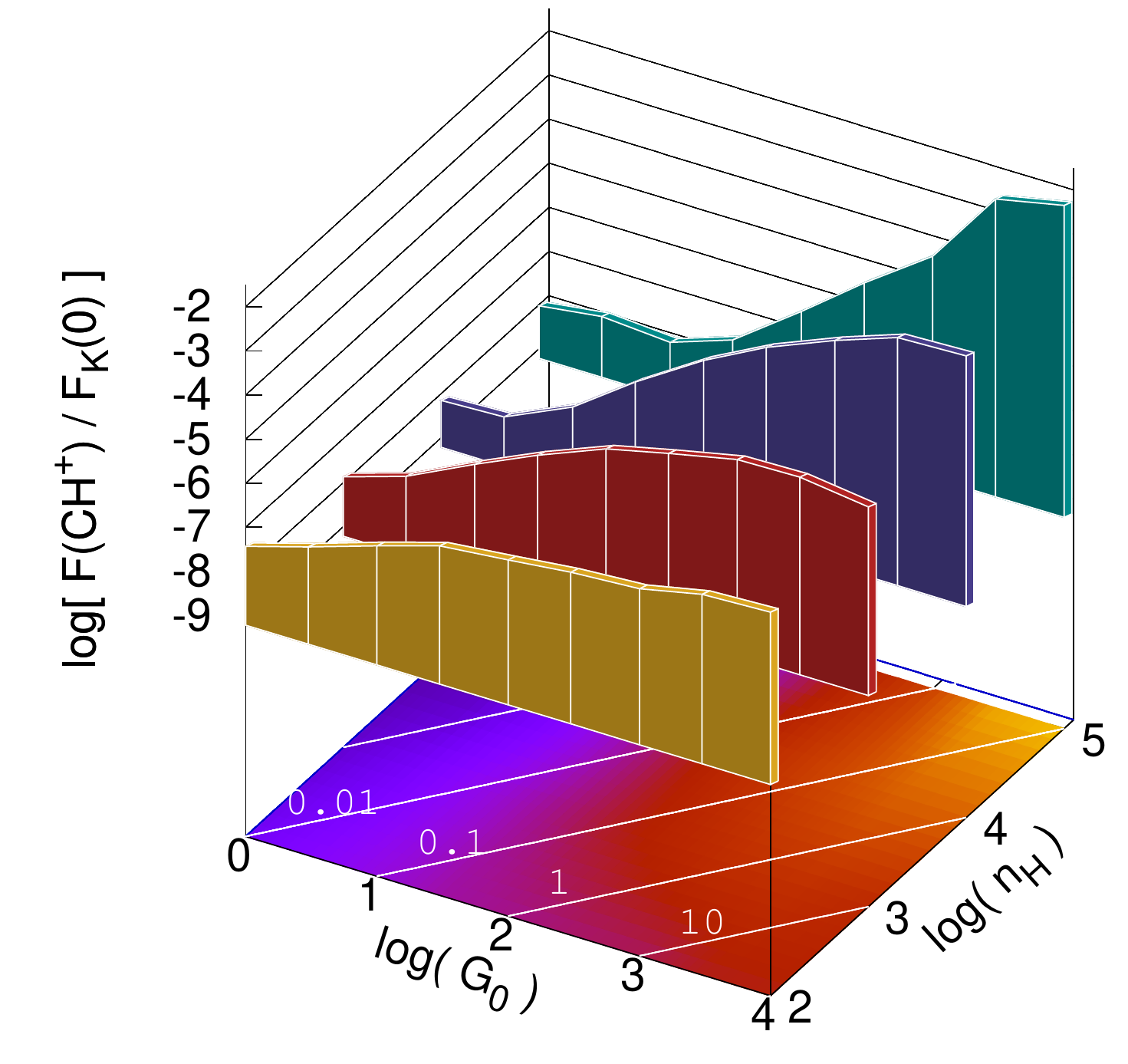}
\includegraphics[width=5.5cm,trim = 3cm 3.5cm 12.0cm 4.0cm, clip,angle=0]{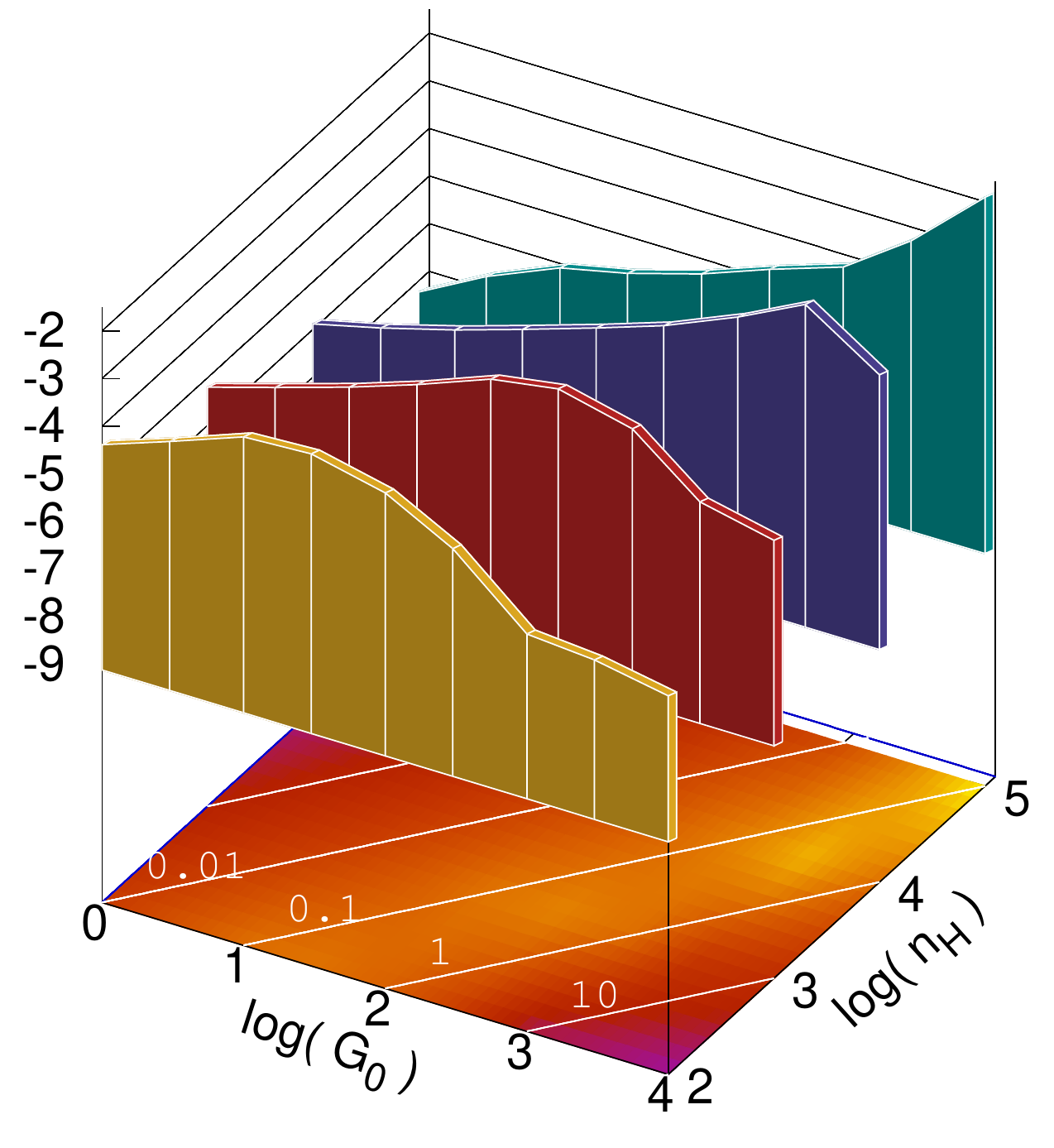}
\includegraphics[width=6.3cm,trim = 3cm 3.5cm 10.0cm 4.0cm, clip,angle=0]{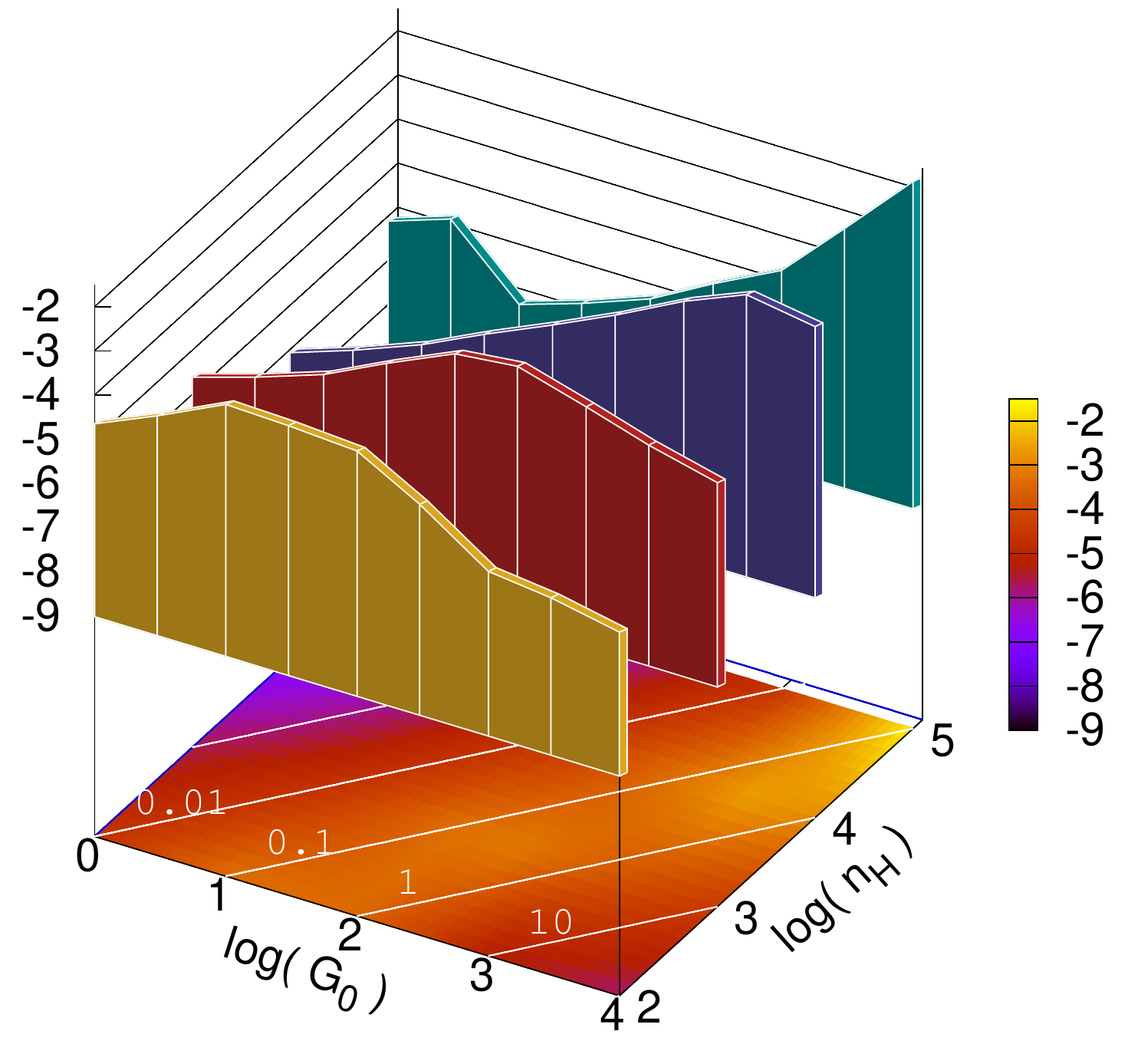}
\caption{Total fluxes of \HH\ rovibrational lines (top panels) and \CHp\ rotational lines 
(bottom panels, summed up to the $J=10 \rightarrow 9$ transition) emerging from shocks 
propagating at 5 (left panels), 10 (middle panels), and 20 (right panels) \kms\ in different 
environments, for a buffer visual extinction $A_V^0=0.1$. The non-varying parameters are 
set to their standard values (see Table \ref{Tab-main}). All fluxes are normalized to the 
input kinetic energy flux of the shock $\mathscr{F}_{\rm K}(0)= \frac{1}{2} \rho V_S^3$. 
The results are displayed 
simultaneously as colour maps and fence plots at fixed densities of $10^2$ (yellow), 
$10^3$ (red), $10^4$ (blue), and $10^5$ (green) \cc. White lines on the colour maps indicate 
contours of constant $G_0/\dens$ ratio.}
\label{Fig-H2-CHp-emis}
\end{center}
\end{figure*}

We recall that the standard model adopted in this section is defined by a 
buffer visual extinction $A_V^0=10^{-1}$.
External sources of UV radiation fields necessarily exert a major impact on the formation 
and excitation of molecules in space. In irradiated shocks, the role of UV photons extends 
far beyond the mere ionization and dissociation processes: the combined effects of radiative 
and mechanical energies modify the structure of shocks, leading to specific chemical 
and radiative tracers. To analyse such effects, the line emission of molecular shocks are 
derived using two different approaches depending on the molecule considered. For \HH, whose 
level populations are self-consistently computed by the code, line emissions are 
calculated in the optically thin limit by summing local emissivities. The level populations 
of other species are computed at statistical equilibrium and their line 
emissions are obtained by postprocessing the output of the shock model with the Large 
Velocity Gradient (LVG) code of \citet{Gusdorf2008a} slightly modified to take into 
account self-absorption in the line radiative transfer. In all cases, both intensities and 
column densities are computed over the shock only, i.e. without the contribution of the
post-shock region identified through Eq. \ref{Eq-size}.

To discuss the results in light of a global energetic budget, 
this section addresses the following questions. What fraction of the input kinetic 
energy flux is radiated away through all the transitions of a given molecular species ? How
is this total intensity spread in different lines ? As a first application, we focus 
on predictions regarding two important species: \HH, whose rovibrational structure
already observed in many galactic sources (e.g. \citealt{Giannini2004,Gillmon2006,
Neufeld2009a,Habart2011}) will soon be unveiled in extragalactic 
environments with the James Webb Space Telescope (JWST), and \CHp, whose recent detection 
in emission in starburst galaxies at high redshift ($z \sim 2-3$) with ALMA may be to 
date the strongest signature of the combined effect of mechanical and radiative energy 
on interstellar chemistry \citep{Falgarone2017}. While other species are briefly 
discussed in Sect. \ref{Sect-CO}, we defer detailed presentations of the impact of
irradiated shocks on the rest of the chemistry to future papers, as motivated by 
emerging observational needs.

\subsection{H$_2$}

\begin{figure*}
\begin{center}
\includegraphics[width=17.0cm,trim = 1.5cm 2cm 1cm 2cm, clip,angle=0]{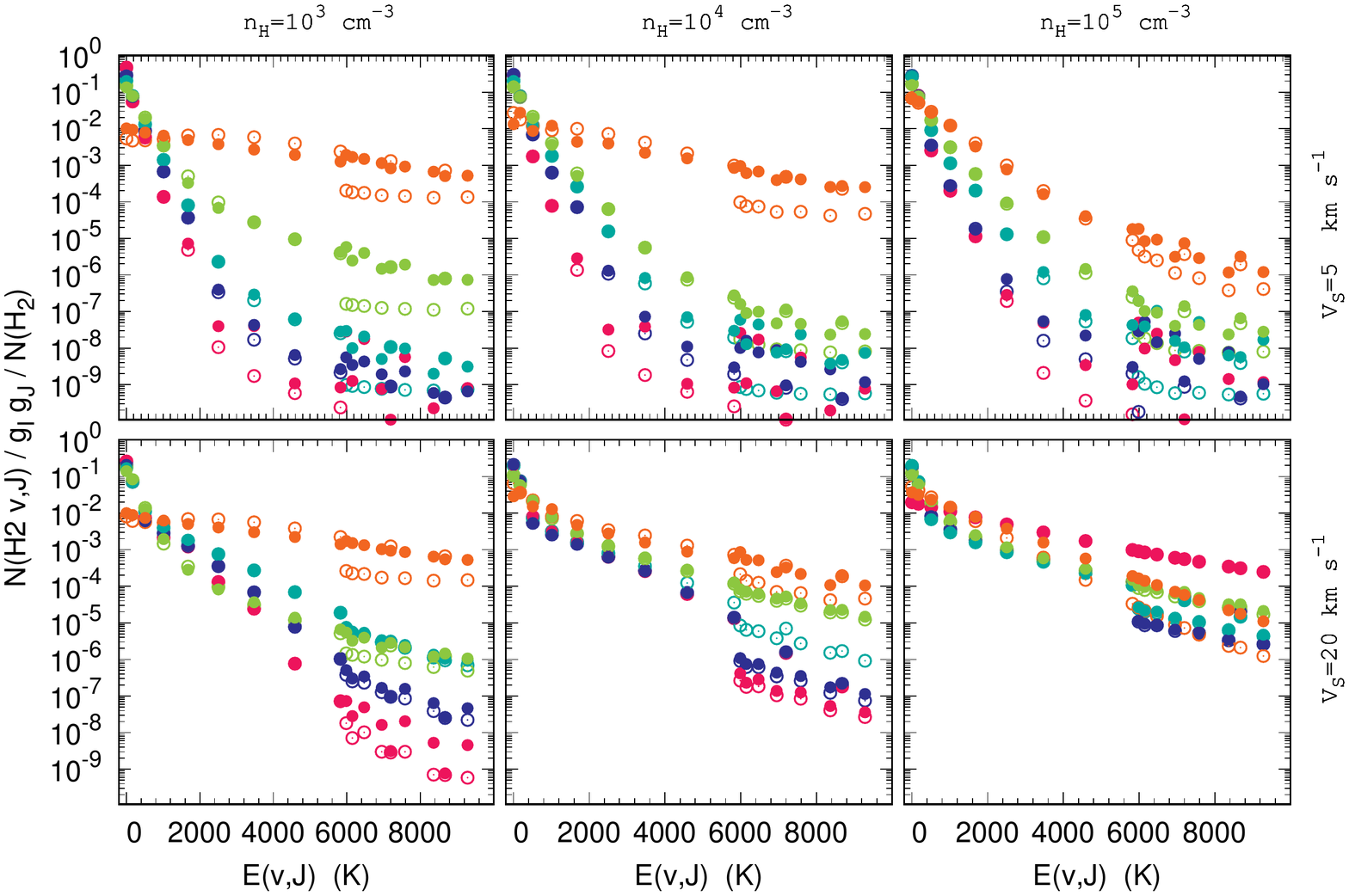}
\caption{\HH\ excitation diagrams normalized to the column density of \HH\ obtained
for shocks propagating at 5 (top panels) and 20 (bottom panels) \kms\ in media with 
densities varying between $10^3$ (left panels) and $10^5$ (right panels) \cc, a scaling 
factor of the UV radiation field of $1$ (red), 10 (dark blue), $10^2$ (light blue), 
$10^3$ (green), and $10^4$ (orange), and a buffer visual extinction 
$A_V^0=0.1$. All non-varying parameters are set to their standard values (see Table 
\ref{Tab-main}). To estimate the effect of the UV pumping of the rovibrational levels 
of \HH, the models have been run with (filled circles) and without (dotted circles) 
this process. We note that only pure rotational levels exist below 5987 K. Above this
limit, both pure rotational and rovibrational levels exist.}
\label{Fig-H2-diag}
\end{center}
\end{figure*}

The energy budget of molecular shocks 
is a complex problem 
that not only depends on the dynamical properties of the shock (velocity and magnetic 
strength) but also on the medium in which it propagates. Interestingly, all studies 
performed in dark and dense clouds with $b=0.1$ and $1$ or in environments illuminated by
a moderate radiation field converge towards similar results, namely the existence of two 
regimes depending mostly on the velocity of the shock. At low velocity ($V_S \leqslant 
5$ \kms) most of the kinetic energy dissipates via magnetic compression and through the 
rotational emission of CO and H$_2$O (in dark environments, see \citealt{Pon2012,Pon2016,
Lehmann2016a}) 
or the fine structure lines of C$^+$ and O (in media with a moderate radiation field; 
see \citealt{Lesaffre2013}). As the velocity increases, the fraction of energy spent in 
magnetic compression decreases and \HH\ becomes the dominant coolant over a wide range
of shock velocities and gas densities (Figs. B1 and B2 of \citealt{Flower2015}, and Fig. 8
of \citealt{Lesaffre2013}).

The results obtained in this work regarding \HH\ emission are in line with those findings, except 
for large UV radiation fields and in CJ-type shocks where additional processes arise. To 
illustrate this, in the top panels of Fig. \ref{Fig-H2-CHp-emis} we show the total
flux emitted in \HH\ by irradiated shocks normalized to the input flux of kinetic energy. 
As expected, \HH\ appears as a major coolant over the entire grid of models: although
not necessarily dominant, the contribution of \HH\ to the total energy budget is never
negligible, varying from about 1\% to 90\% of the shock initial kinetic energy flux. In 
concordance with \citet{Lesaffre2013}, the contribution of \HH\ is minimal for low 
velocity shocks where fine structure lines of O and C$^+$ take over, and reaches 
a maximum at higher velocities regardless of the density of the gas. Interestingly, 
a plateau of \HH\ emission is found over a broad range of radiation field strength, suggesting
that as long as the medium is molecular, \HH\ always emits the same fraction of the input
mechanical energy. This simple picture breaks, however, when the radiation field 
is strong enough to dissociate \HH. At large velocities ($V_S \geqslant 10$ \kms), the 
combination of photodissociation and collisional dissociation in CJ-type shocks induces 
a strong drop in \HH\ abundance. The impact of this drop on \HH\ emission is partly 
compensated by the increase of the gas temperature which enhances the excitation 
of \HH. As a result, the total flux emitted in \HH\ decreases by about a factor of ten 
only. At lower velocities ($V_S \sim 5$ \kms), CJ-type shocks are not strong enough to 
dissociate \HH\ by collisions. In this case, the decrease of \HH\ abundance (due solely 
to photodissociation) combined with the increase of \HH\ excitation leads to slight 
variations of the total flux emitted in \HH.

Two main results therefore emerge from this analysis. Firstly, whatever the physical conditions
or the shock type, \HH\ lines systematically carry a significant amount of the shock 
kinetic energy. Secondly, the radiation field rarely boosts \HH\ emission but may reduce it
while increasing its excitation.

The latter behaviour is shown in Fig. \ref{Fig-H2-diag} which {presents} the normalized \HH\
excitation diagrams computed in different irradiation conditions. At low radiation fields, 
the excitation diagrams of \HH\ predicted in C-type shocks generally break in two parts: 
the low rotational levels which are mainly excited by inelastic collisions, and the 
remaining rovibrational levels whose populations result from a combination of inelastic 
collisions, chemical pumping, and UV pumping. As the radiation field increases, the 
temperature of the gas rises (see Figs. \ref{Fig-velo-profils} \&. \ref{Fig-temp-add}) 
and the contribution of inelastic 
collisions to the excitation of high energy levels increases. This process, combined
with the enhancement of the UV pumping, drastically changes the excitation conditions.
Indeed, we find that varying the input radiation between $G_0=1$ and $G_0=10^4$ almost 
leads to a continuum of excitation temperatures, and amplifies the intensities of the 
rovibrational lines, by several orders of magnitude, for all densities and shock
velocities. Interestingly, the contribution of inelastic collisions may become comparable 
or even larger than that of UV pumping, especially at large density or in high velocity 
CJ-type shocks (e.g. $V_S = 20$ \kms, $\dens > 10^4$ \cc) where the input mechanical 
energy strongly overpowers that of the UV photons (see Sect. \ref{Sect-kin-rad-nrj}).

The impact of shocks on the excitation of \HH\ in outflows has been the subject of many 
studies (e.g. \citealt{Flower2003a,Neufeld2007,Gusdorf2008}) which show that the excitation diagrams produced 
by stationary J-type or non-stationary CJ-type shocks are flatter than those obtained 
in C-type shocks with the same velocities. These studies have concluded that the comparison between 
observations and predictions can be a powerful tool to derive the properties of jets
and the age of the associated astrophysical systems.
The results obtained in this work provide a similar tool for irradiated environments. As the 
radiation field increases and the shock shifts from a C-type to CJ-type structure, the 
excitation diagram of \HH\ flattens. Strong degeneracies appear, implying that the sole
observation of \HH\ excitation, although useful, is not sufficient to circumvent a unique 
set of physical conditions. The clear identification of irradiated shocks requires 
complementary data, such as the total emission of \HH\ and the emission and excitation
diagrams of other molecules, such as \CHp.

\subsection{CH$^+$}

\begin{figure}
\begin{center}
\includegraphics[width=9.0cm,trim = 1.5cm 2.5cm 2.0cm 3.5cm, clip,angle=0]{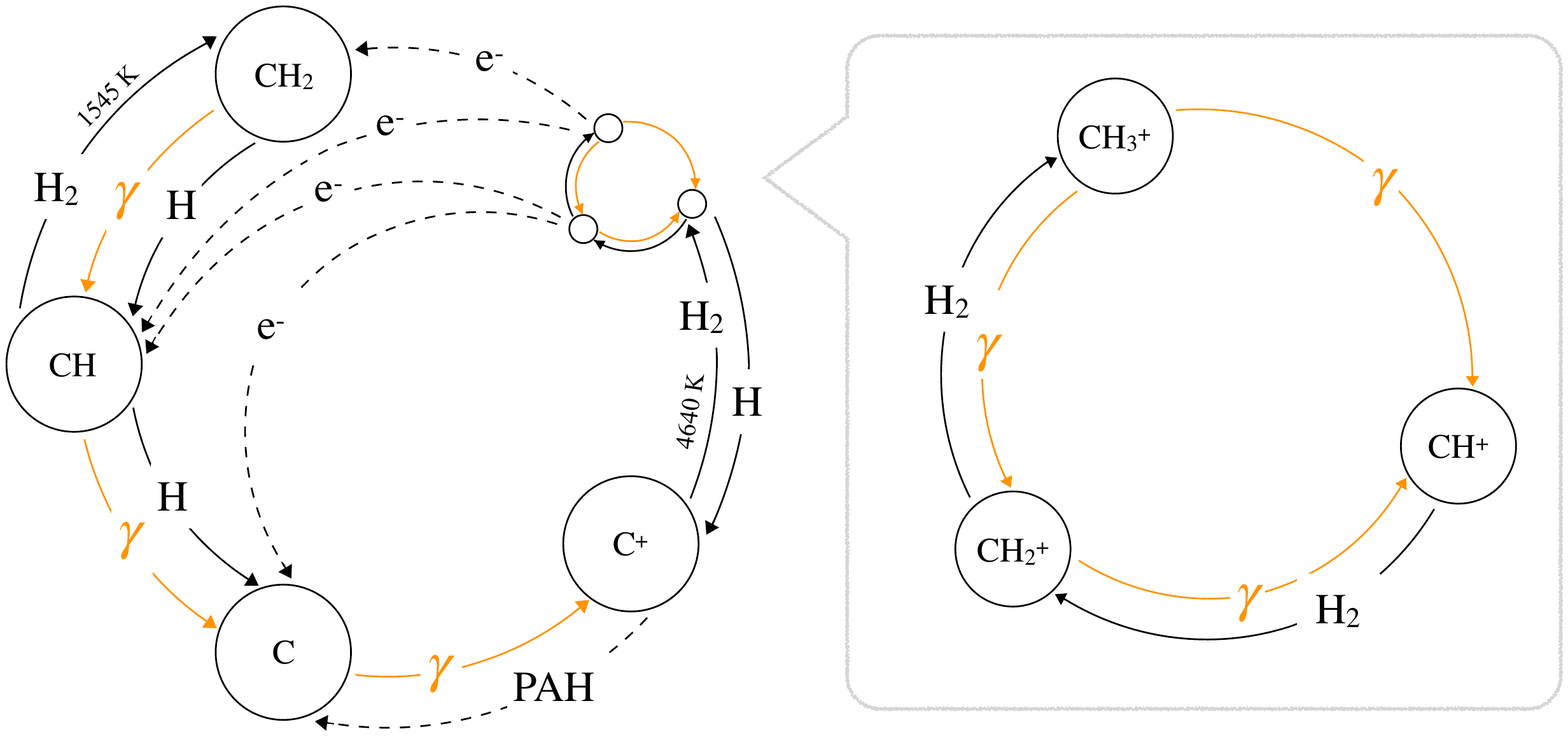}
\includegraphics[width=9.0cm,trim = 1cm 1cm 3.2cm 1cm, clip,angle=0]{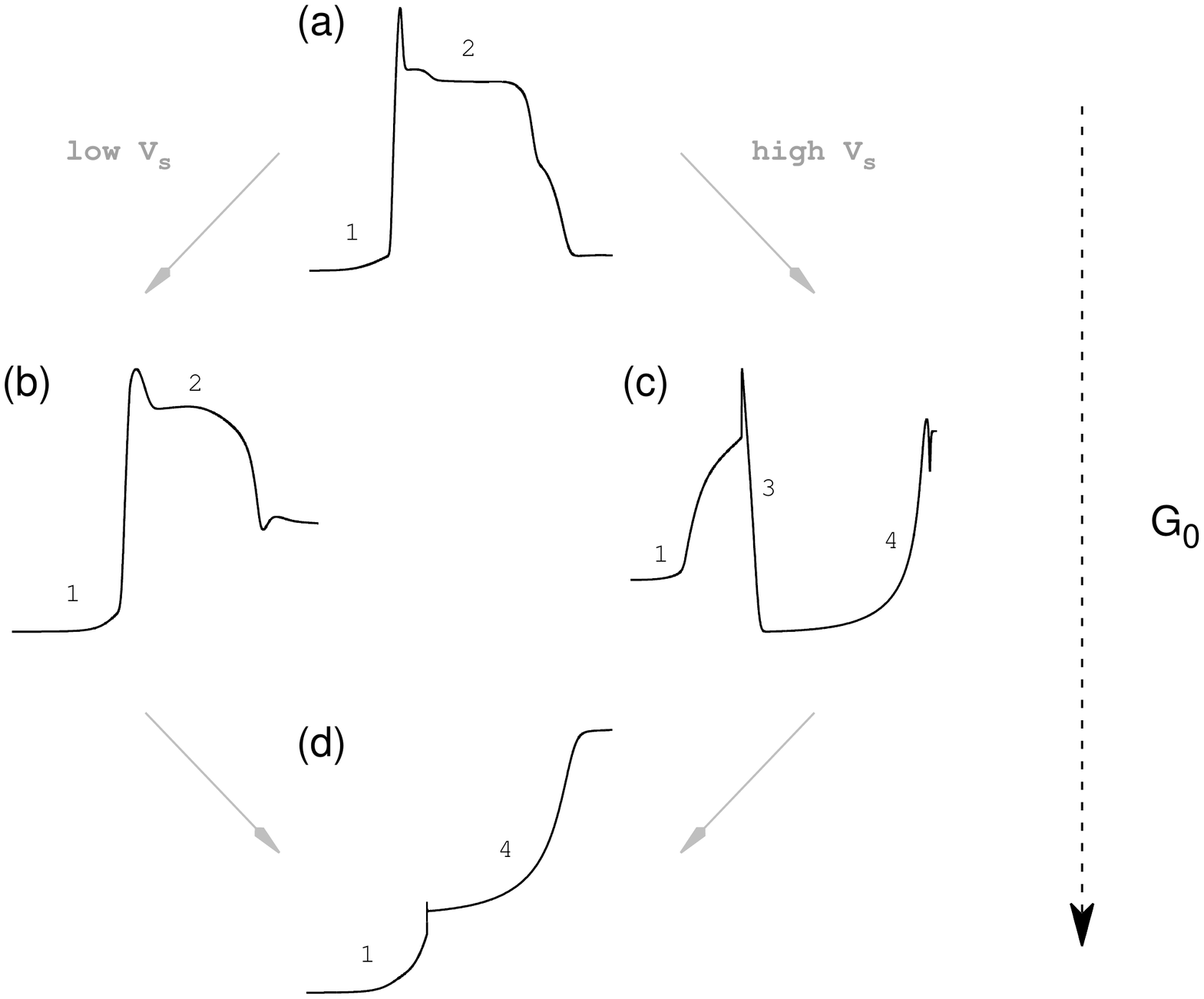}
\caption{{\it Top:} Dominant formation and destruction pathways of the carbon hydrogenation
chain in high temperature media ($T \geqslant 250$ K) or in media irradiated by strong UV 
radiation field. {\it Bottom:} Common forms of the 
density profile of CH$^+$ in molecular shocks irradiated by an external radiation field of 
strength G$_0$. Although the scales are arbitrary, all the density profiles are shown in 
log-log space. Labels highlight parts of the trajectory where the evolution of CH$^+$ 
abundance is driven by a specific thermochemical process (see text).}
\label{Fig-CHp-prof}
\end{center}
\end{figure}

The methylidyne cation \CHp\ is a unique molecule and a golden goose\footnote{see the 
Grimm's fairy tales} for the study of dissipation of mechanical 
energy (e.g. \citealt{Godard2014,Falgarone2017}). This view stems, first and foremost, 
from the simplicity of its chemistry summarized in the top panel of Fig. \ref{Fig-CHp-prof}. 
In irradiated environments, the carbon hydrogenation chain can be seen as a series of 
chemical cycles: a global cycle which controls the hydrogenation of \Cp\ and C and the 
balance between ionized and neutral species, and subcycles which affect how the carbon is 
distributed over the different hydrides. Seen like this, the global production/destruction 
of \CHp, 
\CHdp, and \CHtp\ is driven by only three mechanisms: the hydrogenation of \Cp, 
the destruction of \CHp\ by collision with H, and the dissociative recombinations 
of \CHdp\ 
and \CHtp. The global production rate of \CHp, \CHdp, and \CHtp\ is therefore proportional 
to the abundances of \HH\ and \Cp, and their destruction directly proportional to the 
abundance of electrons. The way carbon is spread over these three hydrides then depends 
on the strength of 
the subcycle highlighted in the top right panel of Fig. \ref{Fig-CHp-prof}, hence on the 
electronic fraction (which favours \CHp), the molecular fraction (which favours \CHtp), and 
the amount of UV photons (which favours \CHp\ through the photodissociation of \CHtp). As 
a result, the 
density profiles of \CHp\ in irradiated shocks can be reduced, over the entire grid of 
models, to only four different categories (bottom panel of Fig. \ref{Fig-CHp-prof}), 
driven by very few thermochemical processes whose importance mostly depends on the 
velocity of the shock and strength of the external radiation field.


In molecular shocks, both the ion-neutral velocity drift and the increase of 
temperature induced by the dissipation of mechanical energy activate the endothermic reaction
\begin{equation} \label{Eq-form-chp}
\Cp + \HH \rightarrow \CHp + {\rm H} \qquad {\frac{\Delta E}{k_B} = 4640\,\,{\rm K}}.
\end{equation}
This induces a rise in the abundance of \CHp\ (label 1 in Fig. \ref{Fig-CHp-prof}) 
and eventually a fall in that of \Cp. At a low radiation field or low velocity (schemes 
a and b of Fig. \ref{Fig-CHp-prof}), the abundance of \CHp\ reaches a plateau (label 2) 
$-$ somehow lower than the maximal abundance (due to the destruction of \Cp) $-$ whose 
size is given by the cooling timescale of the gas. At a large radiation field or large 
velocity (schemes c and d), the collisional dissociation of \HH\ shuts reaction 
(\ref{Eq-form-chp}) off. The abundance of \CHp\ rises then drops (label 3) and reaches 
a plateau whose size is given by the formation timescale of \HH\ (label 4). Because of 
the subcycle described above and the feedback induced by the photodissociation of \CHtp, 
the peak of \CHp\ abundances drastically increases with $G_0$. The column density of 
\CHp\ computed across the shock therefore also increases with $G_0$, but only as long 
as $G_0$ and $V_S$ are small enough to prevent the collisional dissociation of \HH, i.e.
as long as the shock is either C or C*.

\begin{figure}
\begin{center}
\includegraphics[width=9.0cm,trim = 1cm 2.5cm 1cm 1cm, clip,angle=0]{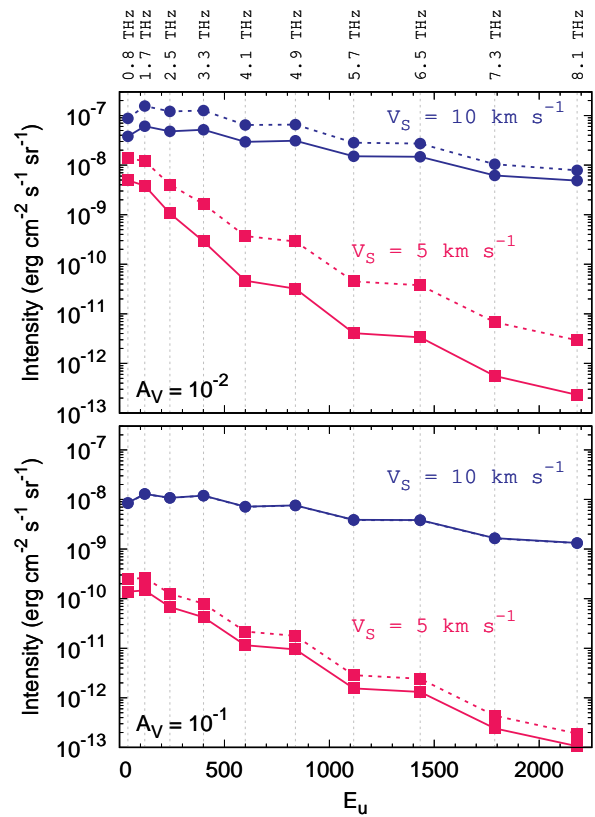}
\caption{Specific intensities of the first ten rotational lines of \CHp\ (from $J=1
\rightarrow 0$ to $J=10 \rightarrow 9$) as functions of the upper level energy expressed
in Kelvin, emitted by two shocks propagating at 5 (red 
squares) and 10 (blue circles) \kms\ in a medium with density of $10^4$ \cc\ and
illuminated by a radiation field $G_0=10^2$ at a visual extinction $A_V^0=0.01$ (top
panel) and 0.1 (bottom panel). All other parameters are set to their standard values 
(see Table \ref{Tab-main}).
The predictions are shown with (dashed) and without (solid) taking into account the 
UV pumping of the rovibrational levels of \HH. The corresponding frequency of
each rotational line is indicated above the figure.}
\label{Fig-CHp-diag}
\end{center}
\end{figure}

As shown in Fig. \ref{Fig-H2-CHp-emis}, this chemical result directly translates
into the efficiency of shocks at producing and exciting \CHp. As the surrounding UV 
radiation field increases by a factor of 1000, the total flux emitted in \CHp\ rises 
by several orders
of magnitude $-$ for a given input of mechanical energy $-$ and reaches a maximum at 
roughly the conditions where CJ-type shocks appear. Because the critical densities of 
the rotational levels of \CHp\ are large ($\geqslant 10^7$ \cc; \citealt{Godard2013}), 
this effect is magnified in high density media where 
the high production rate of \CHp\ is combined with high excitation rates. In these
conditions, \CHp\ is found to radiate as much as several percent of the shock kinetic
energy, a value that is almost comparable to the total flux emitted in \HH. Altogether, 
we find that irradiated C or C*-type shocks propagating in dense media (with $G_0/\dens 
\sim 0.05$ cm$^3$) are the most efficient structures, in terms of energy, for the formation 
and excitation of \CHp, hence the structures most likely to be seen in unresolved 
observations of interstellar environments. Large efficiencies can be 
obtained at all shock velocities, indicating that the sole observation of the total 
emission of \CHp\ is not sufficient to identify specific types of shocks.

This degeneracy could be solved, however, by looking at the excitation diagram of \CHp.
Indeed, as shown in Fig. \ref{Fig-CHp-diag}, increasing the shock velocity from 5 to 10 
\kms\ $-$ hence the kinetic energy flux by a factor of 8 $-$ drastically enhances the emission 
of the mid-$J$ lines of \CHp\ whose intensities raise by several orders of magnitude. 
Observations of the total emission of \CHp\ and of the slope of its excitation diagram 
could therefore be a powerful tool to infer the type and number of shocks in a given 
environment, and thus deduce the total mechanical energy pervading the medium. Given
the line frequencies indicated in Fig. \ref{Fig-CHp-diag}, such measurements could be 
achieved in galactic environments using the GREAT, FIFI-LS, and HIRMES instruments 
on board the SOFIA telescope or in extragalactic sources at high redshift 
using the ALMA interferometer.

\subsection{OH, H$_2$O, CO, and SH$^+$} \label{Sect-CO}

\begin{figure*}[!ht]
\begin{center}
\includegraphics[width=16.5cm,trim = 1cm 2cm 1cm 1cm,clip,angle=0]{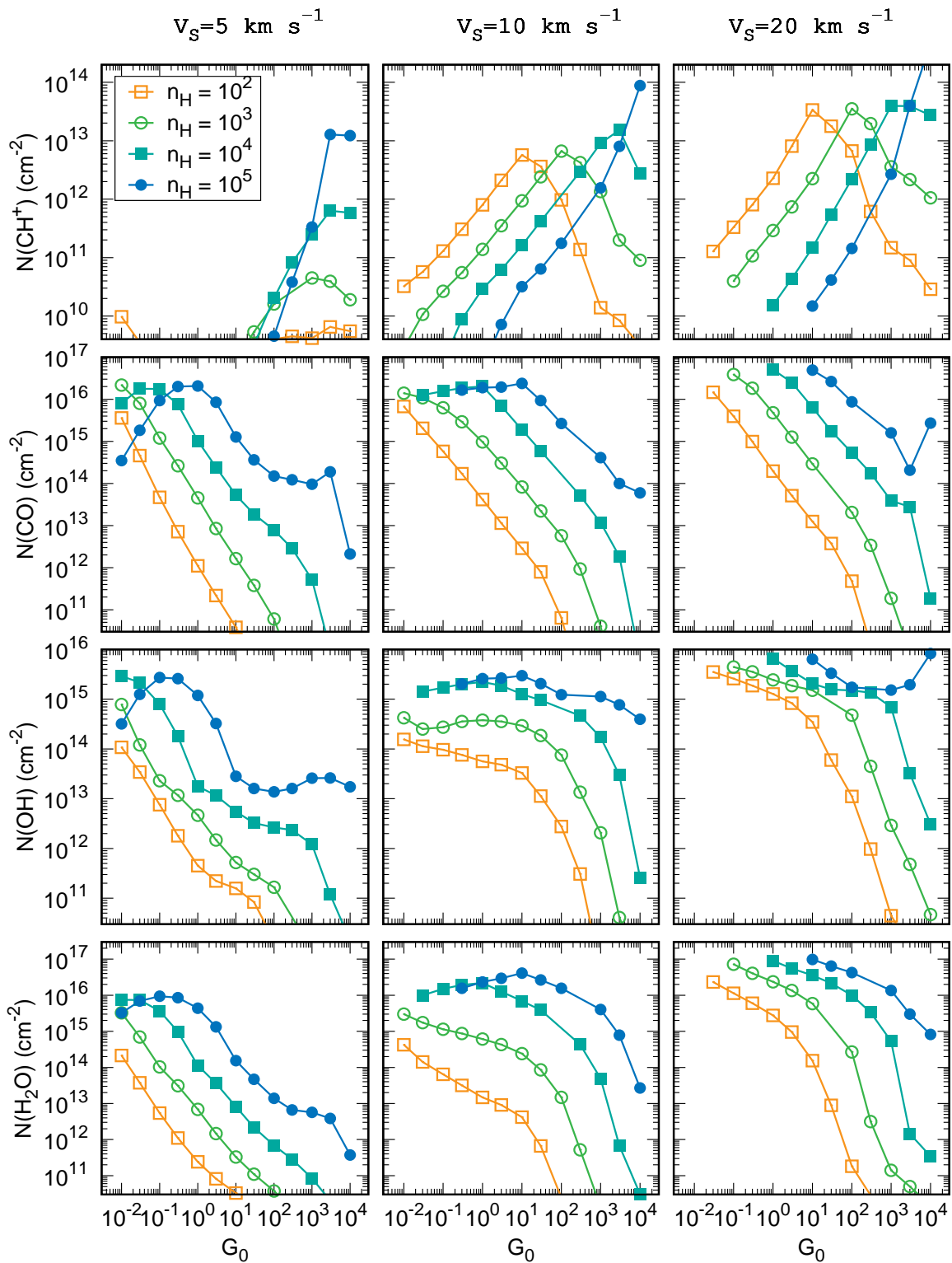}
\caption{Column densities of \CHp, CO, OH, and H$_2$O calculated across shocks 
propagating at 5 (left panels), 10 (middle panels), and 20 (right panels) \kms\ in
media of density $\dens=10^2$ (empty orange squares), $10^3$ (empty green circles),
$10^4$ (filled green squares), and $10^5$ (filled blue circles) \cc, as function
of the impinging radiation field $G_0$, and for a buffer visual extinction $A_V^0=0.1$.
All other parameters are set to their standard values (see Table \ref{Tab-main}).}
\label{Fig-coldens-av1m1}
\end{center}
\end{figure*}

\begin{table*}
\begin{center}
\caption{Total hydrogen column density $N_{\rm H}$ and column densities of several species 
computed across irradiated shocks propagating in a medium of density $\dens=10^4$ \cc\ at a 
visual extinction $A_V^0=0.1$. All other parameters are set to their standard values (Table 
\ref{Tab-main}). Numbers in parenthesis are powers of ten.}
\label{Tab-coldens}
\begin{tabular}{l r r r r r r r r r r }
\hline
  & \multicolumn{1}{c}{$N_{\rm H}$} & \multicolumn{1}{c}{$N({\rm H}_2)$} & \multicolumn{1}{c}{$N(\Cp)$ } & \multicolumn{1}{c}{$N(\CHp)$} & \multicolumn{1}{c}{$N(\SHp)$} & \multicolumn{1}{c}{$N({\rm O})$} & \multicolumn{1}{c}{$N({\rm O}_2)$} & \multicolumn{1}{c}{$N({\rm OH})$} & \multicolumn{1}{c}{$N({\rm H}_2{\rm O})$} & \multicolumn{1}{c}{$N({\rm CO})$} \\
  & \multicolumn{1}{c}{cm$^{-2}$  } & \multicolumn{1}{c}{cm$^{-2}$     } & \multicolumn{1}{c}{cm$^{-2}$} & \multicolumn{1}{c}{cm$^{-2}$} & \multicolumn{1}{c}{cm$^{-2}$} & \multicolumn{1}{c}{cm$^{-2}$   } & \multicolumn{1}{c}{cm$^{-2}$     } & \multicolumn{1}{c}{cm$^{-2}$    } & \multicolumn{1}{c}{cm$^{-2}$            } & \multicolumn{1}{c}{cm$^{-2}$    } \\
\hline
\\
\hline
\multicolumn{11}{c}{$V_S$ = 5 \kms} \\
\hline
$G_0 = 1$    & 5.6 (20) & 2.8 (20) & 4.0 (14) & 2.5 (08) & 1.3 (10) & 1.6 (17) & 2.8 (13) & 1.8 (13) & 1.1 (14) & 1.0 (15) \\
$G_0 = 10$   & 1.3 (20) & 6.5 (19) & 1.1 (15) & 5.5 (08) & 3.9 (08) & 3.9 (16) & 1.1 (12) & 5.5 (12) & 8.2 (12) & 5.3 (13) \\
$G_0 = 100$  & 5.8 (19) & 2.8 (19) & 3.7 (15) & 2.0 (10) & 3.7 (07) & 1.8 (16) & 3.5 (10) & 2.6 (12) & 6.9 (11) & 7.9 (12) \\
$G_0 = 1000$ & 4.3 (19) & 1.2 (19) & 5.7 (15) & 2.5 (11) & 1.7 (07) & 1.3 (16) & 1.4 (09) & 1.2 (12) & 8.4 (10) & 5.2 (11) \\
\\
\hline
\multicolumn{11}{c}{$V_S$ = 10 \kms} \\
\hline
$G_0 = 1$    & 6.6 (20) & 3.3 (20) & 1.4 (14) & 2.9 (10) & 5.4 (11) & 1.4 (17) & 1.6 (15) & 2.2 (15) & 2.2 (16) & 2.0 (16) \\
$G_0 = 10$   & 2.6 (20) & 1.3 (20) & 6.1 (14) & 1.6 (11) & 3.0 (12) & 6.9 (16) & 6.1 (12) & 1.3 (15) & 6.9 (15) & 1.9 (15) \\
$G_0 = 100$  & 3.1 (20) & 1.3 (20) & 4.8 (15) & 1.1 (12) & 1.1 (13) & 9.5 (16) & 1.1 (14) & 7.8 (14) & 1.9 (15) & 3.4 (14) \\
$G_0 = 1000$ & 4.2 (19) & 1.2 (19) & 4.0 (15) & 9.1 (12) & 1.6 (12) & 1.3 (16) & 1.8 (11) & 1.7 (14) & 4.9 (13) & 1.2 (13) \\
\\
\hline
\multicolumn{11}{c}{$V_S$ = 20 \kms} \\
\hline
$G_0 = 1$    & 8.4 (20) & 4.2 (20) & 2.1 (13) & 1.5 (10) & 8.6 (11) & 8.2 (16) & 5.8 (15) & 6.6 (15) & 9.0 (16) & 5.1 (16) \\
$G_0 = 10$   & 7.7 (20) & 3.8 (20) & 7.1 (14) & 1.5 (11) & 3.6 (12) & 1.8 (17) & 3.4 (14) & 2.1 (15) & 3.6 (16) & 6.4 (15) \\
$G_0 = 100$  & 2.9 (20) & 1.3 (20) & 3.1 (15) & 2.2 (12) & 1.9 (13) & 7.5 (16) & 1.2 (13) & 1.5 (15) & 9.9 (15) & 5.3 (14) \\
$G_0 = 1000$ & 6.5 (19) & 1.8 (19) & 4.0 (15) & 4.0 (13) & 3.4 (13) & 1.8 (16) & 7.6 (11) & 7.0 (14) & 5.3 (14) & 3.9 (13) \\
\end{tabular}
\end{center}
\end{table*}

The specificity of \CHp\ chemistry makes irradiated shocks efficient factories of \CHp. The same 
does not hold for all molecules, and in particular those efficiently destroyed 
by UV photons. To illustrate this, we compare in Fig. \ref{Fig-coldens-av1m1} the column 
densities of \CHp\ computed across molecular shocks to those of CO, OH, and H$_2$O. As 
already shown by \citet{Falgarone2010a}, the increase of the column density of \CHp\ with 
increasing values of $G_0$ almost systematically comes with a decrease of the column 
densities of CO, OH, and H$_2$O. The associated drop can cover one to several orders
of magnitude depending on the medium density and velocity of the shock. We find that 
the column density of \CHp\ linearly increases with $G_0$ up to a maximal value, while 
the decrease of the column density of CO, OH, and H$_2$O is steeper, especially at high 
$G_0$ and large $V_S$.

The predicted values obtained for $\dens=10^4$ \cc\ and $1 \leqslant G_0 \leqslant 10^3$ 
are reported in Table \ref{Tab-coldens}, along with the predictions on the total 
hydrogen column density $N_{\rm H}$ and the column densities of \HH, \Cp, \SHp, O, and O$_2$.
The column density ratios $N({\rm H}_2{\rm O})/N({\rm OH})$ and $N({\rm CO})/N({\rm OH})$
appear to decrease strongly with $G_0$, a result in line with \citet{Melnick2015} who 
find similar behaviour in dense environments illuminated by moderate radiation fields 
($\dens = 10^5$ \cc, $G_0 \leqslant 10$). Table \ref{Tab-coldens} reveals that other 
chemical species follow \CHp\ and are strongly enhanced in irradiated shocks. This
is the case, for instance, of the mercapto cation \SHp. For a shock velocity larger than
10 \kms, the predicted column density of \SHp\ is found to increase with $G_0$, although 
with a power-law slope somehow shallower ($N(\SHp) \propto \sqrt{G_0}$) than that found
for \CHp.
All these results indicate that irradiated shocks carry many specific chemical signatures
that could be identified using the current observing facilities.

\section{Discussion} \label{Sect-discussion}

\subsection{Kinetic and radiative energies budget} \label{Sect-kin-rad-nrj}

\begin{figure}
\begin{center}
\includegraphics[width=9.0cm,trim = 1.5cm 2cm 1cm 2cm, clip,angle=0]{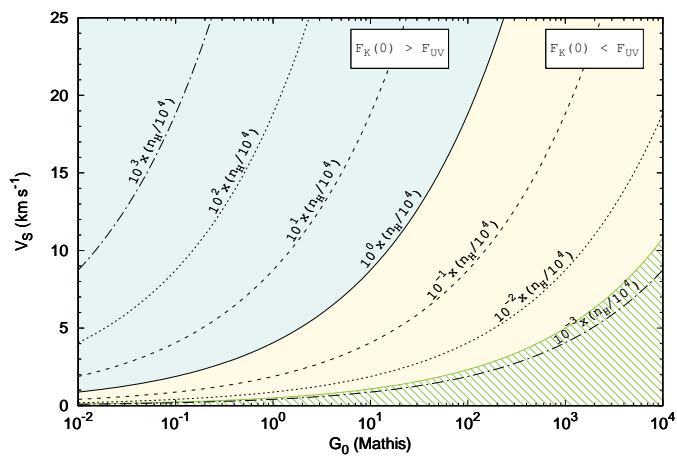}
\caption{Isocontours of the ratio of the initial kinetic energy flux contained in an 
interstellar shock to the radiative energy flux contained in ultraviolet photons, 
$\mathscr{F}_{\rm K}(0) / \mathscr{F}_{\rm UV}$, where $\mathscr{F}_{\rm UV} = 7.7 
\times 10^{-4} G_0$ erg cm$^{-2}$ s$^{-1}$, as a function of $G_0$, the shock 
speed $V_S$, and the pre-shock density $\dens$ (indicated on the isocontours). The
dashed region corresponds to the parameter domain, for $\dens=10^4$ \cc, where the 
amount of UV radiative energy reprocessed into heat within a shock is larger (on 
average) than the amount of mechanical energy reprocessed into heat (Eq. 
\ref{Eq-UV-dominant}).}
\label{Fig-kin-rad-nrj}
\end{center}
\end{figure}

Establishing the energy budget of irradiated molecular shocks is complex. In 
the previous sections, the results on the molecular lines emission were systematically 
presented with respect to the initial kinetic energy flux of the shock (see Fig. 
\ref{Fig-H2-CHp-emis}). The underlying interpretation was that shock radiative emission 
is an outcome of the reprocessing of mechanical energy only. This view could be misleading 
as the matter contained in irradiated shocks also reprocesses part of the input UV 
radiative energy.

To discuss this point, we compare in Fig. \ref{Fig-kin-rad-nrj} the initial kinetic 
energy flux of shocks with the radiative energy flux contained in the 
ambient UV field depending on the model parameters. The grid of physical
conditions explored in this work clearly covers a broad range of scenarios in which the 
input kinetic energy flux is either equal to, greater than, or less than
the flux contained in the UV field\footnote{Amusingly, the amount of kinetic 
energy flux initially contained in a shock propagating at 10 \kms\ in a medium of density 
$\dens=10^4$ \cc\ is equivalent to that contained in UV photons in the ISRF with an 
amplification factor $G_0=10$ (see Fig. \ref{Fig-kin-rad-nrj}).}. In what proportion 
these energy sources are converted into molecular lines over the duration of a shock 
depends on the reprocessing lengths and efficiency of energy conversion.

To simplify, we perform estimations of these quantities considering only 
reprocessing into thermal energy, i.e. assuming that atomic and molecular lines 
primarily result from the gas heating. This approach is valid as long as the
radiative pumping is not the dominant excitation process of the species
considered.

By definition (see Eq. \ref{Eq-size}), the input mechanical energy is entirely reprocessed 
over the size of the shock, shown in Fig. \ref{Fig-shock-size}. For shock velocity
larger than 5 \kms, and standard magnetization ($b=1$), the fraction of energy spent in 
magnetic compression is found to be smaller than a few tens of percent \citep{Lesaffre2013}. 
The mechanical energy is therefore mostly converted into atomic and molecular lines with 
an efficiency of several tens of percent, a value that increases with the speed of the 
shock. In contrast, the radiative energy contained in UV photons is reprocessed by grains 
over about one magnitude of extinction which corresponds to distances, 
\begin{equation}
l_{\rm UV} \sim 12 \times 10^3 \,\, {\rm au} \,\, \left( \frac{\dens}{10^4 \cc} \right)^{-1},
\end{equation}
10 to 100 times larger than the shock sizes. In this case, the amount of radiative energy 
converted into thermal energy of the gas is given by the efficiency of the photoelectric 
effect which ranges between 2 and 7\% (e.g. \citealt{Ingalls2002}). Putting together 
all these characteristics, we find that the amount of radiative energy reprocessed into 
heat dominates, on average, the amount of mechanical energy reprocessed into heat for 
\begin{equation} \label{Eq-UV-dominant}
\frac{G_0}{\dens} \left(\frac{V_S}{5\,\kms}\right)^{-3} \geqslant 0.1 \,\,{\rm cm}^{3},
\end{equation}
i.e. over a small fraction of our grid of models, at low density, low shock velocity, 
and large radiation field intensity (see Fig. \ref{Fig-kin-rad-nrj}).

This result is particularly well illustrated in Fig \ref{Fig-temp-add}. The input 
kinetic energy fluxes of the models explored in this work are found to be always 
sufficient to modify the thermal state of the gas. The relative increase of the gas 
temperature can be, however, quite small in low velocity shocks propagating in strongly 
irradiated environments (e.g. $V_S=5$ \kms\ and $G_0=10^4$) where the maximal temperature 
reached in the shock is almost identical to that found in the post-shock gas. This 
happens because the heating rate induced by UV photons is locally comparable to that 
induced by the dissipation of mechanical energy. In this case, the sole impact of the 
shock is to slightly compress the gas, increasing its thermal pressure and modifying 
its coupling with the external UV field. Except for kinematic signatures, shocks in 
these environments are probably undetectable because they do not impact significantly 
the thermochemical state of the gas.

\subsection{J-type shocks and self-generated UV radiation field} \label{Sect-limits}

The model presented in this work 
contains all the necessary ingredients 
to compute the local emission and absorption of photons by gas and dust particles. It is, 
however, important to stress that neither the feedback of the photons produced by the shock 
itself on the surrounding material nor the radiative transfer of these photons within the
shock are currently treated. This caveat strongly limits the upper range of J-type shocks 
velocities that can be studied with the Paris-Durham shock code.


The adiabatic jump conditions (Eq. \ref{Eq-Tmax}) imply that the temperature 
reached in J-type shocks are often large enough to either ionize chemical species or 
excite their electronic levels \citep{Shull1979}. These mechanisms, followed by radiative 
recombinations and radiative decay, induce the production of high energy photons that heat, 
dissociate, or ionize the gas within the shock, but also in the pre-shock and post-shock
media, leading to the formation of a radiative precursor 
\citep{Neufeld1989,Hollenbach1989,Draine1993}. 

Without magnetic field, J-type shocks with velocities larger than 25 \kms\ emit Ly 
$\alpha$ and Ly $\beta$ photons \citep{Shull1979}. Their energies are strong enough to 
enhance the photoelectric effect, hence the temperature of the pre-shock gas, and 
the dissociation and ionization of many species. For shocks propagating at 30 \kms\ 
in a medium of density
$\dens = 2 \times 10^5$ \cc, \citet{Flower2010a} estimated a UV flux of $10^{11}$ 
photons cm$^{-2}$ s$^{-1}$, corresponding to an equivalent $G_0 = 10^3$. Such
an effect cannot be neglected.

Finding a self-consistent solution requires  building an iterative procedure 
to compute the photon flux emitted by the shock and also to construct a radiative 
code capable of solving the transfer and the scattering of Ly $\alpha$ and Ly $\beta$ 
photons and other optically thick lines. We deliberately chose  to 
limit our analysis to low velocity C-, C*, and CJ-type shocks ($V_S \leqslant 25$ 
\kms) where the production of high energy photons is negligible. All these developments 
are however in progress and will be the subject of a forthcoming paper on self-irradiated
shocks.

\section{Conclusions} \label{Sect-conclusion}

We have built a model designed to compute the dynamics and thermochemistry of 
molecular shocks in environments irradiated by an external UV radiation field. The 
framework is that of a plane-parallel shock at steady-state propagating at a given 
position of a plane-parallel PDR. 
The steady-state 
solutions of magnetohydrodynamics equations are computed through a new and convoluted 
algorithm capable of following the evolution of the gas near and through sonic points and along 
subsonic trajectories of molecular shocks. The resulting code allows the simultaneous 
and self-consistent treatment of gas dynamics and chemistry 
and of the radiative transfer in a wide variety of physical conditions including diffuse 
clouds and dense and hot PDRs. It therefore provides a unique 
tool to study the combined effect of radiative and mechanical energy on the 
thermochemistry of interstellar matter.

Ultraviolet photons exert an impact on molecular shocks that goes far beyond 
ionization and dissociation: they modify the structure of the shock itself. Because 
of the combined effects of dynamics and photoprocesses, irradiated C-type shocks 
shrink with the strength of the radiation field. If the cooling is not strong enough, 
the gas may become subsonic along the trajectory. This induces the formation of
unexplored categories of stationary shocks called 
C*-type or CJ-type, depending on whether the sonic point is crossed continuously or not. 
These types of shocks are dominant in weakly magnetized, 
diffuse, or irradiated environments. We find, for instance, that only low velocity 
C-type shocks ($V_S \leqslant 5$ \kms) can exist in typical PDRs ($\dens \sim 
10^4$ \cc, $G_0 \sim 100$) or in very diffuse clouds ($\dens \sim 30$ \cc, $G_0 \sim 1$) 
at low extinction and for standard magnetic field strengths ($b=1$)~; higher velocity shocks are 
either C* or CJ (for $5 \leqslant V_S \leqslant 25$ \kms), or J (for $V_S \geqslant 
25$ \kms).

The increase of the photodissociation of \HH\ and the diminution of the shock size with
increasing UV field, lead to larger temperature which eventually triggers the collisional 
dissociation of \HH\ even at low velocity. All these processes strongly influence the 
energy budget of shocks but also the chemistry and excitation of atomic and 
molecular lines. Several molecules, such as CO, OH, and H$_2$O, are found to be destroyed
too quickly by UV photons to be efficiently produced in irradiated shocks, while the
formations of others, such as \CHp\ and \SHp, are boosted by the entwined actions of kinetic 
and radiative energies.

The best diagnostics of irradiated shocks are probably the rotational lines of \CHp.
The combination of high production and excitation rates drastically change the 
efficiency of \CHp\ emission which can carry, at large density, as much as several 
percent of the shock kinetic energy. The distribution of \CHp\ emission over its 
different rotational lines is found to be highly sensitive to the shock velocity,
hence providing a unique set of tracers of shock kinetic energy.

One unexpected result is the weak sensitivity of the total \HH\ emission to the 
input UV radiation field, combined with the strong sensitivity of its excitation
diagram. %
This finding extends the results of previous studies on the excitation of \HH\ and 
offers a reliable criterion to infer the irradiation conditions of known interstellar 
shocks that can be tested with the impending launch of the JWST.

The analysis of the shock energy balance reveals the strong interplay of radiative 
and mechanical energies. In most cases, molecular line emissions 
result from the conversion of the shock mechanical energy while this conversion 
is affected by the impinging UV field. This ceases to be true, however, at low velocity, low density, or 
large radiation fields where the input UV radiative energy reprocessed by the shock 
itself becomes dominant.

The work presented in this paper sheds new light on the interpretation 
of line emission and provides new templates for the analysis of observations.
It calls, however, for several developments. Deliberately limited to a handful of species, 
it would be interesting to extend our analysis to a more systematic study of the 
chemistry and address, for instance, the formation and excitation of other known tracers of 
shocks and turbulent dissipation, such as SiO, \SHp, and \HCOp. Similarly, it will soon 
be essential to broaden the exploration of the physical conditions to unveil 
the signatures of high velocity irradiated J-type shocks and those of shocks propagating 
in highly and weakly magnetized media. One of the most promising applications is the 
disentanglement of the contribution of stellar radiation and mechanical feedback in the 
emission bulk of galaxies in order to establish their energy budget and understand the
nature of their chemical complexity.

\begin{acknowledgements}

The research leading to these results has received fundings from the European Research 
Council, under the European Community's Seventh framework Programme, through the 
Advanced Grant MIST (FP7/2017-2022, No 787813). We would also like to acknowledge the 
support from the Programme National ``Physique et Chimie du Milieu Interstellaire'' 
(PCMI) of CNRS/INSU with INC/INP co-funded by CEA and CNES.

\end{acknowledgements}

\bibliographystyle{aa} 
\bibliography{mybib}

\begin{thebibliography}{113}
\expandafter\ifx\csname natexlab\endcsname\relax\def\natexlab#1{#1}\fi

\bibitem[{{Abgrall} {et~al.}(1992){Abgrall}, {Le Bourlot}, {Pineau des
  For\^ets}, {Roueff}, {Flower}, \& {Heck}}]{Abgrall1992}
{Abgrall}, H., {Le Bourlot}, J., {Pineau des For\^ets}, G., {et~al.} 1992,
  \aap, 253, 525

\bibitem[{{Aikawa} {et~al.}(1996){Aikawa}, {Miyama}, {Nakano}, \&
  {Umebayashi}}]{Aikawa1996}
{Aikawa}, Y., {Miyama}, S.~M., {Nakano}, T., \& {Umebayashi}, T. 1996, \apj,
  467, 684

\bibitem[{{Anderl} {et~al.}(2013){Anderl}, {Guillet}, {Pineau des For{\^e}ts},
  \& {Flower}}]{Anderl2013}
{Anderl}, S., {Guillet}, V., {Pineau des For{\^e}ts}, G., \& {Flower}, D.~R.
  2013, \aap, 556, A69

\bibitem[{{Appleton} {et~al.}(2006){Appleton}, {Xu}, {Reach}, {Dopita}, {Gao},
  {Lu}, {Popescu}, {Sulentic}, {Tuffs}, \& {Yun}}]{Appleton2006}
{Appleton}, P.~N., {Xu}, K.~C., {Reach}, W., {et~al.} 2006, \apjl, 639, L51

\bibitem[{{Bakes} \& {Tielens}(1994)}]{Bakes1994}
{Bakes}, E.~L.~O. \& {Tielens}, A.~G.~G.~M. 1994, \apj, 427, 822

\bibitem[{{Balsara}(1996)}]{Balsara1996}
{Balsara}, D.~S. 1996, \apj, 465, 775

\bibitem[{{Barlow}(1978)}]{Barlow1978}
{Barlow}, M.~J. 1978, \mnras, 183, 367

\bibitem[{{Burton} {et~al.}(1990){Burton}, {Hollenbach}, {Haas}, \&
  {Erickson}}]{Burton1990}
{Burton}, M.~G., {Hollenbach}, D.~J., {Haas}, M.~R., \& {Erickson}, E.~F. 1990,
  \apj, 355, 197

\bibitem[{{Cabrit} {et~al.}(2004){Cabrit}, {Lefloch}, {Cernicharo}, {Pineau Des
  For{\^e}ts}, {Lebourlot}, \& {Flower}}]{Cabrit2004}
{Cabrit}, S., {Lefloch}, B., {Cernicharo}, J., {et~al.} 2004, in The Dense
  Interstellar Medium in Galaxies, ed. S.~{Pfalzner}, C.~{Kramer},
  C.~{Staubmeier}, \& A.~{Heithausen}, Vol.~91, 587

\bibitem[{{Chapman} \& {Wardle}(2006)}]{Chapman2006}
{Chapman}, J.~F. \& {Wardle}, M. 2006, \mnras, 371, 513

\bibitem[{{Chernoff}(1987)}]{Chernoff1987}
{Chernoff}, D.~F. 1987, \apj, 312, 143

\bibitem[{{Chi\`eze} {et~al.}(1998){Chi\`eze}, {Pineau des For\^ets}, \&
  {Flower}}]{Chieze1998}
{Chi\`eze}, J.-P., {Pineau des For\^ets}, G., \& {Flower}, D.~R. 1998, \mnras,
  295, 672

\bibitem[{{Ciolek} \& {Roberge}(2002)}]{Ciolek2002}
{Ciolek}, G.~E. \& {Roberge}, W.~G. 2002, \apj, 567, 947

\bibitem[{{Ciolek} {et~al.}(2004){Ciolek}, {Roberge}, \&
  {Mouschovias}}]{Ciolek2004}
{Ciolek}, G.~E., {Roberge}, W.~G., \& {Mouschovias}, T.~C. 2004, \apj, 610, 781

\bibitem[{{Dopita} \& {Sutherland}(2003)}]{Dopita2003}
{Dopita}, M.~A. \& {Sutherland}, R.~S. 2003, Astrophysics of the Diffuse
  Universe (Springer, Berlin, Heidelberg)

\bibitem[{{Draine}(1980)}]{Draine1980}
{Draine}, B.~T. 1980, \apj, 241, 1021

\bibitem[{{Draine}(2002)}]{Draine2002}
{Draine}, B.~T. 2002, The Cold Universe, ed. A.~W. {Blain}, F.~{Combes}, B.~T.
  {Draine}, D.~{Pfenniger}, \& Y.~{Revaz}, Vol.~32 (Swiss Society for
  Astrophysics and Astronomy, Saas-Fee Advanced Course)

\bibitem[{{Draine}(2011)}]{Draine2011}
{Draine}, B.~T. 2011, {Physics of the Interstellar and Intergalactic Medium}
  (Princeton University Press)

\bibitem[{{Draine} \& {Bertoldi}(1996)}]{Draine1996}
{Draine}, B.~T. \& {Bertoldi}, F. 1996, \apj, 468, 269

\bibitem[{{Draine} \& {Lee}(1984)}]{Draine1984}
{Draine}, B.~T. \& {Lee}, H.~M. 1984, \apj, 285, 89

\bibitem[{{Draine} \& {Li}(2007)}]{Draine2007}
{Draine}, B.~T. \& {Li}, A. 2007, \apj, 657, 810

\bibitem[{{Draine} \& {McKee}(1993)}]{Draine1993}
{Draine}, B.~T. \& {McKee}, C.~F. 1993, \araa, 31, 373

\bibitem[{{Draine} \& {Sutin}(1987)}]{Draine1987}
{Draine}, B.~T. \& {Sutin}, B. 1987, \apj, 320, 803

\bibitem[{{Falgarone} {et~al.}(2010){Falgarone}, {Ossenkopf}, {Gerin},
  {Lesaffre}, {Godard}, {Pearson}, {Cabrit}, {Joblin}, {Benz}, {Boulanger},
  {Fuente}, {G{\"u}sten}, {Harris}, {Klein}, {Kramer}, {Lord}, {Martin},
  {Martin-Pintado}, {Neufeld}, {Phillips}, {R{\"o}llig}, {Simon}, {Stutzki},
  {van der Tak}, {Teyssier}, {Yorke}, {Erickson}, {Fich}, {Jellema}, {Marston},
  {Risacher}, {Salez}, \& {Schm{\"u}lling}}]{Falgarone2010a}
{Falgarone}, E., {Ossenkopf}, V., {Gerin}, M., {et~al.} 2010, \aap, 518, L118

\bibitem[{{Falgarone} {et~al.}(2017){Falgarone}, {Zwaan}, {Godard}, {Bergin},
  {Ivison}, {Andreani}, {Bournaud}, {Bussmann}, {Elbaz}, {Omont}, {Oteo}, \&
  {Walter}}]{Falgarone2017}
{Falgarone}, E., {Zwaan}, M.~A., {Godard}, B., {et~al.} 2017, \nat, 548, 430

\bibitem[{{Federman} {et~al.}(1979){Federman}, {Glassgold}, \&
  {Kwan}}]{Federman1979}
{Federman}, S.~R., {Glassgold}, A.~E., \& {Kwan}, J. 1979, \apj, 227, 466

\bibitem[{{Fitzpatrick} \& {Massa}(1986)}]{Fitzpatrick1986}
{Fitzpatrick}, E.~L. \& {Massa}, D. 1986, \apj, 307, 286

\bibitem[{{Flower}(2010)}]{Flower2010}
{Flower}, D. 2010, in Lecture Notes in Physics, Berlin Springer Verlag, Vol.
  793, Lecture Notes in Physics, Berlin Springer Verlag, ed. P.~J.~V. {Garcia}
  \& J.~M. {Ferreira}, 161

\bibitem[{{Flower} {et~al.}(2003){Flower}, {Le Bourlot}, {Pineau des
  For{\^e}ts}, \& {Cabrit}}]{Flower2003a}
{Flower}, D.~R., {Le Bourlot}, J., {Pineau des For{\^e}ts}, G., \& {Cabrit}, S.
  2003, \mnras, 341, 70

\bibitem[{{Flower} \& {Pineau des For{\^e}ts}(2003)}]{Flower2003}
{Flower}, D.~R. \& {Pineau des For{\^e}ts}, G. 2003, \mnras, 343, 390

\bibitem[{{Flower} \& {Pineau des For{\^e}ts}(2010)}]{Flower2010a}
{Flower}, D.~R. \& {Pineau des For{\^e}ts}, G. 2010, \mnras, 406, 1745

\bibitem[{{Flower} \& {Pineau des For{\^e}ts}(2013)}]{Flower2013}
{Flower}, D.~R. \& {Pineau des For{\^e}ts}, G. 2013, \mnras, 436, 2143

\bibitem[{{Flower} \& {Pineau des For{\^e}ts}(2015)}]{Flower2015}
{Flower}, D.~R. \& {Pineau des For{\^e}ts}, G. 2015, \aap, 578, A63

\bibitem[{{Flower} {et~al.}(1985){Flower}, {Pineau des For\^ets}, \&
  {Hartquist}}]{Flower1985}
{Flower}, D.~R., {Pineau des For\^ets}, G., \& {Hartquist}, T.~W. 1985, \mnras,
  216, 775

\bibitem[{{Flower} {et~al.}(2005){Flower}, {Pineau des For{\^e}ts}, \&
  {Walmsley}}]{Flower2005}
{Flower}, D.~R., {Pineau des For{\^e}ts}, G., \& {Walmsley}, C.~M. 2005, \aap,
  436, 933

\bibitem[{{Giannini} {et~al.}(2004){Giannini}, {McCoey}, {Caratti o Garatti},
  {Nisini}, {Lorenzetti}, \& {Flower}}]{Giannini2004}
{Giannini}, T., {McCoey}, C., {Caratti o Garatti}, A., {et~al.} 2004, \aap,
  419, 999

\bibitem[{{Gibb} {et~al.}(2000){Gibb}, {Whittet}, {Schutte}, {Boogert},
  {Chiar}, {Ehrenfreund}, {Gerakines}, {Keane}, {Tielens}, {van Dishoeck}, \&
  {Kerkhof}}]{Gibb2000}
{Gibb}, E.~L., {Whittet}, D.~C.~B., {Schutte}, W.~A., {et~al.} 2000, \apj, 536,
  347

\bibitem[{{Gillmon} {et~al.}(2006){Gillmon}, {Shull}, {Tumlinson}, \&
  {Danforth}}]{Gillmon2006}
{Gillmon}, K., {Shull}, J.~M., {Tumlinson}, J., \& {Danforth}, C. 2006, \apj,
  636, 891

\bibitem[{{Godard} \& {Cernicharo}(2013)}]{Godard2013}
{Godard}, B. \& {Cernicharo}, J. 2013, \aap, 550, A8

\bibitem[{{Godard} {et~al.}(2014){Godard}, {Falgarone}, \& {Pineau des
  For{\^e}ts}}]{Godard2014}
{Godard}, B., {Falgarone}, E., \& {Pineau des For{\^e}ts}, G. 2014, \aap, 570,
  A27

\bibitem[{{Gorti} \& {Hollenbach}(2002)}]{Gorti2002}
{Gorti}, U. \& {Hollenbach}, D. 2002, \apj, 573, 215

\bibitem[{{Guillard} {et~al.}(2009){Guillard}, {Boulanger}, {Pineau des
  For{\^e}ts}, \& {Appleton}}]{Guillard2009}
{Guillard}, P., {Boulanger}, F., {Pineau des For{\^e}ts}, G., \& {Appleton},
  P.~N. 2009, \aap, 502, 515

\bibitem[{{Guillet}(2008)}]{Guillet2008}
{Guillet}, V. 2008, PhD thesis, Universit{\'e} Paris XI

\bibitem[{{Guillet} {et~al.}(2009){Guillet}, {Jones}, \& {Pineau des
  For{\^e}ts}}]{Guillet2009}
{Guillet}, V., {Jones}, A.~P., \& {Pineau des For{\^e}ts}, G. 2009, \aap, 497,
  145

\bibitem[{{Guillet} {et~al.}(2007){Guillet}, {Pineau des For{\^e}ts}, \&
  {Jones}}]{Guillet2007}
{Guillet}, V., {Pineau des For{\^e}ts}, G., \& {Jones}, A.~P. 2007, \aap, 476,
  263

\bibitem[{{Guillet} {et~al.}(2011){Guillet}, {Pineau des For{\^e}ts}, \&
  {Jones}}]{Guillet2011}
{Guillet}, V., {Pineau des For{\^e}ts}, G., \& {Jones}, A.~P. 2011, \aap, 527,
  A123

\bibitem[{{Gusdorf} {et~al.}(2017){Gusdorf}, {Anderl}, {Lefloch}, {Leurini},
  {Wiesemeyer}, {G{\"u}sten}, {Benedettini}, {Codella}, {Godard},
  {G{\'o}mez-Ruiz}, {Jacobs}, {Kristensen}, {Lesaffre}, {Pineau des
  For{\^e}ts}, \& {Lis}}]{Gusdorf2017}
{Gusdorf}, A., {Anderl}, S., {Lefloch}, B., {et~al.} 2017, \aap, 602, A8

\bibitem[{{Gusdorf} {et~al.}(2008{\natexlab{a}}){Gusdorf}, {Cabrit}, {Flower},
  \& {Pineau des For{\^e}ts}}]{Gusdorf2008a}
{Gusdorf}, A., {Cabrit}, S., {Flower}, D.~R., \& {Pineau des For{\^e}ts}, G.
  2008{\natexlab{a}}, \aap, 482, 809

\bibitem[{{Gusdorf} {et~al.}(2008{\natexlab{b}}){Gusdorf}, {Pineau des
  For{\^e}ts}, {Cabrit}, \& {Flower}}]{Gusdorf2008}
{Gusdorf}, A., {Pineau des For{\^e}ts}, G., {Cabrit}, S., \& {Flower}, D.~R.
  2008{\natexlab{b}}, \aap, 490, 695

\bibitem[{{Habart} {et~al.}(2011){Habart}, {Abergel}, {Boulanger}, {Joblin},
  {Verstraete}, {Compi{\`e}gne}, {Pineau Des For{\^e}ts}, \& {Le
  Bourlot}}]{Habart2011}
{Habart}, E., {Abergel}, A., {Boulanger}, F., {et~al.} 2011, \aap, 527, A122

\bibitem[{{Hasegawa} \& {Herbst}(1993)}]{Hasegawa1993}
{Hasegawa}, T.~I. \& {Herbst}, E. 1993, \mnras, 263, 589

\bibitem[{{Heays} {et~al.}(2017){Heays}, {Bosman}, \& {van
  Dishoeck}}]{Heays2017}
{Heays}, A.~N., {Bosman}, A.~D., \& {van Dishoeck}, E.~F. 2017, \aap, 602, A105

\bibitem[{{Hennebelle} \& {Falgarone}(2012)}]{Hennebelle2012}
{Hennebelle}, P. \& {Falgarone}, E. 2012, \aapr, 20, 55

\bibitem[{{Hewitt} {et~al.}(2009){Hewitt}, {Yusef-Zadeh}, \&
  {Wardle}}]{Hewitt2009}
{Hewitt}, J.~W., {Yusef-Zadeh}, F., \& {Wardle}, M. 2009, \apjl, 706, L270

\bibitem[{{Hily-Blant} {et~al.}(2018){Hily-Blant}, {Faure}, {Rist}, {Pineau des
  For{\^e}ts}, \& {Flower}}]{Hily-Blant2018}
{Hily-Blant}, P., {Faure}, A., {Rist}, C., {Pineau des For{\^e}ts}, G., \&
  {Flower}, D.~R. 2018, \mnras, 477, 4454

\bibitem[{{Hindmarsh}(1983)}]{Hindmarsh1983}
{Hindmarsh}, A.~C. 1983, {Scientific Computing} (Amsterdam: North-Holland)

\bibitem[{{Hollenbach} {et~al.}(2009){Hollenbach}, {Kaufman}, {Bergin}, \&
  {Melnick}}]{Hollenbach2009}
{Hollenbach}, D., {Kaufman}, M.~J., {Bergin}, E.~A., \& {Melnick}, G.~J. 2009,
  \apj, 690, 1497

\bibitem[{{Hollenbach} \& {McKee}(1979)}]{Hollenbach1979}
{Hollenbach}, D. \& {McKee}, C.~F. 1979, \apjs, 41, 555

\bibitem[{{Hollenbach} \& {McKee}(1989)}]{Hollenbach1989}
{Hollenbach}, D. \& {McKee}, C.~F. 1989, \apj, 342, 306

\bibitem[{{Hollenbach} \& {Tielens}(1999)}]{Hollenbach1999}
{Hollenbach}, D.~J. \& {Tielens}, A.~G.~G.~M. 1999, Reviews of Modern Physics,
  71, 173

\bibitem[{{Ingalls} {et~al.}(2002){Ingalls}, {Reach}, \& {Bania}}]{Ingalls2002}
{Ingalls}, J.~G., {Reach}, W.~T., \& {Bania}, T.~M. 2002, \apj, 579, 289

\bibitem[{{Jones} {et~al.}(1996){Jones}, {Tielens}, \&
  {Hollenbach}}]{Jones1996}
{Jones}, A.~P., {Tielens}, A.~G.~G.~M., \& {Hollenbach}, D.~J. 1996, \apj, 469,
  740

\bibitem[{{Karska} {et~al.}(2014){Karska}, {Kristensen}, {van Dishoeck},
  {Drozdovskaya}, {Mottram}, {Herczeg}, {Bruderer}, {Cabrit}, {Evans},
  {Fedele}, {Gusdorf}, {J{\o}rgensen}, {Kaufman}, {Melnick}, {Neufeld},
  {Nisini}, {Santangelo}, {Tafalla}, \& {Wampfler}}]{Karska2014}
{Karska}, A., {Kristensen}, L.~E., {van Dishoeck}, E.~F., {et~al.} 2014, \aap,
  572, A9

\bibitem[{{Kaufman} \& {Neufeld}(1996)}]{Kaufman1996}
{Kaufman}, M.~J. \& {Neufeld}, D.~A. 1996, \apj, 456, 611

\bibitem[{{Kennel} {et~al.}(1990){Kennel}, {Blandford}, \& {Wu}}]{Kennel1990}
{Kennel}, C.~F., {Blandford}, R.~D., \& {Wu}, C.~C. 1990, Physics of Fluids B,
  2, 253

\bibitem[{{Kristensen} {et~al.}(2013){Kristensen}, {van Dishoeck}, {Benz},
  {Bruderer}, {Visser}, \& {Wampfler}}]{Kristensen2013}
{Kristensen}, L.~E., {van Dishoeck}, E.~F., {Benz}, A.~O., {et~al.} 2013, \aap,
  557, A23

\bibitem[{{Kristensen} {et~al.}(2017){Kristensen}, {van Dishoeck}, {Mottram},
  {Karska}, {Y{\i}ld{\i}z}, {Bergin}, {Bjerkeli}, {Cabrit}, {Doty}, {Evans},
  {Gusdorf}, {Harsono}, {Herczeg}, {Johnstone}, {J{\o}rgensen}, {van Kempen},
  {Lee}, {Maret}, {Tafalla}, {Visser}, \& {Wampfler}}]{Kristensen2017}
{Kristensen}, L.~E., {van Dishoeck}, E.~F., {Mottram}, J.~C., {et~al.} 2017,
  \aap, 605, A93

\bibitem[{{Laor} \& {Draine}(1993)}]{Laor1993}
{Laor}, A. \& {Draine}, B.~T. 1993, \apj, 402, 441

\bibitem[{{Le Bourlot} {et~al.}(2012){Le Bourlot}, {Le Petit}, {Pinto},
  {Roueff}, \& {Roy}}]{Le-Bourlot2012}
{Le Bourlot}, J., {Le Petit}, F., {Pinto}, C., {Roueff}, E., \& {Roy}, F. 2012,
  \aap, 541, A76

\bibitem[{{Le Bourlot} {et~al.}(2002){Le Bourlot}, {Pineau des For{\^e}ts},
  {Flower}, \& {Cabrit}}]{Le-Bourlot2002}
{Le Bourlot}, J., {Pineau des For{\^e}ts}, G., {Flower}, D.~R., \& {Cabrit}, S.
  2002, \mnras, 332, 985

\bibitem[{{Le Bourlot} {et~al.}(1995){Le Bourlot}, {Pineau des For\^ets},
  {Roueff}, \& {Flower}}]{Le-Bourlot1995}
{Le Bourlot}, J., {Pineau des For\^ets}, G., {Roueff}, E., \& {Flower}, D.~R.
  1995, \aap, 302, 870

\bibitem[{{Le Gal} {et~al.}(2014){Le Gal}, {Hily-Blant}, {Faure}, {Pineau des
  For{\^e}ts}, {Rist}, \& {Maret}}]{Le-Gal2014}
{Le Gal}, R., {Hily-Blant}, P., {Faure}, A., {et~al.} 2014, \aap, 562, A83

\bibitem[{{Le Petit} {et~al.}(2006){Le Petit}, {Nehm{\'e}}, {Le Bourlot}, \&
  {Roueff}}]{Le-Petit2006}
{Le Petit}, F., {Nehm{\'e}}, C., {Le Bourlot}, J., \& {Roueff}, E. 2006, \apjs,
  164, 506

\bibitem[{{Le Petit} {et~al.}(2002){Le Petit}, {Roueff}, \& {Le
  Bourlot}}]{Le-Petit2002}
{Le Petit}, F., {Roueff}, E., \& {Le Bourlot}, J. 2002, \aap, 390, 369

\bibitem[{{Lee} {et~al.}(1996){Lee}, {Herbst}, {Pineau des For\^ets}, {Roueff},
  \& {Le Bourlot}}]{Lee1996}
{Lee}, H.-H., {Herbst}, E., {Pineau des For\^ets}, G., {Roueff}, E., \& {Le
  Bourlot}, J. 1996, \aap, 311, 690

\bibitem[{{Lee} {et~al.}(2016){Lee}, {Madden}, {Lebouteiller}, {Gusdorf},
  {Godard}, {Wu}, {Galametz}, {Cormier}, {Le Petit}, {Roueff}, {Bron},
  {Carlson}, {Chevance}, {Fukui}, {Galliano}, {Hony}, {Hughes}, {Indebetouw},
  {Israel}, {Kawamura}, {Le Bourlot}, {Lesaffre}, {Meixner}, {Muller}, {Nayak},
  {Onishi}, {Roman-Duval}, \& {Sewi{\l}o}}]{Lee2016}
{Lee}, M.-Y., {Madden}, S.~C., {Lebouteiller}, V., {et~al.} 2016, \aap, 596,
  A85

\bibitem[{{Lehmann}(2017)}]{Lehmann2017}
{Lehmann}, A. 2017, PhD thesis, Macquarie University

\bibitem[{{Lehmann} {et~al.}(2016){Lehmann}, {Federrath}, \&
  {Wardle}}]{Lehmann2016}
{Lehmann}, A., {Federrath}, C., \& {Wardle}, M. 2016, \mnras, 463, 1026

\bibitem[{{Lehmann} \& {Wardle}(2016)}]{Lehmann2016a}
{Lehmann}, A. \& {Wardle}, M. 2016, \mnras, 455, 2066

\bibitem[{{Lesaffre} {et~al.}(2004){Lesaffre}, {Chi{\`e}ze}, {Cabrit}, \&
  {Pineau des For{\^e}ts}}]{Lesaffre2004}
{Lesaffre}, P., {Chi{\`e}ze}, J.-P., {Cabrit}, S., \& {Pineau des For{\^e}ts},
  G. 2004, \aap, 427, 147

\bibitem[{{Lesaffre} {et~al.}(2013){Lesaffre}, {Pineau des For{\^e}ts},
  {Godard}, {Guillard}, {Boulanger}, \& {Falgarone}}]{Lesaffre2013}
{Lesaffre}, P., {Pineau des For{\^e}ts}, G., {Godard}, B., {et~al.} 2013, \aap,
  550, A106

\bibitem[{{Leurini} {et~al.}(2014){Leurini}, {Gusdorf}, {Wyrowski}, {Codella},
  {Csengeri}, {van der Tak}, {Beuther}, {Flower}, {Comito}, \&
  {Schilke}}]{Leurini2014}
{Leurini}, S., {Gusdorf}, A., {Wyrowski}, F., {et~al.} 2014, \aap, 564, L11

\bibitem[{{Leurini} {et~al.}(2015){Leurini}, {Wyrowski}, {Wiesemeyer},
  {Gusdorf}, {G{\"u}sten}, {Menten}, {Gerin}, {Levrier}, {H{\"u}bers},
  {Jacobs}, {Ricken}, \& {Richter}}]{Leurini2015}
{Leurini}, S., {Wyrowski}, F., {Wiesemeyer}, H., {et~al.} 2015, \aap, 584, A70

\bibitem[{{Louvet} {et~al.}(2016){Louvet}, {Motte}, {Gusdorf}, {Nguy{\^e}n
  Luong}, {Lesaffre}, {Duarte-Cabral}, {Maury}, {Schneider}, {Hill}, {Schilke},
  \& {Gueth}}]{Louvet2016}
{Louvet}, F., {Motte}, F., {Gusdorf}, A., {et~al.} 2016, \aap, 595, A122

\bibitem[{{Mathis} {et~al.}(1983){Mathis}, {Mezger}, \& {Panagia}}]{Mathis1983}
{Mathis}, J.~S., {Mezger}, P.~G., \& {Panagia}, N. 1983, \aap, 128, 212

\bibitem[{{Mathis} {et~al.}(1977){Mathis}, {Rumpl}, \&
  {Nordsieck}}]{Mathis1977}
{Mathis}, J.~S., {Rumpl}, W., \& {Nordsieck}, K.~H. 1977, \apj, 217, 425

\bibitem[{{Melnick} \& {Kaufman}(2015)}]{Melnick2015}
{Melnick}, G.~J. \& {Kaufman}, M.~J. 2015, \apj, 806, 227

\bibitem[{{Monteiro} {et~al.}(1988){Monteiro}, {Flower}, {Pineau des For\^ets},
  \& {Roueff}}]{Monteiro1988}
{Monteiro}, T.~S., {Flower}, D.~R., {Pineau des For\^ets}, G., \& {Roueff}, E.
  1988, \mnras, 234, 863

\bibitem[{{Mullan}(1971)}]{Mullan1971}
{Mullan}, D.~J. 1971, \mnras, 153, 145

\bibitem[{{Neufeld} \& {Dalgarno}(1989)}]{Neufeld1989}
{Neufeld}, D.~A. \& {Dalgarno}, A. 1989, \apj, 340, 869

\bibitem[{{Neufeld} {et~al.}(2015){Neufeld}, {Godard}, {Gerin}, {Pineau des
  For{\^e}ts}, {Bernier}, {Falgarone}, {Graf}, {G{\"u}sten}, {Herbst},
  {Lesaffre}, {Schilke}, {Sonnentrucker}, \& {Wiesemeyer}}]{Neufeld2015}
{Neufeld}, D.~A., {Godard}, B., {Gerin}, M., {et~al.} 2015, \aap, 577, A49

\bibitem[{{Neufeld} {et~al.}(2007){Neufeld}, {Hollenbach}, {Kaufman}, {Snell},
  {Melnick}, {Bergin}, \& {Sonnentrucker}}]{Neufeld2007}
{Neufeld}, D.~A., {Hollenbach}, D.~J., {Kaufman}, M.~J., {et~al.} 2007, \apj,
  664, 890

\bibitem[{{Neufeld} \& {Kaufman}(1993)}]{Neufeld1993}
{Neufeld}, D.~A. \& {Kaufman}, M.~J. 1993, \apj, 418, 263

\bibitem[{{Neufeld} {et~al.}(2009){Neufeld}, {Nisini}, {Giannini}, {Melnick},
  {Bergin}, {Yuan}, {Maret}, {Tolls}, {G{\"u}sten}, \&
  {Kaufman}}]{Neufeld2009a}
{Neufeld}, D.~A., {Nisini}, B., {Giannini}, T., {et~al.} 2009, \apj, 706, 170

\bibitem[{{Nisini} {et~al.}(2015){Nisini}, {Santangelo}, {Giannini},
  {Antoniucci}, {Cabrit}, {Codella}, {Davis}, {Eisl{\"o}ffel}, {Kristensen},
  {Herczeg}, {Neufeld}, \& {van Dishoeck}}]{Nisini2015}
{Nisini}, B., {Santangelo}, G., {Giannini}, T., {et~al.} 2015, \apj, 801, 121

\bibitem[{{Omont} {et~al.}(2013){Omont}, {Yang}, {Cox}, {Neri}, {Beelen},
  {Bussmann}, {Gavazzi}, {van der Werf}, {Riechers}, {Downes}, {Krips}, {Dye},
  {Ivison}, {Vieira}, {Wei{\ss}}, {Aguirre}, {Baes}, {Baker}, {Bertoldi},
  {Cooray}, {Dannerbauer}, {De Zotti}, {Eales}, {Fu}, {Gao}, {Gu{\'e}lin},
  {Harris}, {Jarvis}, {Lehnert}, {Leeuw}, {Lupu}, {Menten}, {Micha{\l}owski},
  {Negrello}, {Serjeant}, {Temi}, {Auld}, {Dariush}, {Dunne}, {Fritz},
  {Hopwood}, {Hoyos}, {Ibar}, {Maddox}, {Smith}, {Valiante}, {Bock},
  {Bradford}, {Glenn}, \& {Scott}}]{Omont2013}
{Omont}, A., {Yang}, C., {Cox}, P., {et~al.} 2013, \aap, 551, A115

\bibitem[{{Pon} {et~al.}(2016){Pon}, {Johnstone}, {Caselli}, {Fontani},
  {Palau}, {Butler}, {Kaufman}, {Jim{\'e}nez-Serra}, \& {Tan}}]{Pon2016}
{Pon}, A., {Johnstone}, D., {Caselli}, P., {et~al.} 2016, \aap, 587, A96

\bibitem[{{Pon} {et~al.}(2012){Pon}, {Johnstone}, \& {Kaufman}}]{Pon2012}
{Pon}, A., {Johnstone}, D., \& {Kaufman}, M.~J. 2012, \apj, 748, 25

\bibitem[{{Reach} {et~al.}(2005){Reach}, {Rho}, \& {Jarrett}}]{reach2005}
{Reach}, W.~T., {Rho}, J., \& {Jarrett}, T.~H. 2005, \apj, 618, 297

\bibitem[{{Roberge} \& {Draine}(1990)}]{Roberge1990}
{Roberge}, W.~G. \& {Draine}, B.~T. 1990, \apj, 350, 700

\bibitem[{{R{\"o}llig} {et~al.}(2007){R{\"o}llig}, {Abel}, {Bell}, {Bensch},
  {Black}, {Ferland}, {Jonkheid}, {Kamp}, {Kaufman}, {Le Bourlot}, {Le Petit},
  {Meijerink}, {Morata}, {Ossenkopf}, {Roueff}, {Shaw}, {Spaans}, {Sternberg},
  {Stutzki}, {Thi}, {van Dishoeck}, {van Hoof}, {Viti}, \&
  {Wolfire}}]{Rollig2007}
{R{\"o}llig}, M., {Abel}, N.~P., {Bell}, T., {et~al.} 2007, \aap, 467, 187

\bibitem[{{Shen} {et~al.}(2004){Shen}, {Greenberg}, {Schutte}, \& {van
  Dishoeck}}]{Shen2004}
{Shen}, C.~J., {Greenberg}, J.~M., {Schutte}, W.~A., \& {van Dishoeck}, E.~F.
  2004, \aap, 415, 203

\bibitem[{{Shull} \& {McKee}(1979)}]{Shull1979}
{Shull}, J.~M. \& {McKee}, C.~F. 1979, \apj, 227, 131

\bibitem[{{Snell} {et~al.}(2005){Snell}, {Hollenbach}, {Howe}, {Neufeld},
  {Kaufman}, {Melnick}, {Bergin}, \& {Wang}}]{Snell2005}
{Snell}, R.~L., {Hollenbach}, D., {Howe}, J.~E., {et~al.} 2005, \apj, 620, 758

\bibitem[{{St{\"o}rzer} \& {Hollenbach}(1998)}]{Storzer1998}
{St{\"o}rzer}, H. \& {Hollenbach}, D. 1998, \apj, 495, 853

\bibitem[{{Swinbank} {et~al.}(2010){Swinbank}, {Smail}, {Longmore}, {Harris},
  {Baker}, {De Breuck}, {Richard}, {Edge}, {Ivison}, {Blundell}, {Coppin},
  {Cox}, {Gurwell}, {Hainline}, {Krips}, {Lundgren}, {Neri}, {Siana},
  {Siringo}, {Stark}, {Wilner}, \& {Younger}}]{Swinbank2010}
{Swinbank}, A.~M., {Smail}, I., {Longmore}, S., {et~al.} 2010, \nat, 464, 733

\bibitem[{{Tafalla} {et~al.}(2015){Tafalla}, {Bachiller}, {Lefloch},
  {Rodr{\'{\i}}guez-Fern{\'a}ndez}, {Codella}, {L{\'o}pez-Sepulcre}, \&
  {Podio}}]{Tafalla2015}
{Tafalla}, M., {Bachiller}, R., {Lefloch}, B., {et~al.} 2015, \aap, 573, L2

\bibitem[{{Tielens} \& {Hollenbach}(1985)}]{Tielens1985}
{Tielens}, A.~G.~G.~M. \& {Hollenbach}, D. 1985, \apj, 291, 722

\bibitem[{{Trinchieri} {et~al.}(2003){Trinchieri}, {Sulentic}, {Breitschwerdt},
  \& {Pietsch}}]{Trinchieri2003}
{Trinchieri}, G., {Sulentic}, J., {Breitschwerdt}, D., \& {Pietsch}, W. 2003,
  \aap, 401, 173

\bibitem[{{van Dishoeck}(1988)}]{van-Dishoeck1988}
{van Dishoeck}, E.~F. 1988, in Rate Coefficients in Astrochemistry. Editors,
  T.J. Millar, D.A. Williams; Publisher, Kluwer Academic Publishers., ed.
  {T.~J.~Millar \& D.~A.~Williams}, 49

\bibitem[{{van Dishoeck} {et~al.}(2006){van Dishoeck}, {Jonkheid}, \& {van
  Hemert}}]{van-Dishoeck2006}
{van Dishoeck}, E.~F., {Jonkheid}, B., \& {van Hemert}, M.~C. 2006, Faraday
  Discussions, 133, 231

\bibitem[{{Walmsley} {et~al.}(2005){Walmsley}, {Pineau des For{\^e}ts}, \&
  {Flower}}]{Walmsley2005}
{Walmsley}, M., {Pineau des For{\^e}ts}, G., \& {Flower}, D. 2005, in IAU
  Symposium, Vol. 231, Astrochemistry: Recent Successes and Current Challenges,
  ed. D.~C. {Lis}, G.~A. {Blake}, \& E.~{Herbst}, 135--140

\bibitem[{{Wardle}(1999)}]{Wardle1999}
{Wardle}, M. 1999, \apjl, 525, L101

\end{thebibliography}

\newpage

\appendix

\section{Grain size distribution} \label{Append-grains}

\begin{figure}
\begin{center}
\includegraphics[width=8.0cm,angle=0]{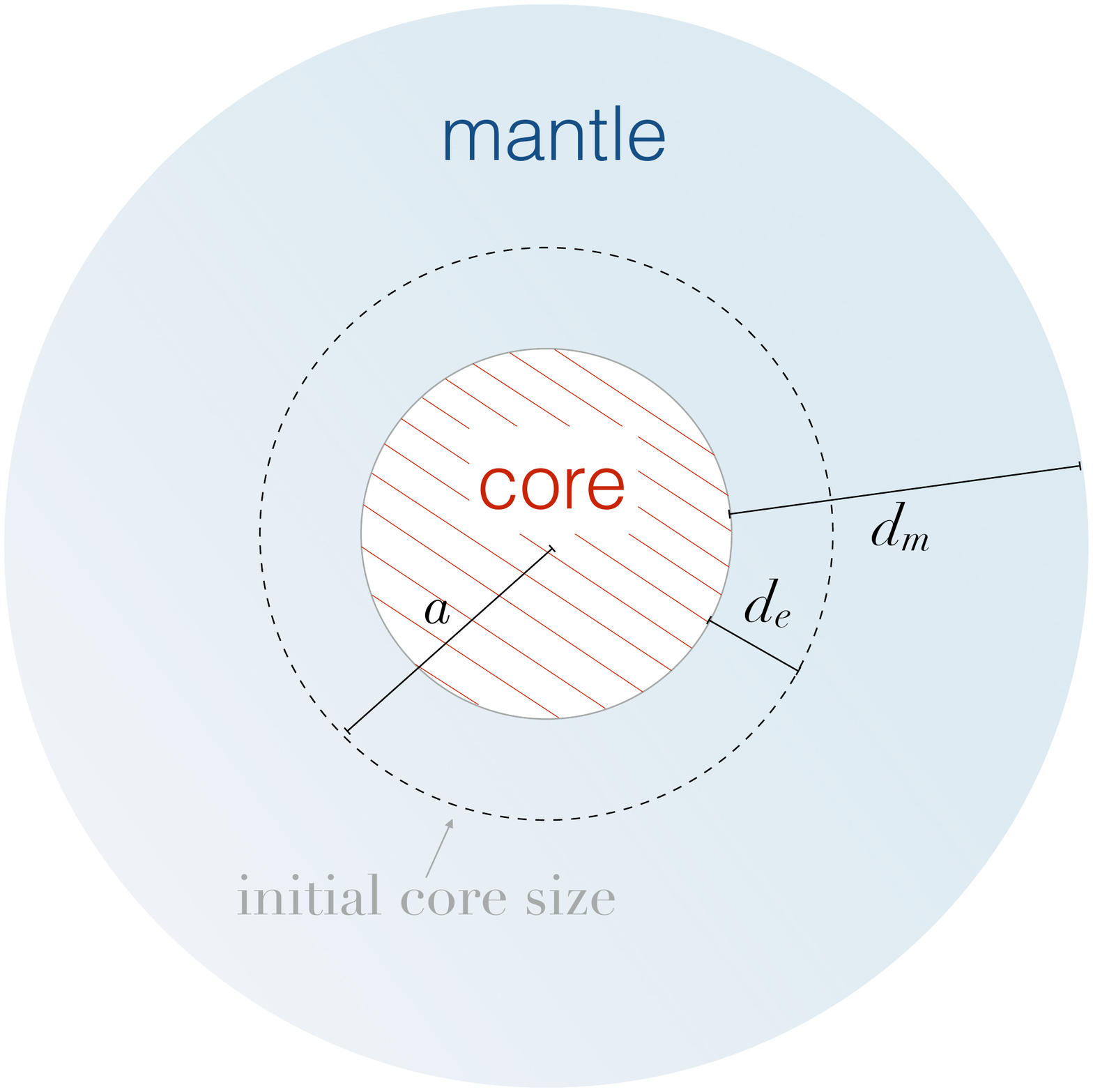}
\caption{Schematic view of erosion and adsorption on grains.}
\label{Fig-scheme-grain}
\end{center}
\end{figure}

All processes involving grains are treated using a prescription similar to that proposed 
by \citet{Le-Bourlot1995}. As drawn on Figure \ref{Fig-scheme-grain}, we assume spherical 
grains composed of mineral cores and surrounded by ice mantles. Shattering and coagulations
processes are neglected. Erosion processes are supposed to reduce evenly the initial size 
of the cores by a thickness $d_e$, while adsorption and desorption mechanisms are supposed 
to build evenly mantles on cores over a thickness $d_m$.

We assume that the grain cores follow a power-law size distribution from a minimal
size $a_{\rm m}$ to a maximal size $a_{\rm M}$. At the ionization front (see Fig. 
\ref{Fig-scheme-shock}), the initial number of cores per unit volume of gas 
with radii between $a$ and $a+da$, $dn_g^0(a)$, is
\begin{equation}
dn_g^0(a) = A a^{-\alpha} da.
\end{equation}
With the initial condition $d_e = d_m = 0$, the integration of the grain mass over 
the size distribution leads to the normalization coefficient
\begin{equation}
A = \frac{3}{4 \pi \rho_c f_4} \sum_i m({\rm X}_i^{**}) n^0({\rm X}_i^{**}),
\end{equation}
where
\begin{equation}
f_n = \frac{a_{\rm M}^{n-\alpha}-a_{\rm m}^{n-\alpha}}{n-\alpha}.
\end{equation}
$m({\rm X}_i^{**})$ and $n^0({\rm X}_i^{**})$ are the masses and initial densities 
of the chemical species ${\rm X}_i^{**}$ which compose the core, and 
${\rho_c}$ is the core mass density, assumed to be homogeneous and identical for all 
grains. The initial number of grains per unit volume of gas therefore
writes
\begin{equation}
n_g^0 = Af_1.
\end{equation}

Starting from these initial conditions, the evolution of dust grains is computed
in the buffer and in the shock taking into account several processes: i.e. adsorption,
desorption, and erosion, which modify the thicknesses $d_e$ and $d_m$, and the compression
induced by the shock. Defining $\mathscr{C}_g$ as the compression factor of the 
grain fluid\footnote{Assuming that all grains are coupled to the magnetic field
(see Sect. \ref{Sect-gr-pahs}), $\mathscr{C}_g$ is governed by the simple differential 
equation $d\mathscr{C}_g/dz = - \mathscr{C}_g / u_i \times du_i/dz$, where $u_i$ is 
the ion fluid velocity, and $\mathscr{C}_g(z=0)=1$.}, the number of grains $n_g$ per unit volume of gas writes
\begin{equation}
n_g = A f_1 \mathscr{C}_g.
\end{equation}

If we assume that erosion, adsorption, and desorption mechanisms only depend on geometrical 
cross sections, the integration of the corresponding rates over the size distribution reduces 
to the calculation of the rate for a single grain of radius $\langle r_c^2 \rangle^{1/2}$ 
(for grain core processes) and $\langle r_m^2 \rangle^{1/2}$ (for grain mantle processes),
where
\begin{equation}
\langle r_c^2 \rangle  = \frac{\int_{a_{\rm m}}^{a_{\rm M}} (a-d_e)^2 a^{-\alpha} da}{\int_{a_{\rm m}}^{a_{\rm M}} a^{-\alpha} da}
\end{equation}
is the mean square radius of eroded grain cores and
\begin{equation}
\langle r_m^2 \rangle  = \frac{\int_{a_{\rm m}}^{a_{\rm M}} (a-d_e+d_m)^2 a^{-\alpha} da}{\int_{a_{\rm m}}^{a_{\rm M}} a^{-\alpha} da}
\end{equation}
is the mean square radius of grains including mantles. The integration of those equations 
yields analytical expressions relating $\langle r_c^2 \rangle$ and $\langle r_m^2 
\rangle$ to $d_e$ and $d_m$, i.e.
\begin{equation} \label{Eq-rc}
\langle r_c^2 \rangle  = \frac{1}{f_1} \left[ f_3 - 2 f_2 de + f_1 d_e^2 \right]
\end{equation}
and
\begin{equation} \label{Eq-rm}
\langle r_m^2 \rangle  = \frac{1}{f_1} \left[ f_3 + 2 f_2 (d_m - de) - 2 f_1 d_e d_m + f_1 (d_e^2 + d_m^2) \right].
\end{equation}
According to the geometrical and chemical models, the volume of grain cores removed by 
the erosion mechanisms (per unit volume of gas) is
\begin{equation}
\mathcal{V}_e = \mathscr{C}_g \int_{a_{\rm m}}^{a_{\rm M}} A \frac{4 \pi}{3} \left[ a^3-(a-d_e)^3 \right] a^{-\alpha} da
\end{equation}
and
\begin{equation}
\mathcal{V}_e = \frac{1}{\rho_c} \sum_i m({\rm X}_i^{**}) \left( \mathscr{C}_g n^0({\rm X}_i^{**}) - n({\rm X}_i^{**}) \right)
,\end{equation}
which leads to a cubic equation for $d_e$
\begin{equation} \label{Eq-de}
3 f_3 d_e - 3 f_2 d_e^2 + f_1 d_e^3 = f_4 \left( 1 - \frac{\sum_i m({\rm X}_i^{**}) n({\rm X}_i^{**})}{\sum_i m({\rm X}_i^{**}) n^0({\rm X}_i^{**}) \mathscr{C}_g} \right).
\end{equation}
Similarly, the volume of grain mantles writes
\begin{equation}
\mathcal{V}_m = \mathscr{C}_g \int_{a_{\rm m}}^{a_{\rm M}} A \frac{4 \pi}{3} \left[ (a-d_e+d_m)^3-(a-d_e)^3 \right] a^{-\alpha} da
\end{equation}
and
\begin{equation}
\mathcal{V}_m = \frac{1}{\rho_m} \sum_i m({\rm X}_i^{*}) n({\rm X}_i^{*})
,\end{equation}
where $\rho_m$ is the mantle mass density and $m({\rm X}_i^{*})$ and $n({\rm X}_i^{*})$ 
are the masses and densities (per unit volume of gas) of the chemical species 
${\rm X}_i^{*}$ which compose the mantle. The development of those equations 
finally leads to a cubic expression for $d_m$
\begin{dmath} \label{Eq-dm}
3 f_3 d_m - 6 f_2 d_e d_m + 3 f_1 d_e^2 d_m + 3 f_2 d_m^2 - 3 f_1 d_e d_m^2 + f_1 d_m^3 = 
f_4 \frac{\rho_c}{\rho_m} \frac{\sum_i m({\rm X}_i^{*}) n({\rm X}_i^{*})}{\sum_i m({\rm X}_i^{**}) n^0({\rm X}_i^{**}) \mathscr{C}_g}.
\end{dmath}

\begin{table}
\begin{center}
\caption{Parameters of the grain size distribution model and initial abundances 
$x({\rm X}_i^{**})$ of grain core compounds. Numbers in parenthesis are powers of 
ten.}
\begin{tabular}{l r@{.}l l l}
\hline
Parameter & \multicolumn{2}{c}{value} & unit & ref \\
\hline
$\alpha$             & -3 & 5        &             & a \\
$a_{\rm m}$          &  1 & 0   (-6) & cm          & a \\
$a_{\rm M}$          &  3 & 0   (-5) & cm          & a \\
$\rho_c$             &  2 & 0        & g cm$^{-3}$ & b \\
$\rho_m$             &  1 & 0        & g cm$^{-3}$ & b \\
$x^0({\rm O}^{**})$  &  1 & 40 (-4)  &             & c \\
$x^0({\rm Si}^{**})$ &  3 & 37 (-5)  &             & c \\
$x^0({\rm Mg}^{**})$ &  3 & 70 (-5)  &             & c \\
$x^0({\rm Fe}^{**})$ &  3 & 23 (-5)  &             & c \\
$x^0({\rm C}^{**})$  &  1 & 63 (-4)  &             & c \\
\hline
\end{tabular}
\begin{list}{}{}
$a$ $-$ \citet{Mathis1977}, $b$ $-$ \citet{Hily-Blant2018}, $c$ $-$ \citet{Flower2003}
\end{list}
\label{Tab-grain}
\end{center}
\end{table}

In the Paris-Durham shock code, equations \ref{Eq-de} and \ref{Eq-dm} are solved at
each time step using a Newton-Raphson scheme. The values of $d_e$ and $d_m$ are then
injected in equations \ref{Eq-rc} and \ref{Eq-rm} to compute the mean square radii of
the cores and mantles and the subsequent erosion, adsorption, and desorption rates,
which control the evolution of the ${\rm X}_i^{**}$ and ${\rm X}_i^{*}$ abundances 
over the next time step.
All the parameters used in this work are given in Table \ref{Tab-grain}. The initial
abundances of the grain core compounds correspond to an initial dust-to-gas mass 
ratio of 0.56 \%.

Erosion of grain mantles, adsorption, and 
photodesorption processes for grains of size $a_g$ and relative abundance 
$x_g$ occur over characteristic timescales, roughly given by 
\citep{Barlow1978,Flower2005,Hollenbach2009}
\begin{multline}
\mathscr{T}_{\rm ero} \sim 5 \times 10^{-1} \left( \frac{10^4\,\cc}{\dens} \right) 
\left( \frac{10^{-6}\,{\rm cm}}{a_g} \right)^2 \\
\left( \frac{10^{6}\,{\rm cm\,\,s^{-1}}}{\Delta v} \right)
{\rm exp}\left[ 3 \left(\frac{10^{6}\,{\rm cm\,\,s^{-1}}}{\Delta v} \right)^2 \right]
\,\, {\rm yr},
\end{multline}
\begin{multline}
\mathscr{T}_{\rm ads} \sim 3 \times 10^{6} \left( \frac{10^4\,\cc}{\dens} \right) 
\left(\frac{10^{-11}}{x_g} \right) \left( \frac{10^{-6}\,{\rm cm}}{a_g} \right)^2 
\,\, {\rm yr}, \\
\end{multline}
and
\begin{multline}
\mathscr{T}_{\rm phd} \sim 2 \times 10^{4} 
\left( G_0 + 4\times 10^{-5} \frac{\zeta_{\HH}}{3\times 10^{-17} s^{-1}} \right)^{-1}
\left( \frac{10^{-6}\,{\rm cm}}{a_g} \right)^2
\,\, {\rm yr},
\end{multline}
where $\Delta v$ is the velocity difference between the ions and neutrals.
Those values compared with the dynamical timescales given in Fig. \ref{Fig-shock-size}
show that adsorption never occurs over the duration of a shock, while photodesorption
and the erosion of grain mantles may occur within the shock depending on the density of 
the gas, strength of the radiation field, and ion-neutral velocity drift. It 
follows that the adsorption processes are important only for the computation of the 
equilibrium composition of the pre-shock and the post-shock gas, that the erosion processes 
only arise within the shock where the ions and the neutrals decouple, and that photodesorption 
may be important everywhere depending on the irradiation conditions.

\section{Comparison with PDR models} \label{Append-PDR}

\begin{figure}
\begin{center}
\includegraphics[width=8.0cm,trim = 2cm 2cm 1cm 1cm, clip,angle=0]{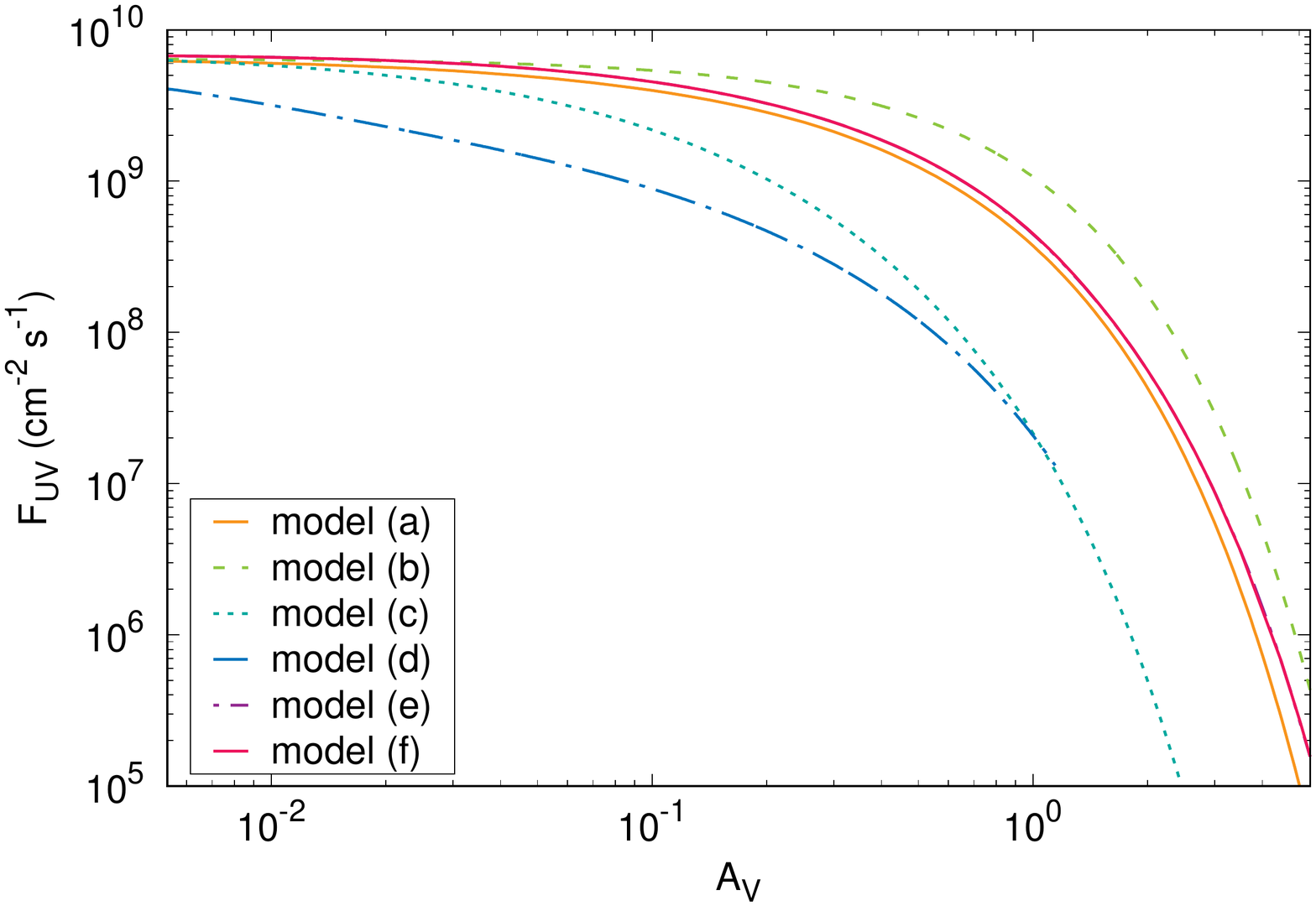}
\includegraphics[width=8.0cm,trim = 2cm 2cm 1cm 1cm, clip,angle=0]{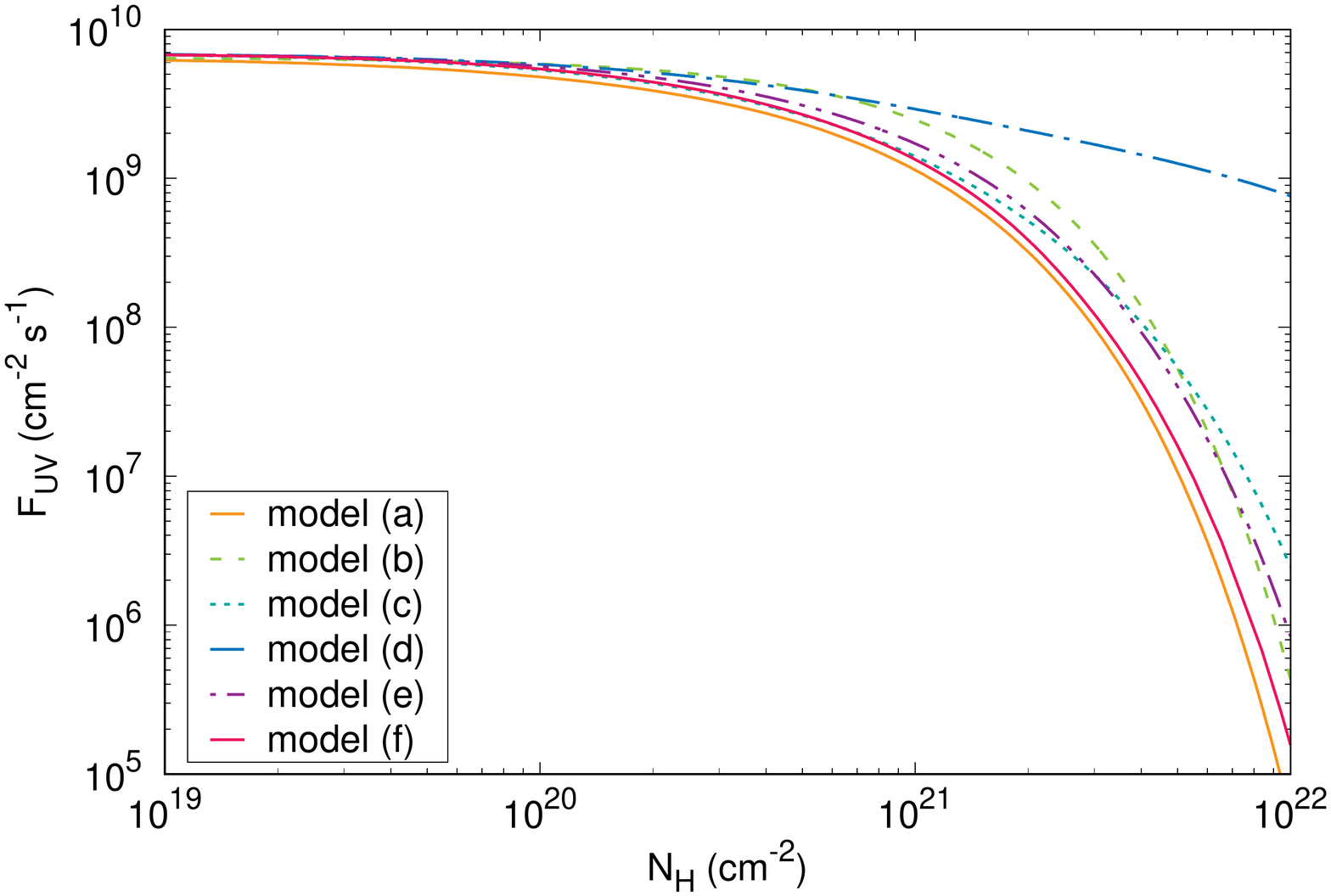}
\caption{Ultraviolet photon flux (integrated from 911 to 2400 \AA) as a function of the total 
hydrogen column density (bottom panel) and the visual extinction (top panel) calculated 
with the Meudon PDR code (a), by \citet{Tielens1985} (b), and with the Paris-Durham 
shock model set in advection mode (c-f) for an isochoric gas of density $\dens=10^4$ \cc\ 
with an incident radiation field $G_0=10^2$. The predictions of the Paris-Durham model 
are shown for pure graphite grains with a size distribution defined by Table 
\ref{Tab-grain} (standard model, labelled c), for silicate grains with the same size 
distribution (d), graphite grains with a minimal radius set to $3\times 10^{-6}$ cm 
(e), and graphite grains with a minimal radius set to $3\times 10^{-6}$ cm where the 
grain absorption coefficients are scaled to impose a $N_{\rm H}/A_V$ ratio of $1.87\times 
10^{21}$ cm$^{-2}$ (f).}
\label{Fig-radiation-flux}
\end{center}
\end{figure}

To validate the treatment of the radiative transfer and of the multiple interactions of 
far-UV photons with interstellar matter, we perform comparisons of state-of-the-art PDR 
models with the predictions of the Paris-Durham shock code set in advection mode. In this mode, 
the code follows the isochoric or isobaric thermo-chemical evolution of a single fluid 
moving away from the ionization front with a constant velocity $v_{\rm adv}$. For
simplicity, we consider a prototypical isochoric PDR of density $\dens = 10^4$ \cc, 
illuminated on one side by an isotropic radiation field with a scaling factor $G_0 = 100$,
and moving at a speed $v_{\rm adv} = 0.01$ \kms. Unless indicated otherwise, all 
adsorption and desorption mechanisms are switched off.

\begin{figure*}[!ht]
\begin{center}
\includegraphics[width=16.0cm,trim = 1.5cm 2cm 1cm 1cm, clip,angle=0]{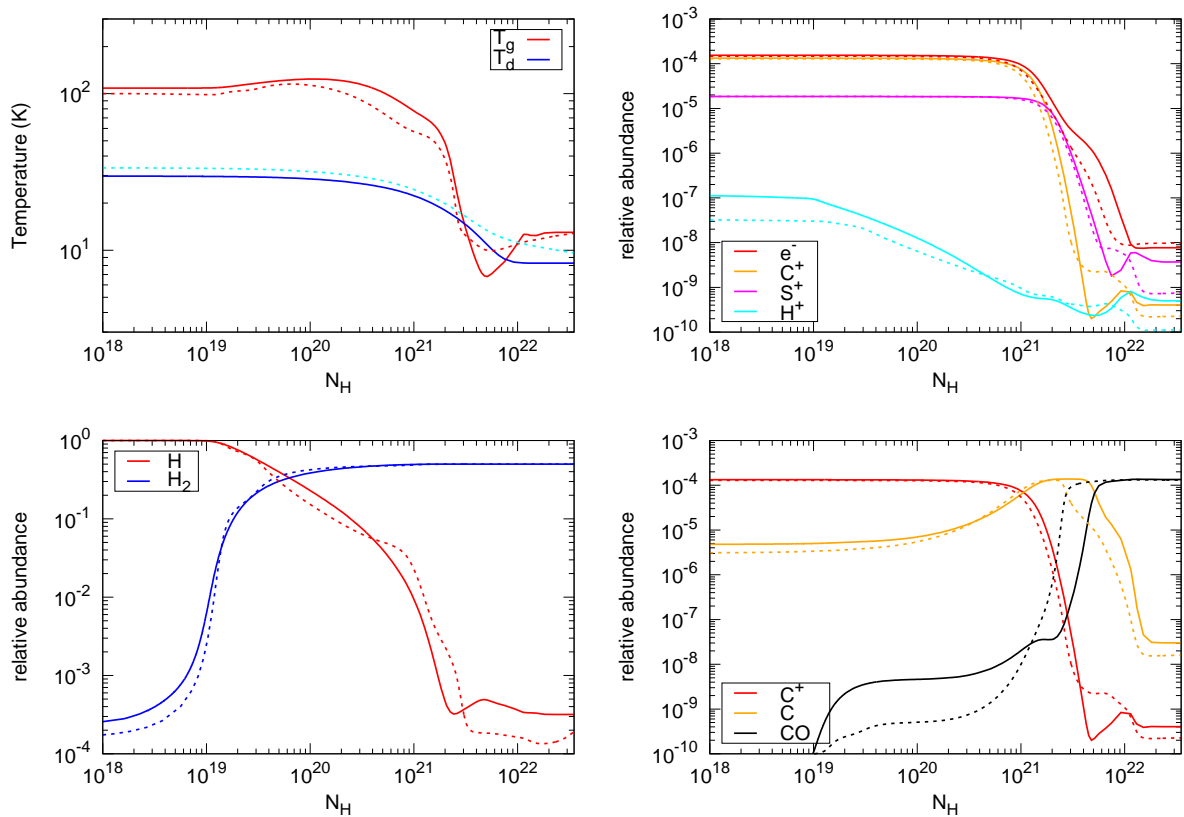}
\caption{Gas and dust (radius = 0.05 $\mu$m) temperatures (top left panel) and abundances 
of the main ions, atomic, and molecular compounds (other panels) as functions of the
total hydrogen column density computed with the Meudon PDR code (dashed lines) and the 
Paris-Durham code (continuous lines) for an isochoric PDR of density $\dens=10^4$ \cc\ 
with an incident radiation field $G_0=10^2$.}
\label{Fig-profils-pdr}
\end{center}
\end{figure*}

Fig. \ref{Fig-radiation-flux} shows model predictions of the photon flux (integrated 
between 911 and 2400 \AA) as a function of the visual extinction and total 
hydrogen column density for different models of grains. By default, most of the existing 
PDR codes tune the grain absorption and scattering coefficients at all wavelengths in 
order to match the standard galactic extinction curve observed by \citet{Fitzpatrick1986} 
and to reproduce a fixed $N_{\rm H}/{A_V}$ ratio of $1.87 \times 10^{21}$ cm$^{-2}$. This 
is not the case in the shock model where the extinction curve and colour indices are 
derived from grain properties alone and may differ from one model to the next. Hence, 
before comparing the predictions we need to choose which of $A_V$ or $N_{\rm H}$ is 
the best natural variable to display PDR profiles.
Analysis of the absorption coefficients shows that the ratios of the visible extinction 
to the extinctions at other wavelengths vary considerably depending on the grain size and 
composition. As a result, $A_V$ is not only weakly representative of the absorption of photons 
over the entire UV range when the grain model is changed, but is also a poor tracer of the
amount of dust in the cloud. In contrast, since the abundance of grains is considered as
constant, $N_{\rm H}$ is a natural proxy of the dust column density regardless of the grain 
size and composition. All the comparisons below will thus be discussed as functions of 
$N_{\rm H}$.

The comparison of the shock results with classical PDR models in the bottom panel of Fig. 
\ref{Fig-radiation-flux} shows that almost all the prescriptions of grains adopted in this work, with the exception of pure silicate grains, 
give satisfactory predictions of the absorption of UV flux as a function of the depth 
into the cloud. This discrepancy for the
pure silicate grain model is due to the absorption properties of dust at large wavelengths. 
While graphite and silicate have similar absorption coefficients around 1000 \AA, that of 
silicate is up to 100 times smaller than that of graphite at 2000 \AA\ for a grain with 
radius between 0.01 and 0.1 $\mu$m. The radiation field is thus absorbed over distances 
considerably larger when silicate grains are used. The best agreement with the Meudon PDR 
code is obtained for pure graphite grains with a minimal radius set to $3 \times 10^{-6}$ 
cm and where the grain absorption coefficients are scaled homogeneously to impose a 
$N_{\rm H}/A_V$ ratio of $1.87 \times 10^{21}$ cm$^{-2}$ (model f in Fig. 
\ref{Fig-radiation-flux}).

\begin{figure}
\begin{center}
\includegraphics[width=8.0cm,trim = 2cm 2cm 1cm 1cm, clip,angle=0]{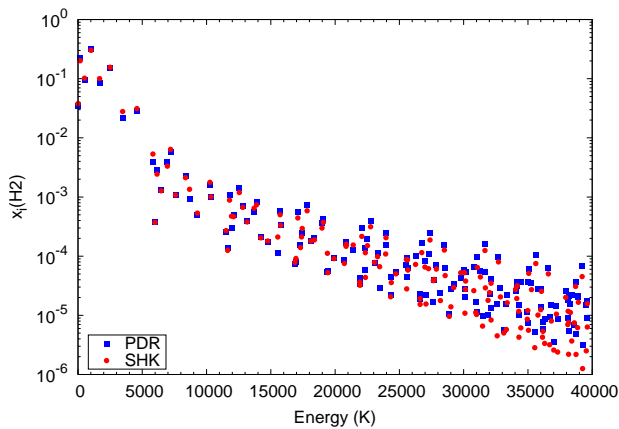}
\caption{\HH\ level populations as functions of their energies computed with the Meudon PDR
code (blue squares) and the Paris-Durham model (red circles) at $N_{\rm H}=1.87 \times 10^{15}$
cm$^{-2}$, in an isochoric PDR of density $\dens=10^4$ \cc\ with an incident radiation 
field $G_0=10^2$.}
\label{Fig-H2pop-pdr}
\end{center}
\end{figure}

Keeping this last configuration, we compare the predictions of the Paris-Durham shock 
model with those of the Meudon PDR code regarding the main physical properties of PDRs, 
i.e. the temperatures of gas and dust, ionization fraction, H/\HH\ and \Cp/C/CO 
transitions, and excitation of the rovibrational levels of \HH. As a whole, Fig. 
\ref{Fig-profils-pdr} shows a satisfactory agreement between the results of the two 
models. In particular, the temperature profiles, H/\HH\ transition, and ionization
fraction are found to follow the same trends with values almost identical at the border 
and in the interior of the cloud. All the discrepancies observed, i.e. the difference of 
about a factor of five in the calculation of the \Sp\ abundance in the environment well 
shielded from the radiation field and the C/CO transition that arises at distances three 
times larger than in the Meudon PDR code, are due to the different chemical networks and prescriptions of PAHs (e.g. size distribution, calculation of ionization, and 
recombination processes) used in each model. The populations of the first 200 
rovibrational levels of \HH\ computed at the border of the cloud and shown in Fig. 
\ref{Fig-H2pop-pdr} are found to be analogous in both models; there is a maximum discrepancy 
of about a factor of three for the most energetic levels. In summary, these results not 
only validate the treatments of the radiative processes now performed in the Paris-Durham 
shock code but also justify the approximation of using a single size grain in the computation 
of the UV radiative transfer.

To check the treatments of the adsorption and desorption processes, we finally compare 
the solid-phase and gas-phase abundances of a few species (namely CO, CH$_4$ and H$_2$O) 
predicted by the code (when adsorption and desorption are switched on) with those obtained 
by \citet{Hollenbach2009} (Figs. 3 \& 4). In 
both models, mantles become the main carrier of all heavy elements at similar distances 
($\sim 6 \times 10^{17}$ cm). Comparable results are also found for gas-phase abundances. 
Because of the lack of efficient desorption processes in the interior of the cloud ($> 2 
\times 10^{18}$ cm), the two models converge towards a chemical equilibrium with gas-phase 
abundances below $10^{-7}$ for both carbon and oxygen bearing species. The fact that
both models give virtually identical results is surprising in the light of the treatment 
of photodesorption by secondary UV photons. Indeed, according to \citet{Hollenbach2009}, 
the secondary photon flux associated with a primary cosmic-ray ionization rate of $10^{-17}$ 
s$^{-1}$ in a medium with a density of $10^4$ \cc\ is equivalent to $G_0 \sim 10^{-3}$. 
In comparison, the treatment we propose in Sect. \ref{Sect-ChemNet} leads to an equivalent 
radiation field of $G_0 \sim 10^{-5}$ for the same conditions, close to the value derived 
by \citet{Shen2004} and that used in \citet{Heays2017}. The agreement between the gas-phase 
abundances computed in both models indicates that this apparent discrepancy is likely to be 
a typo in Hollenbach's paper not implemented in their chemical code, which probably uses the 
correct value, i.e. a secondary photon flux of $2 \times 10^3$ photons cm$^{-2}$ s$^{-1}$
for $\zeta_{\HH} = 10^{-17}$ s$^{-1}$ and $\dens=10^4$ \cc.

\section{Computation of stationary solutions} \label{Append-method}

The governing equations of partially ionized, plane-parallel, and magnetized molecular shocks 
have been derived in several papers (e.g. \citealt{Draine1980, Flower1985}). Under the 
assumption of stationarity, they organize into a set of coupled, first-order, ordinary 
differential equations that fully describe the exchanges of mass, momentum, and energy 
that occur between the different chemical flows. Together with the initial conditions, 
the system thus forms an initial value problem which, in principle, may be integrated using 
a classical forward integration technique (e.g. DVODE, \citealt{Hindmarsh1983}).
As proven with the Paris-Durham shock code (e.g. \citealt{Flower2003,Flower2015}), such 
a method is highly robust and efficient in computing the steady states of J-type shocks 
(single fluid) or that of C-type shocks, i.e. shocks where the neutrals and ions 
are decoupled and where the flow stays supersonic over its entire trajectory. In the
case of C-type shocks, this method fails, however, if the cooling is not efficient enough 
so that the neutral flow becomes subsonic at some point along the trajectory. Indeed, 
these conditions trigger the emergence of another category of stationary structures, 
called CJ-type and C*-type shocks \citep{Chernoff1987,Roberge1990}. Because these 
structures require crossing critical points, forward integration techniques become 
unstable and highly sensitive to numerical errors.



The method we apply to treat CJ-type and C*-type shocks is based on the work
of \citet{Roberge1990} who thoroughly described the physics of shocks transiting from a 
supersonic to a subsonic regime. The trajectory of these kinds of shocks can be divided
in three different parts (see Figs. 5 and 6 of \citealt{Roberge1990}) : (i) an upstream 
supersonic regime where the magnetic precursor develops inducing a decoupling between the 
ionized and neutral fluids; (ii) a subsonic excursion where the ions and the neutrals
are still decoupled ; and (iii) a downstream supersonic flow where the ions and 
neutrals recouple and relax down to the speed and temperature of the post-shock
medium\footnote{In general, the downstream flow can be either supersonic or subsonic.
However, the solutions we explore in this work mostly lead to supersonic post-shock media.}.
Downstream, the transition from the subsonic to supersonic regime (from ii to 
iii) can only occur continuously, passing through the downstream sonic point (labelled 
C' in Fig. 1 of \citealt{Roberge1990}). In contrast, the upstream transition from the 
supersonic to the subsonic regime (from i to ii) can potentially happen through an
infinite number of trajectories, which involve either an adiabatic jump or a continuous 
transition via the upstream sonic point (labelled C in Fig. 1 of \citealt{Roberge1990}). 
As shown by \citet{Chernoff1987}, most of these trajectories lead to unphysical solutions
where either $d v_n/dz$ or $d v_i/dz$ change sign. For a given set of physical conditions,
only one amongst all trajectories leads to the downstream sonic point, hence to a full 
stationary solution that satisfies all the equations of a plane-parallel molecular shock. This solution is called a CJ-type shock if it involves an upstream adiabatic jump, and a 
C*-type shock if it involves a continuous upstream transition between the supersonic and subsonic regimes.

\begin{figure*}[!ht]
\begin{center}
\includegraphics[width=9.0cm,trim = 2cm 2cm 3cm 1cm, clip,angle=0]{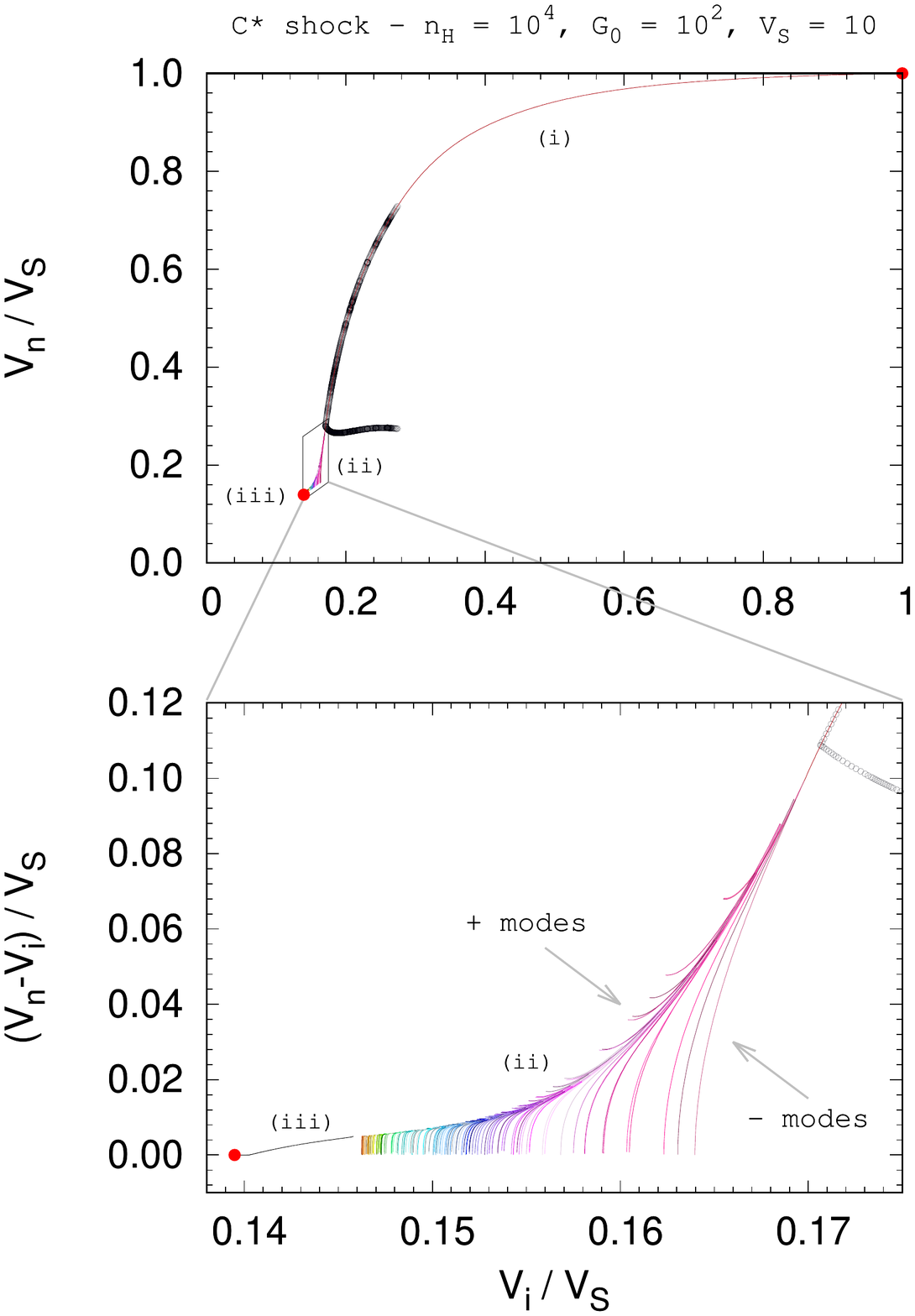}
\includegraphics[width=9.0cm,trim = 2cm 2cm 3cm 1cm, clip,angle=0]{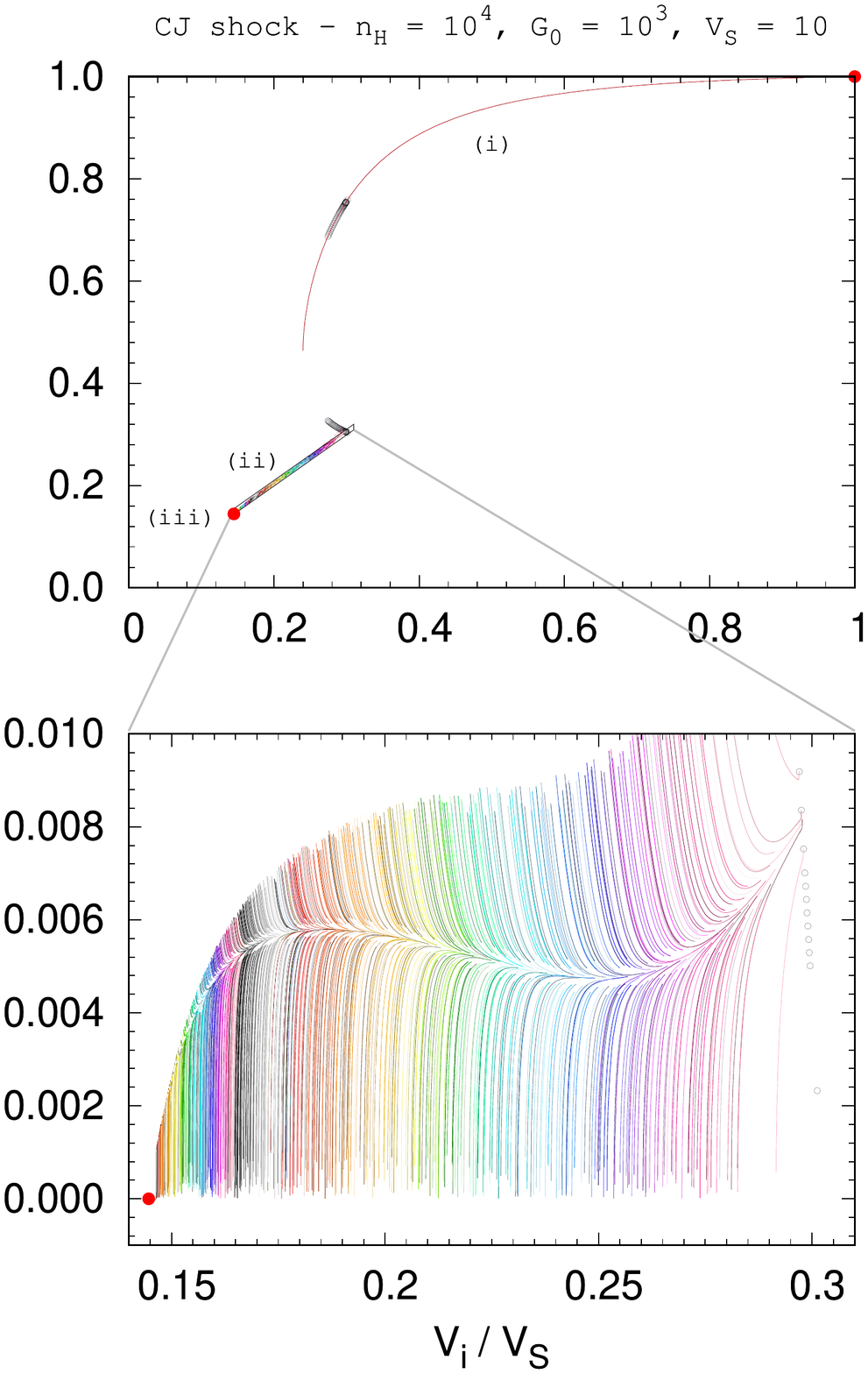}
\caption{Phase diagrams of stationary shocks propagating at a speed of $10$ \kms\ in a 
pre-shock medium with a density of $10^4$ \cc, located at 0.6 mpc of the ionization front 
illuminated by an isotropic radiation field of $G_0=10^2$ (left) and $10^3$ (right). All
other parameters are set to their standard values (see Table \ref{Tab-main}). 
The coloured curves correspond to the potential trajectories computed by the 
algorithm described in Appendix \ref{Append-method}. The true stationary solution connects the 
upstream and downstream trajectories (i) and (iii). This solution is bounded in the 
subsonic excursion (labelled ii) by the $+$ and $-$ modes (see main text). The top
panels show the neutral and ion velocities relative to the velocity of the shock.
The bottom panels are zooms of the top panels rotated to highlight the ion-neutral
velocity drift. Initial and final kinematic states and potential jump conditions are 
indicated with red points and black circles, respectively.}
\label{Fig-phase-diag}
\end{center}
\end{figure*}

Examples of CJ and C* shocks are shown in Fig. \ref{Fig-phase-diag}, which shows 
the typical evolutions of the neutral and ion velocities ($v_n$ and $v_i$) set by
the steady-state evolution equations. The initial (pre-shock) and final (post-shock) 
kinematic states of the gas are indicated with red points.
Assuming that their initial conditions are known, both the upstream and downstream 
supersonic parts (i and iii in Fig. \ref{Fig-phase-diag}) can be easily computed 
using the forward integration scheme of the Paris-Durham code. The real difficulty 
lies in calculating the subsonic excursion (part ii in Fig. \ref{Fig-phase-diag})
connecting the two, which corresponds to a unique solution among an infinite set of 
trajectories and which provides the initial conditions of the downstream flow. As 
shown by \citet{Roberge1990}, the solution can be found considering its two main 
properties: firstly, it necessarily ends at the downstream sonic point; secondly,
it separates the phase diagram into two spaces where any perturbation on either the
neutral or the ion velocities grows exponentially towards an unphysical solution
(see Fig. \ref{Fig-phase-diag}). In the following we define as $+$ modes the 
trajectories leading to a change of sign of $d v_n/dz$ and as $-$ modes those 
leading to a change of sign of $d v_i/dz$.

With all these considerations, we built a convoluted algorithm based on a single 
precision parameter $\varepsilon$ which works as follows:
\begin{enumerate}
\item The upstream trajectory (i) is computed starting with the initial conditions of 
the flow. If the neutral fluid stays supersonic, the structure is a C-type shock 
and the computation is carried out as usual. If the neutral velocity gets close to the 
sound speed, the structure is either a CJ-type or C*-type shock and step 2 is activated.
\item The upstream trajectory is scanned to identify all the possible positions of an
adiabatic jump (black points in Fig. \ref{Fig-phase-diag}), i.e. the positions where 
applying the Rankine-Hugoniot relations conserves the signs of $d v_n/dz$ and 
$d v_i/dz$ before and after the jump (see \citealt{Roberge1990} for more details). 
Potential trajectories are then computed, starting from different jump positions to find a $+$ and a $-$ mode. The jump positions 
are then refined by dichotomy such that the $+$ and $-$ modes converge towards 
the true stationary solution. This solution, identified when the relative velocities 
of the $+$ and $-$ modes differ by less than $\varepsilon$, is calculated as the 
average of the $+$ and $-$ trajectories. The flow then starts its subsonic excursion (ii).
\item During the subsonic excursion a shooting technique is applied where both the
ions and neutral velocities are perturbed in order to identify again a $+$ and 
a $-$ mode. These perturbations are done in the $(v_n,v_i)$ phase space, in the 
direction perpendicular to the trajectory identified in step 2 as the true trajectory,
with typical relative amplitudes smaller than $\varepsilon$. Once a $+$ 
and a $-$ modes are found, the true trajectory is computed as their average, up until
the time where they differ by more than $\varepsilon$. At which point we start again 
step 3.
\item The previous algorithm stops either when $|v_n-v_i|/V_S \leqslant \varepsilon$ or
$|v_n-c_n|/V_S \leqslant \varepsilon$, where $c_n$ is the sound speed and $V_S$ the
shock speed. In the first case we assume that the neutrals and ions recouple~;
this happens for instance in the bottom right panel of Fig. \ref{Fig-phase-diag}. 
In the second case, we estimate the distance $d$ left to reach the downstream sonic 
point and then extrapolate all variables (dynamical, thermal, and chemical) at twice 
that distance to continuously jump this point (see for instance the jump
between parts ii and iii in the bottom left panel of Fig. \ref{Fig-phase-diag}). Once 
one of these two options is selected, the downstream trajectory (iii) is calculated 
by forward integration.
\end{enumerate}

By default and for all the models presented in this paper, we use $\varepsilon = 10^{-2}$, 
with the exception of Fig. \ref{Fig-phase-diag} where a value of $10^{-3}$ is adopted.
It is worth stressing that the success or failure of the algorithm is independent of
$\varepsilon$. Changing this parameter impacts only the precision of the results and  
the computational time\footnote{For $\varepsilon = 10^{-2}$, the average computational
time of stationary CJ or C* shocks is about 30 minutes on a typical 2.7 GHz processor
with 256 kb and 6 Mb of level 1 and 2 cache memory.}. Tests of this algorithm over the 
entire grid of models (see Sect. \ref{Sect-dynamics}) shows that the procedure is stable and highly 
reliable with a success rate of about 97\%. Most of the failures correspond to the
parameter domain where stationary structures shift from C to C* shocks (i.e. at 
the interface between green and yellow panels in Fig. \ref{Fig-velo-profils}) and where
computing the solution requires an adaptative selection of $\varepsilon$.

\section{Exploration of the parameter space} \label{Append-add-figures}

\begin{figure*}[!ht]
\begin{center}
\includegraphics[width=16.0cm,trim = 2cm 11cm 2cm 2cm,angle=0]{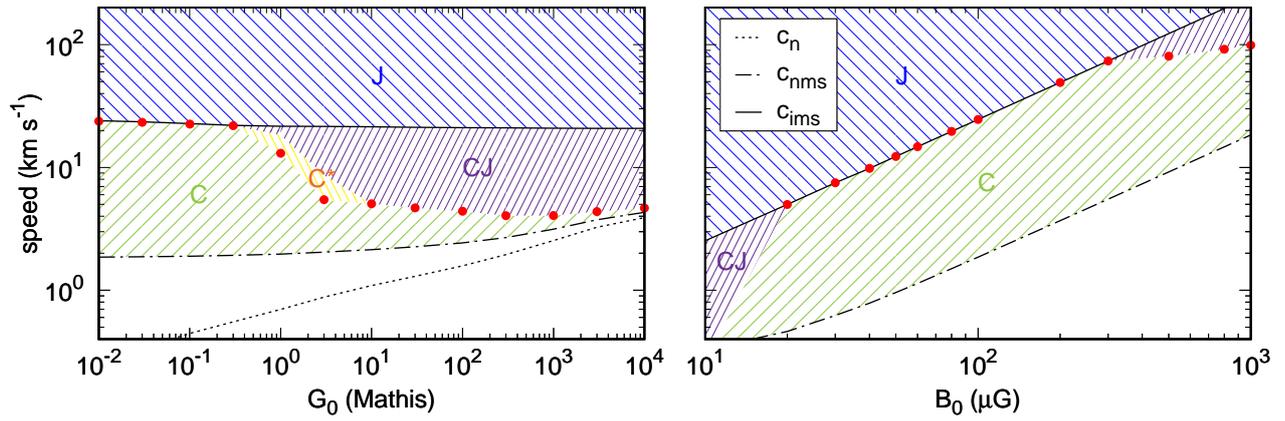}
\caption{Neutral sound speed (dotted) and neutral (dot-dashed) and ion (solid) magnetosonic 
speeds as functions of the UV radiation field in a medium of density $\dens=10^2$ \cc\ 
and a visible extinction $A_V^0 = 10^{-2}$ (upper left), and as functions of the transverse 
magnetic field intensity in a medium of density $\dens=10^4$ \cc\ and $G_0=0$ (upper right).
All other parameters are set to their standard values (see Table \ref{Tab-main}). 
The critical velocities above which C-type shocks cannot exist are shown with red points, 
while coloured areas highlight the range of velocities at which C- (green), C*- (yellow), 
CJ- (violet), and J-type (blue) shocks develop.}
\label{Fig-zermatt-add}
\end{center}
\end{figure*}

\begin{figure*}[!ht]
\begin{center}
\includegraphics[width=16.0cm,trim = 1cm 1.5cm 1cm 1.5cm,clip,angle=0]{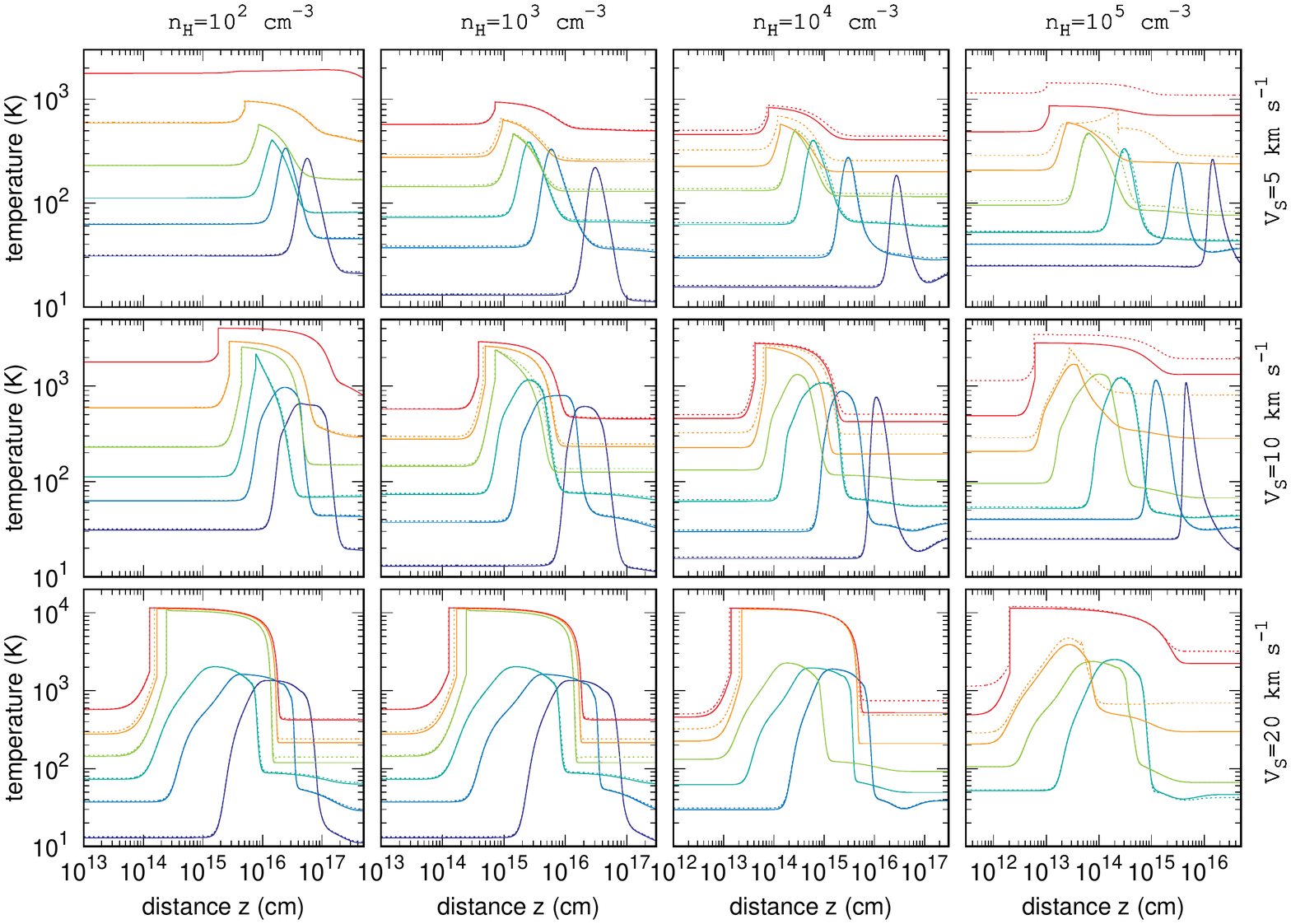}
\includegraphics[width=16.0cm,trim = 1cm 1.5cm 1cm 1.5cm,clip,angle=0]{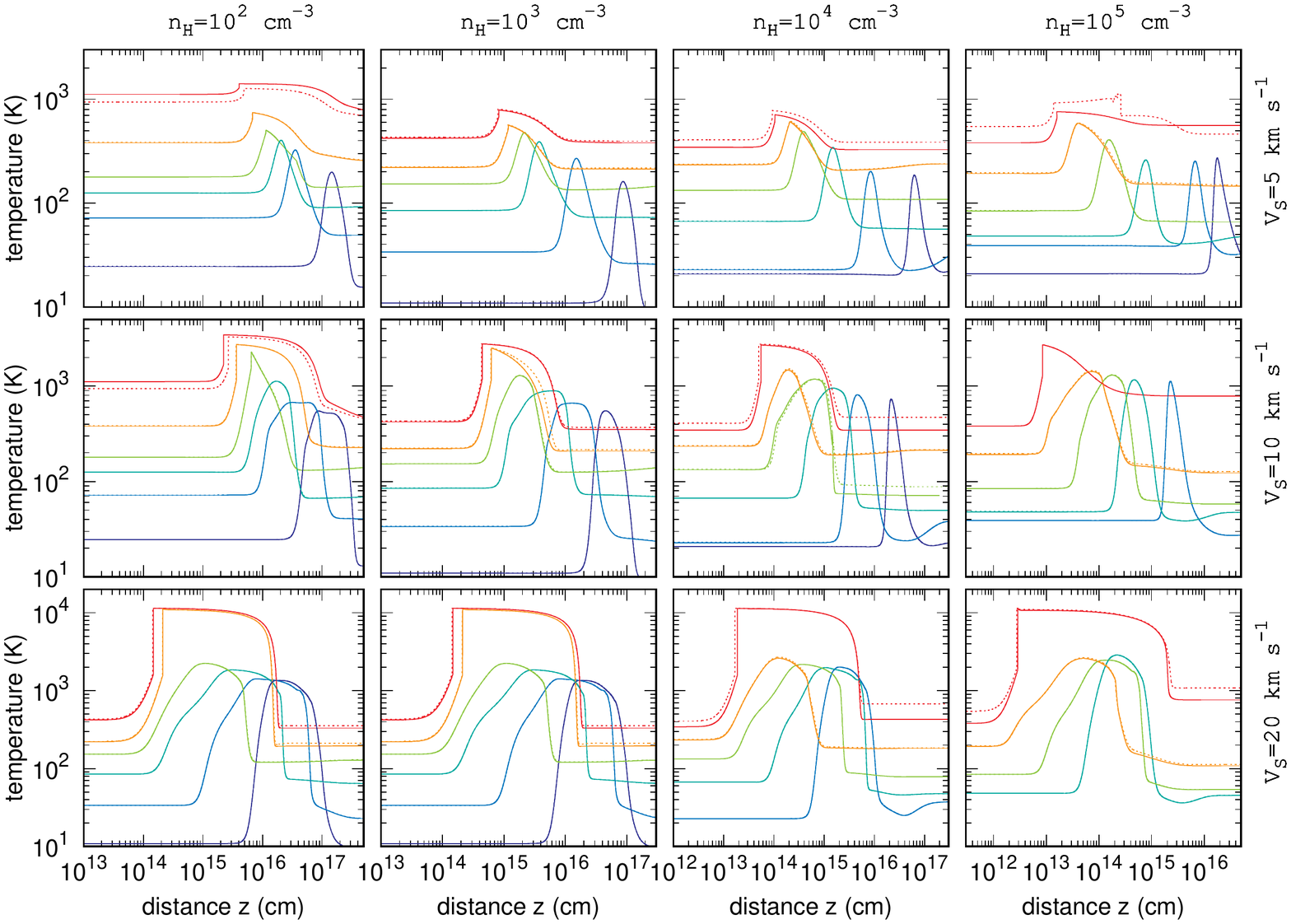}
\caption{Temperature profiles of shocks from 5 to 20 \kms\ (from top to bottom panels)
propagating in media with density ranging from $10^2$ to $10^5$ \cc\ (from left to right 
panels) and illuminated by a UV radiation field ranging from $10^{-1}$ to $10^4$ (from
colder to warmer pre-shock conditions on each panel), assuming a buffer visual extinction 
$A_V^0=10^{-2}$ (top 12 panels) and $10^{-1}$ (bottom 12 panels). All other parameters 
are set to their standard values 
(see Table \ref{Tab-main}). Models run with and without taking the UV 
pumping of the rovibrational levels of \HH\ into account
are shown in dashed and solid curves, 
respectively.}
\label{Fig-temp-add}
\end{center}
\end{figure*}

\begin{figure*}[!ht]
\begin{center}
\includegraphics[width=17.0cm,trim = 1cm 2cm 1cm 1cm,clip,angle=0]{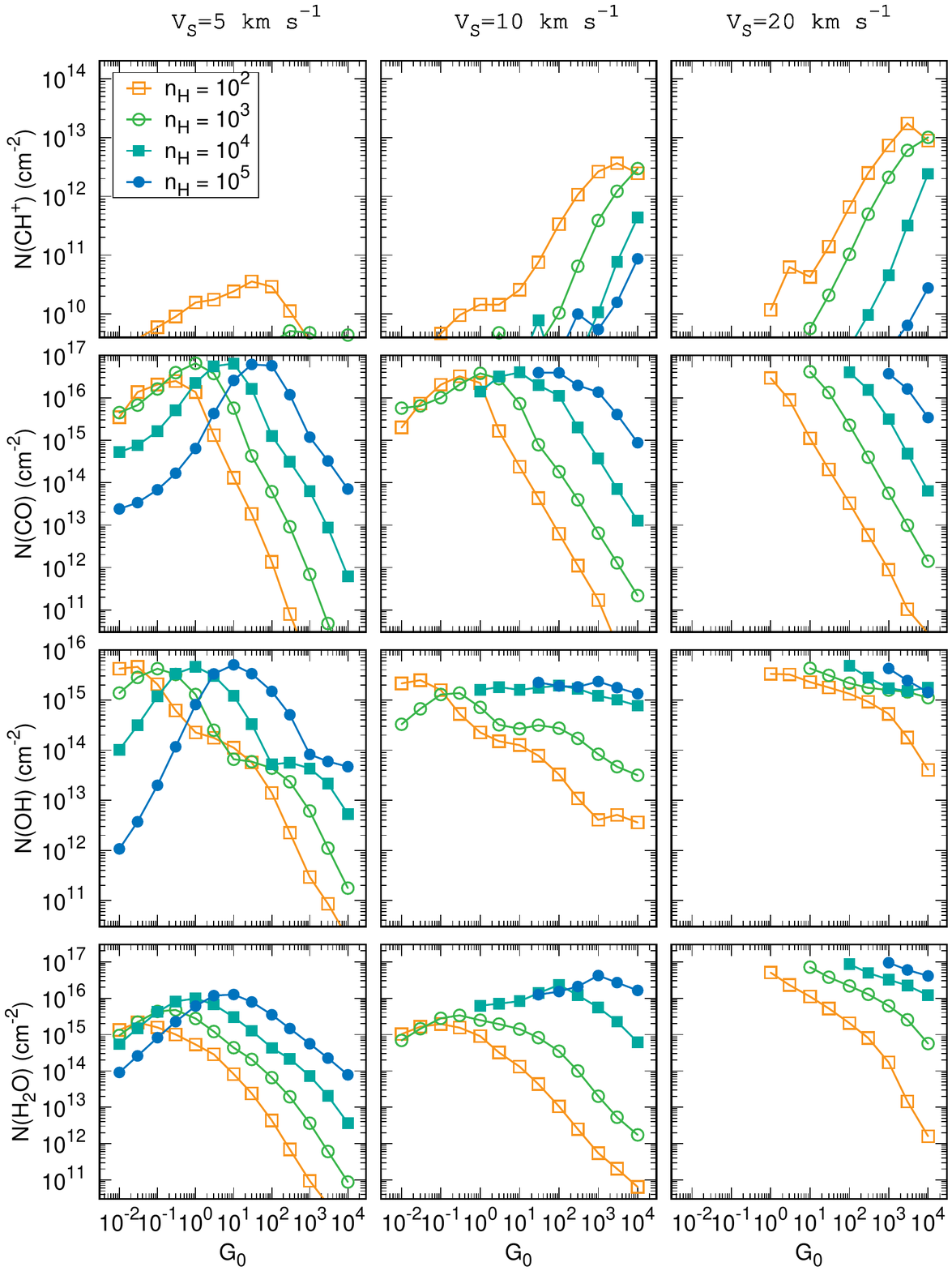}
\caption{Same as Fig. \ref{Fig-coldens-av1m1} assuming a buffer visual extinction 
$A_V^0=1$.}
\label{Fig-coldens-av1p0}
\end{center}
\end{figure*}

To be exhaustive, we regroup a few additional figures and results that 
complement those already discussed in the main text of the paper. 

Fig. \ref{Fig-zermatt-add} shows the domains of existence of C, C*, CJ, and J-type 
shocks in irradiated diffuse interstellar media (left panel) and in dense and
totally shielded environments (right panel). We recall that if no radiation field is
considered ($G_0=0$), as it is the case in the right panel of Fig. \ref{Fig-zermatt-add},
adsorption and desorption processes are switched off for the computation of the pre-shock 
properties during which we adopt the chemical composition of grain mantles given in 
Table 2 of \citet{Flower2003}.

We find that the domain of existence of C-type shocks considerably reduces in diffuse 
environments compared with that computed in dense clouds with similar irradiation 
conditions (see Fig. \ref{Fig-zermatt}). Interestingly, C-type shocks with velocity
larger than 13 \kms\ do not exist in typical diffuse clouds of density $\dens=10^2$
\cc\ with mild irradiation conditions ($G_0$=1), a limit that quickly drops with 
the strength of the radiation field. In these conditions, C-type shocks are replaced 
by CJ-type shocks, which clearly become a dominant dissipative structure in highly 
irradiated diffuse gas. The limits inferred in completely shielded environments are
found to be in good agreement with those derived with the previous version of
the Paris-Durham shock code \citep{Cabrit2004}, except at low magnetization where we 
highlight a new domain of existence of CJ-type shocks.

As a complement to Fig. \ref{Fig-velo-profils}, we represent in Fig. \ref{Fig-temp-add} 
the temperature profiles obtained in 288 models of molecular shocks covering roughly 
the entire grid of parameters explored in this work. As described in the main text,
the external UV radiation field reduces the size of the shock, which induces an
increase of the temperature peak and reveals the entwinement of the impacts of 
radiative and mechanical energies on the thermal evolution of the gas. 
We find that low velocity shocks ($V_S \leqslant 5$ \kms) propagating in highly 
illuminated environments ($G_0=10^4$) have almost no impact on the temperature of 
the gas. This result unveils the limit at which the dissipation of mechanical 
energy is virtually undetectable compared to the amount of radiative energy locally 
reprocessed by interstellar matter.

In addition to Fig. \ref{Fig-coldens-av1m1}, we finally indicate in Fig. 
\ref{Fig-coldens-av1p0} the column densities of \CHp, CO, OH, and H$_2$O  
computed across molecular shocks in well-shielded environments ($A_V^0=1$). 
The results obtained in these conditions are found to be similar to those
obtained at lower visual extinction for lower irradiation conditions (Fig. 
\ref{Fig-coldens-av1m1}). In particular, the strong increase of \CHp\
column densities and the strong decrease of those of CO, OH, and H$_2$O 
in highly irradiated environments is a general result that does not depend 
on the extinction considered.

\end{document}